\documentclass[fleqn,usenatbib]{mnras}

\usepackage[T1]{fontenc}

\DeclareRobustCommand{\VAN}[3]{#2}
\let\VANthebibliography\thebibliography
\def\thebibliography{\DeclareRobustCommand{\VAN}[3]{##3}\VANthebibliography}

\usepackage{graphicx}	
\usepackage{amsmath}	
\usepackage{amssymb}	

\usepackage{color}

\usepackage{url}

\usepackage{float}

\usepackage{placeins}

\title[Six New Triply Eclipsing \textit{TESS} Triples]{Six New Compact Triply Eclipsing Triples Found With \textit{TESS}}

\author[Rappaport et al.]{
S.\,A~Rappaport$^1$\thanks{E-mail: sar@mit.edu},
T. Borkovits$^{2,3,4,5}$,
R.~Gagliano$^{6}$,
T.\,L.~Jacobs$^{7}$,
V.\,B. Kostov$^{8,9}$,
\newauthor
B.\,P. Powell$^8$,
I.~Terentev$^{10}$,
M.\,Omohundro$^{10}$,
G.\,Torres$^{11}$,
A.\,Vanderburg$^{1}$,
\newauthor
T.\,Mitnyan$^{2,12}$,
M.\,H.\,Kristiansen$^{13,14}$,
D.\,LaCourse$^{15}$,
H.\,M.\,Schwengeler$^{10}$,
\newauthor
T. G. Kaye$^{16}$, 
A.\,P\'al$^{3,17}$,
T. Pribulla$^{18}$,
I. B. B\'\i r\'o$^2$, 
I. Cs\'anyi$^2$,
Z. Garai$^{4,5,18}$, 
\newauthor
P.~Zasche$^{19}$,
P.\,F.\,L.\,Maxted$^{20}$,
J.\,E.\,Rodriguez$^{21}$,
D.\,J.\,Stevens$^{22,23}$ \\
$^1$ Department of Physics, Kavli Institute for Astrophysics and Space Research, M.I.T., Cambridge, MA 02139, USA\\
$^{2}$Baja Astronomical Observatory of University of Szeged, H-6500 Baja, Szegedi \'ut, Kt. 766, Hungary\\
$^{3}$Konkoly Observatory, Research Centre for Astronomy and Earth Sciences,  H-1121 Budapest, Konkoly Thege Mikl\'os \'ut 15-17, Hungary\\
$^4$ ELTE E{\"o}tv{\"o}s Lor\'and University, Gothard Astrophysical Observatory, Szent Imre h. u. 112, 9700 Szombathely, Hungary \\
$^{5}$ MTA-ELTE Exoplanet Research Group, H-9700 Szombathely, Szent Imre h. u. 112, Hungary \\
$^{6}$ Amateur Astronomer, Glendale, AZ 85308 \\
$^{7}$ Amateur Astronomer, 12812 SE 69th Place Bellevue, WA 98006, USA \\
$^8$ NASA Goddard Space Flight Center, 8800 Greenbelt Road, Greenbelt, MD 20771, USA \\
$^9$ SETI Institute, 189 Bernardo Avenue, Suite 200, Mountain View, CA 94043, USA\\
$^{10}$ Citizen Scientist, c/o Zooniverse, Dept,~of Physics, University of Oxford, Denys Wilkinson Building, Keble Road, Oxford, OX1 3RH, UK \\
$^{11}$ Center for Astrophysics $|$ Harvard \& Smithsonian, 60 Garden St., Cambridge, MA 02138, USA \\
$^{12}$ Department of Experimental Physics, University of Szeged, 6720 Szeged, 
D\'om t\'er 9, Hungary \\
$^{13}$ Brorfelde Observatory, Observator Gyldenkernes Vej 7, DK-4340 T\o ll\o se, Denmark \\
$^{14}$ National Space Institute, Technical University of Denmark, Elektrovej 327, DK-2800 Lyngby, Denmark \\
$^{15}$ Amateur Astronomer, 7507 52nd Place NE Marysville, WA 98270, USA \\
$^{16}$ Foundation for Scientific Advancement, Patterson Observatory, AZ \\
$^{17}$ Kavli Institute for Astrophysics and Space Research, M.I.T., Cambridge, MA 02139, USA \\
$^{18}$ Astronomical Institute, Slovak Academy of Sciences, 05960 Tatransk\'a Lomnica, Slovakia \\
$^{19}$ Astronomical Institute, Charles University, Faculty of Mathematics and Physics, V Hole\v{s}ovi\v{c}k\'{a}ch 2, CZ-180 00, Praha 8, Czech Republic\\
$^{20}$ Astrophysics Group, Keele University, Staffordshire, ST5 5BG, UK\\ 
$^{21}$ Department of Physics and Astronomy, Michigan State University, East Lansing, MI 48824, USA \\
$^{22}$ Department of Astronomy \& Astrophysics, The Pennsylvania State University, 525 Davey Lab, University Park, PA 16802, USA \\
$^{23}$ Center for Exoplanets and Habitable Worlds, The Pennsylvania State University, 525 Davey Lab, University Park, PA 16802, USA\\ }
\date{Accepted XXX. Received YYY; in original form ZZZ}

\pubyear{2022}

\begin{document}
\label{firstpage}
\pagerange{\pageref{firstpage}--\pageref{lastpage}}
\maketitle

\begin{abstract}
 In this work we report the discovery and analysis of six new compact triply eclipsing triple star systems found with the {\em TESS} mission: TICs 37743815, 42565581, 54060695, 178010808, 242132789, and 456194776.  All of these exhibit distinct third body eclipses where the inner eclipsing binary (EB) occults the third (`tertiary') star, or vice versa.  We utilized the {\it TESS} photometry, archival photometric data, and available archival spectral energy distribution curves (SED) to solve for the properties of all three stars, as well as many of the orbital elements.  We describe in detail our SED fits, search of the archival data for the outer orbital period, and the final global photodynamical analyses.  From these analyses we find that all six systems are coplanar to within $0^\circ - 5^\circ$, and are viewed nearly edge on (i.e., within a couple of degrees).  The outer orbital periods and eccentricities of the six systems are \{$P_{\rm out}$ (days),\,$e$\}:  \{68.7, 0.36\}, \{123, 0.16\}, \{60.7, 0.01\}, \{69.0, 0.29\}, \{41.5, 0.01\}, \{93.9, 0.29\}, respectively, in the order the sources are listed above. The masses of all 12 EB stars were in the range of 0.7-1.8 M$_\odot$ and were situated near the main sequence.  By contrast, the masses and radii of the tertiary stars ranged from 1.5-2.3 M$_\odot$ and 2.9-12 R$_\odot$, respectively. We use this information to estimate the occurrence rate of compact flat triple systems.
\end{abstract}




\section{Introduction}
\label{sect_intro}

With the advent of long-term, wide field, precision photometry from space with such missions as {\it Kepler} \citep{borucki10}, {\it K2} \citep{howell14}, and {\it TESS} \citep{ricker15}, it has become relatively easy to discover triply eclipsing triple star systems.  These are often found when an extra, isolated pair of eclipses appear in the lightcurve of an ordinary eclipsing binary (EB), or a long exotic-looking extra eclipse appears that cannot be produced in a simple binary (see the recent extensive review of \citealt{borkovits22}).  We refer to these as `third-body' events where either the EB occults the third star (hereafter, the `tertiary') in its outer orbit, or vice versa.  Typical eclipse periods for the inner EBs are days, while the period for the extra third-body eclipses range from a month to about a year.  When one of these triples is found, no further vetting of the object is typically needed before concluding that this is a bound triple system (or possible a higher stellar multiple).  By contrast, when a pair of EBs is found in the same photometric aperture, it is not immediately clear whether the two EBs are physically bound or are simply close together in the sky by chance (see the quadruples catalog of \citealt{kostov22}), and further vetting in the form of, e.g., radial velocity measurements (RVs) or eclipse timing variations (ETVs) is required.

Once a compact triple system has been identified via its third body eclipses, several additional characteristics can quickly become apparent about the system.  First, if more than one outer eclipse of the same type\footnote{As in ordinary EBs, outer eclipses come in two flavors---primary and secondary eclipses.} are seen in succession, then the outer orbital period of the triple is immediately revealed.  If both the primary and secondary outer eclipses are seen, then, just as in an ordinary EB, the quantity $e_{\rm out} \cos \omega_{\rm out}$ can be measured, where $e_{\rm out}$ and $\omega_{\rm out}$ are the eccentricity and argument of periastron of the outer orbit, respectively.  Finally, the presence of both inner and outer eclipses gives some good indication that the binary orbital plane and the outer orbital plane are at least somewhat aligned (i.e., the systems tend to be `flat') or else the likelihood of detecting both sets of eclipses is relatively lower.

As we and others have shown in a number of previous papers \citep[see, e.~g.][]{carter11,borkovitsetal13,masudaetal15,orosz15,alonso15,borkovitsetal19a,borkovitsetal20b,borkovitsetal22}, the outer eclipses have encoded in them a substantial amount of information which, when combined with supplementary data, can ultimately lead to a complete description of the stellar properties (masses, radii, $T_{\rm eff}$, age, and metallicity) as well as the complete orbital configuration of the system.  The supplemental material can involve the EB lightcurve itself, the extracted ETV curve, spectral energy distribution (SED) measurements from archival surveys, ground-based photometric surveys, and RVs.

Another feature of compact triple star systems that makes them fascinating objects to study is the relatively short timescales for dynamical interactions.  These can occur over a year, a few months, or even weeks.  Interesting effects to look for include dynamical as well as light-travel time delays in the ETVs of the EBs, forced apsidal motion in the EB, orbital plane precession if the two orbital planes are misaligned, and even large amplitude von Zeipel-Kozai-Lidov cycles (\citealt{vonzeipel910}; \citealt{lidov962}; \citealt{kozai962}) in the case of strongly misaligned orbital planes.

Triple star systems are also interesting in terms of the insight they provide about the formation and subsequent evolution of multistellar systems (see, e.g., \citealt{tokovinin21}; \citealt{borkovits22}; Sect.~\ref{sec:summary}).  They are the next simplest entity after binary star systems.  In some ways they are analogous to studying He atoms after mastering H atoms. However, while there are more than a million eclipsing binary systems known (see, e.g., Sect.~\ref{sec:discovery}; \citealt{powell21}; E. Kruse, 2022 in preparation), the number of triply eclipsing triple systems in the literature, is currently under 20.

Here we present the discovery and detailed analyses of six new compact triply eclipsing triple star systems.  In Section \ref{sec:discovery} we discuss how the discovery of the third body events were made using the {\em TESS} data, and present plots of the third body events.  Archival spectral energy distributions (SED) are then used in Section \ref{sec:SED} to make first estimates of the constituent stellar masses, radii, and $T_{\rm eff}$.  We then use archival photometric data from a number of ground-based surveys to determine the outer orbital period of the triples via the third-body eclipses (see Sect.~\ref{sec:outer_orbit}). The detailed photodynamical model by which we analyze jointly the photometric lightcurves, eclipse timing variations, and spectral energy distributions is reviewed in Section \ref{sec:photodynamical}.  The system parameters for each of the six triple systems are presented in Section \ref{sec:results} in the form of  comprehensive tables, including  extracted masses, radii, and effective temperatures, as well as the orbital parameters for both the inner and outer orbits.  We summarize our results and discuss a few of the salient findings from our study in Section \ref{sec:summary}. We also discuss how our compact triple systems may inform us about multistellar formation and evolution.

\begin{table*}
\centering
\caption{Main properties of the six triple systems from different catalogs}
\begin{tabular}{lcccccc}
\hline
\hline
Parameter & TIC 37743815 & TIC 42565581 & TIC 54060695 & TIC 178010808 & TIC 242132789 & TIC 456194776  \\
\hline
RA (J2000) & $06:15:28.89$ & $06:26:37.58$ & $06:56:14.83$ & $07:33:05.27$ & $06:14:26.95$ & $03:28:29.41$ \\  
Dec (J2000)& $-29:39:12.14$ & $-03:23:50.66$ & $-25:25:14.49$ & $-04:23:20.36$ & $-04:08:12.48$ & $43:36:44.56$ \\  
$T^a$ & $12.928\pm0.007$  & $12.867\pm0.014$ & $12.132\pm0.022$ & $12.132\pm0.007$  & $12.636\pm0.009$ & $11.766\pm0.008$ \\
$G^b$ & $13.479\pm0.001$  & $13.433\pm0.001$ & $12.679\pm0.001$ & $12.531\pm0.000$  & $13.492\pm0.001$ & $12.243\pm0.001$ \\
$G_{\rm BP}^b$ & $13.939\pm0.003$ & $13.955\pm0.005$ & $13.123\pm0.003$ & $12.837\pm0.001$ & $14.325\pm0.004$ & $12.620\pm0.002$ \\
$G_{\rm RP}^b$ & $12.859\pm0.002$ & $12.748\pm0.002$ & $12.060\pm0.002$ & $12.062\pm0.001$ & $12.582\pm0.003$ & $11.690\pm0.001$ \\
B$^a$ & $14.639 \pm 0.035$ & $14.626\pm0.044$ & $13.755\pm0.112$ & $13.320\pm0.071$ & $15.454\pm0.073$ & $12.801\pm0.396$ \\
V$^c$ & $13.767 \pm 0.126$ & $14.010\pm0.183$ & $12.884\pm0.080$ & $12.649\pm0.069$ & $13.869\pm0.103$ & $11.997\pm0.029$ \\
J$^d$ & $12.075 \pm 0.023$ & $11.794\pm0.023$ & $11.254\pm0.023$ & $11.544\pm0.024$ & $11.211\pm0.021$ & $11.069\pm0.025$ \\
H$^d$ & $11.664 \pm 0.027$ & $11.335\pm0.025$ & $10.836\pm0.023$ & $11.305\pm0.025$ & $10.560\pm0.023$ & $10.777\pm0.029$ \\
K$^d$ & $11.527 \pm 0.021$ & $11.158\pm0.021$ & $10.735\pm0.023$ & $11.224\pm0.025$ & $10.347\pm0.022$ & $10.705\pm0.022$ \\
W1$^e$ & $11.476\pm0.023$  & $11.102\pm0.023$ & $10.661\pm0.023$ & $11.221\pm0.023$  & $10.223\pm0.022$ & $10.644\pm0.023$  \\
W2$^e$ & $11.497\pm0.021$  & $11.149\pm0.020$ & $10.706\pm0.020$ & $11.241\pm0.021$  & $10.259\pm0.020$ & $10.647\pm0.021$ \\
W3$^e$ & $11.463\pm0.149$  & $11.266\pm0.131$ & $10.597\pm0.087$ & $11.247\pm0.140$  & $10.270\pm0.089$ & $10.617\pm0.099$ \\
$T_{\rm eff}$ (K)$^b$ & $5039\pm100$ & $4895\pm175$ & $5029\pm220$ & $5862\pm30$ & $4015\pm135$ & $5375\pm550$ \\
$T_{\rm eff}$ (K)$^a$ & $5135\pm125$ & $5374\pm135$ & $5660\pm125$ & $6090\pm126$ & $4593\pm123$ & $6690\pm62$ \\
$R$ $({\rm R}_\odot)^b$ & $4.73\pm0.18$ & $8.43\pm0.60$ & $8.96\pm0.68$ & $4.12\pm0.04$ & $14.5\pm0.9$ & $7.06\pm1.30$ \\
$R$ $({\rm R}_\odot)^a$ & $4.46\pm NA$ & $7.97\pm NA$ & $7.79\pm NA$ & $3.85\pm0.28$ & $13.3\pm NA$ & $5.14\pm NA$ \\
Distance (pc)$^f$ & $1857\pm39$ & $3281\pm160$ & $2221\pm50$ & $1464\pm30$ & $3258\pm165$ & $1590\pm40$ \\ 
$E(B-V)^a$ & $0.036\pm 0.006$ & $0.213\pm0.017$ & $0.168\pm 0.033$ & $0.046 \pm0.006$ & $0.336\pm0.011$ & $0.167\pm NA$ \\
$\mu_\alpha$ (mas ~${\rm yr}^{-1}$)$^b$ & $-0.13\pm0.01$ & $+0.81\pm0.02$ & $-2.32\pm0.01$ & $-2.03\pm0.01$ & $0.63\pm0.02$ & $-0.18\pm0.02$ \\ 
$\mu_\delta$ (mas ~${\rm yr}^{-1}$)$^b$ &  $+7.58\pm0.01$ & $-1.03\pm0.02$ & $+3.26\pm0.01$ &  $-1.17\pm0.01$ & $-1.31\pm0.02$ & $-2.74\pm0.01$ \\ 
\hline
\label{tbl:mags}  
\end{tabular}

\textit{Notes.}  (a) {\it TESS} Input Catalog (TIC v8.2) \citep{TIC8}. (b) Gaia EDR3 \citep{GaiaEDR3}; the uncertainty in $T_{\rm eff}$ and $R$ listed here is 1.5 times the geometric mean of the upper and lower error bars cited in DR2. (c) AAVSO Photometric All Sky Survey (APASS) DR9, \citep{APASS}, \url{http://vizier.u-strasbg.fr/viz-bin/VizieR?-source=II/336/apass9}. (d) 2MASS catalog \citep{2MASS}.  (e) WISE point source catalog \citep{WISE}. (f) \citet{bailer-jonesetal21}.  (g) \url{http://argonaut.skymaps.info/query}\\
\end{table*}

\begin{table}
\centering
\caption{{\it TESS} Observation Sectors for the Triples}
\begin{tabular}{lcc}
\hline
\hline
Object & Sectors Observed & Third Body Events  \\
\hline
TIC 37743815 & S6 \& S33  & S6 \& S33  \\
TIC 42565581 & S6 \& S33 & S6 \& S33  \\ 
TIC 54060695 & S6 \& S7 \& S33 & S6 \& S7 \& S33  \\
TIC 178010808 & S7 \& S34 & S7  \\
TIC 242132789 & S6 \& S33 & S6 \& S33  \\
TIC 456194776 & S18  & S18  \\
\hline
\label{tbl:sectors}  
\end{tabular}

\end{table}

\section{Discovery of Triply Eclipsing Triples with {\it TESS}}
\label{sec:discovery}

Our `Visual Survey Group' (VSG; \citealt{kristiansen22}) continues to search for multi-stellar systems in the \textit{TESS} lightcurves.  We estimate that, thus far, we have visually inspected some 10 million lightcurves from {\em TESS}.  Such visual searches are a complement to more automated ones using machine learning algorithms (see, e.g., \citealt{powell21}; \citealt{kostovetal21}; \citealt{kostov22}).  The lightcurves are displayed with Allan Schmitt's {\tt LcTools} and {\tt LcViewer} software \citep{schmitt19}, which allows for an inspection of a typical lightcurve in just $\sim$5 seconds.  It is important to note that 9 million of the studied lightcurves were of anonymous stars, while 1 million lightcurves were of preselected eclipsing binaries that were found in the {\it TESS} data via machine learning searches (see \citealt{powell21}; E. Kruse, 2022 in preparation).

For our survey work we largely made use of lightcurves from the following sources: Science Processing Operations Center (SPOC, \citealt{jenkins16}); the Difference Imaging Pipeline \citep{oelkers18}; the PSF-based Approach to TESS High quality data Of Stellar clusters (PATHOS, \citealt{nardiello19}); the Cluster Difference Imaging Photometric Survey (CDIPS, \citealt{bouma19}); the MIT Quick Look Pipeline (QLP, \citealt{huang20}); the TESS Image CAlibrator Full Frame Images (TICA, \citealt{fausnaugh20}); and the Goddard Space Flight Center (GSFC, see Sect. 2, \citealt{powell21}). 

The first signatures that are looked for in terms of identifying triply eclipsing triples are an eclipsing binary lightcurve with an additional strangely shaped extra eclipse or rapid succession of isolated eclipses.  One gratifying aspect of finding triply eclipsing triples is that they are in a sense `self-vetted'.  In particular, there is no way for a single binary, or sets of independent stars or binaries to produce such `extra' eclipsing events.  Therefore, additional vetting becomes largely unnecessary in proving that these are indeed triples (or possibly higher order multiples).

While searching through the lightcurves obtained from the first three full years of {\em TESS} observations we have found more than $\sim$50 of these triply eclipsing triples.  Of these we have determined the outer orbital period for 20 of them. We have previously reported on four of these systems (\citealt{borkovitsetal20b}; \citealt{borkovitsetal22}).  He we present the discovery and analysis of six new triply eclipsing triples from among this set: TIC 37743815, TIC 4256558, TIC 54060695, TIC 178010808, TIC 242132789, and TIC 456194776. (See Table~\ref{tbl:mags} for the main catalog data of the targets.)

\begin{figure*}
\begin{center}
\includegraphics[width=0.45 \textwidth]{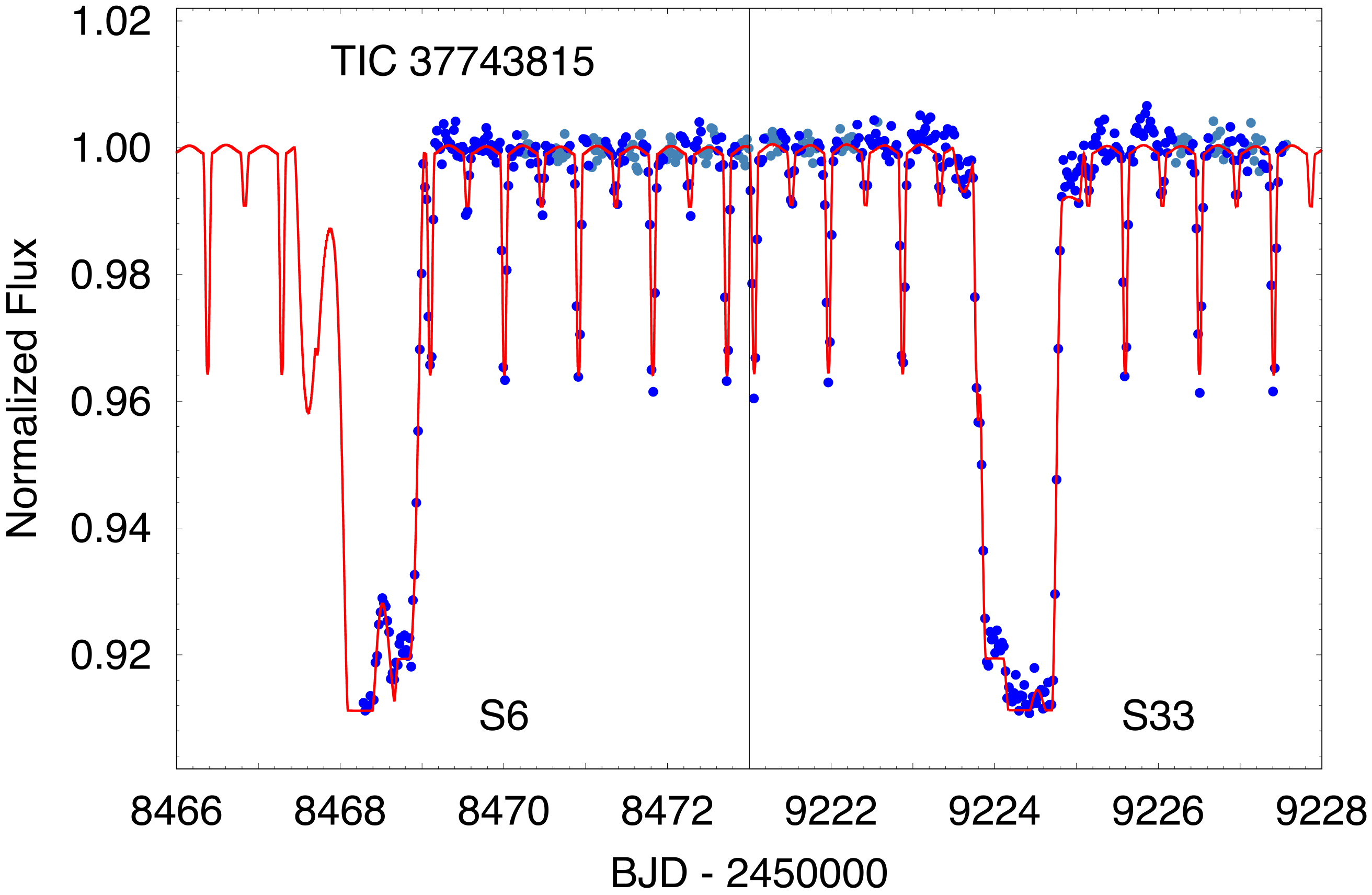} \hglue0.3cm
\includegraphics[width=0.45 \textwidth]{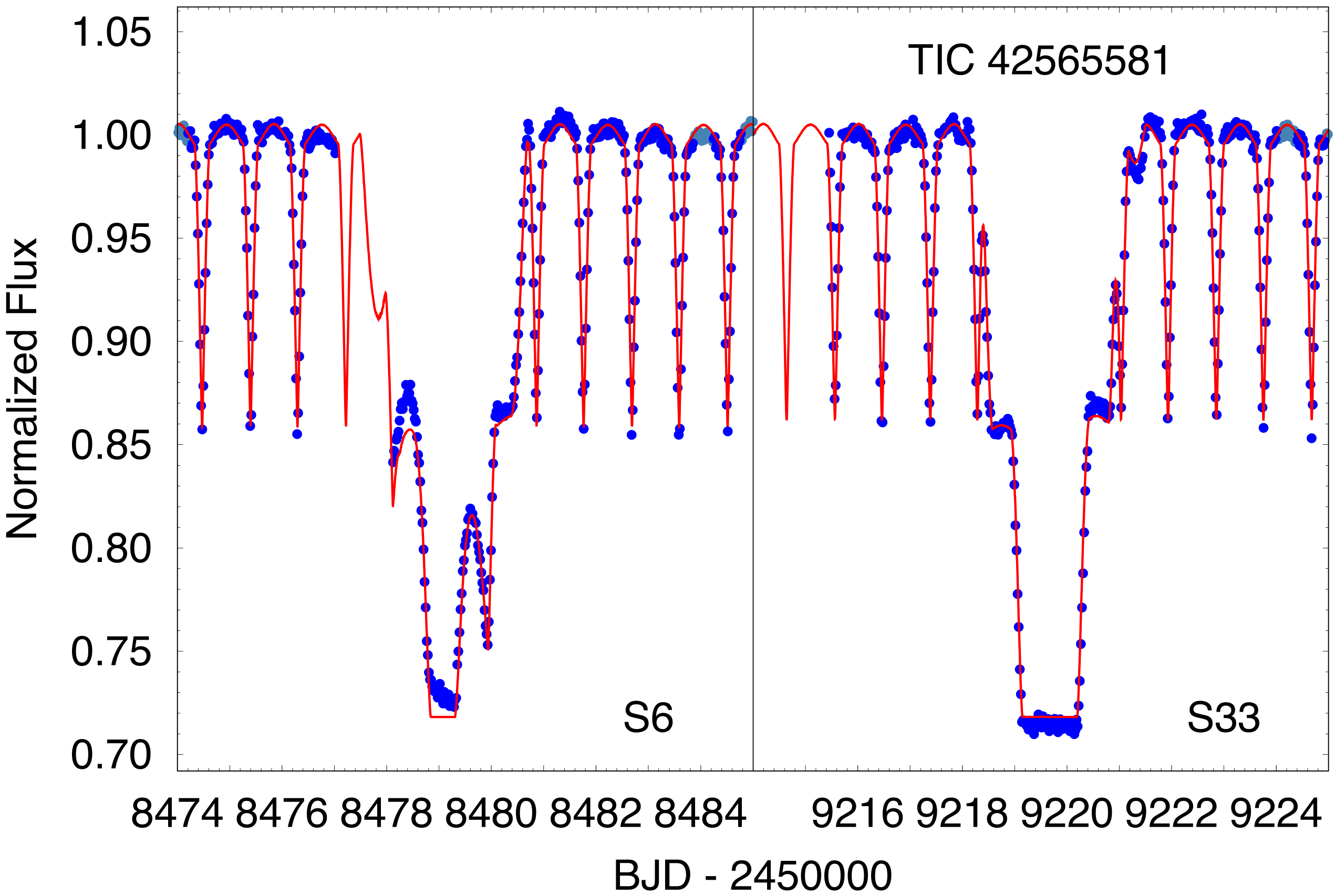}  \vglue0.2cm  
\includegraphics[width=0.44 \textwidth]{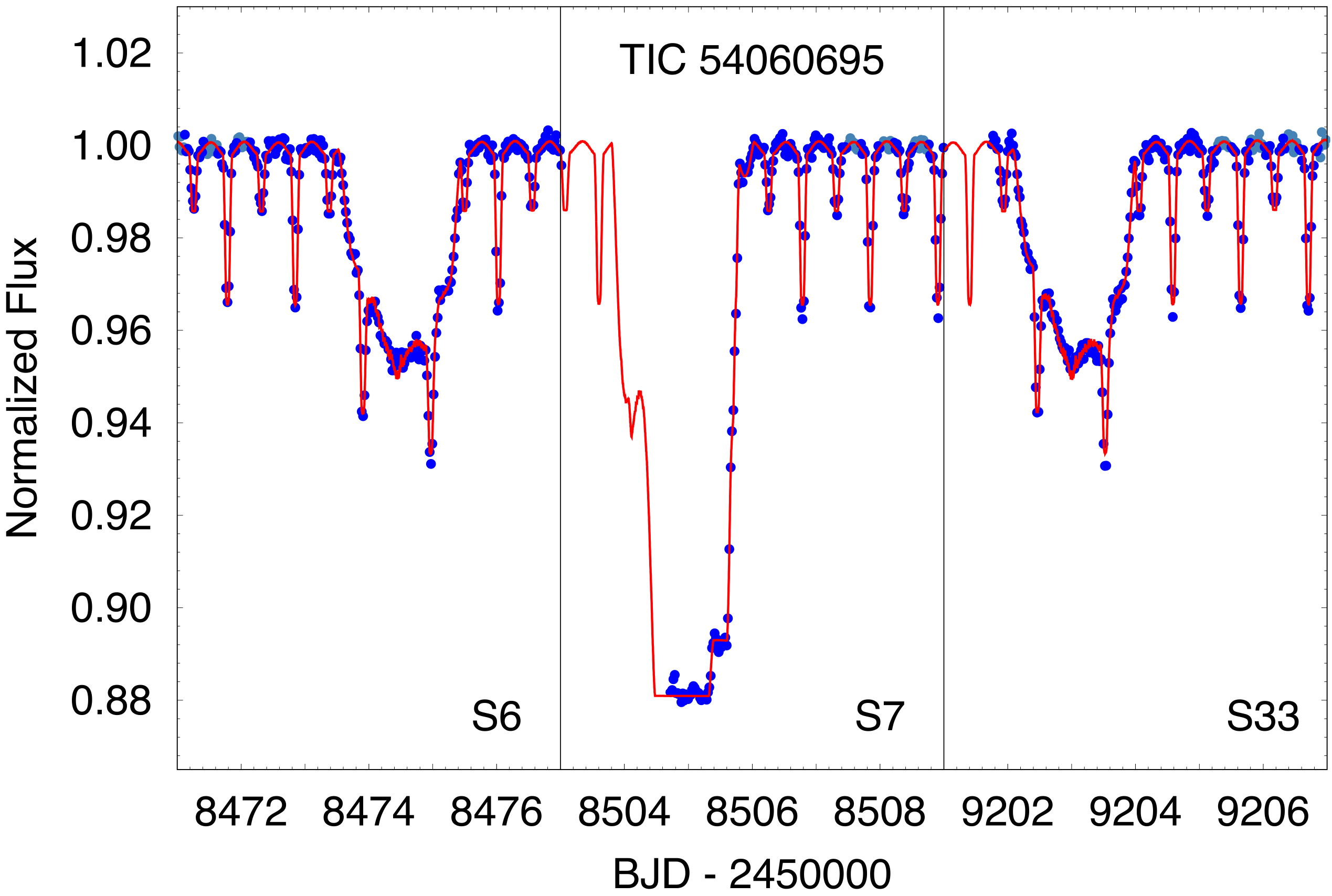} \hglue0.3cm
\includegraphics[width=0.47 \textwidth]{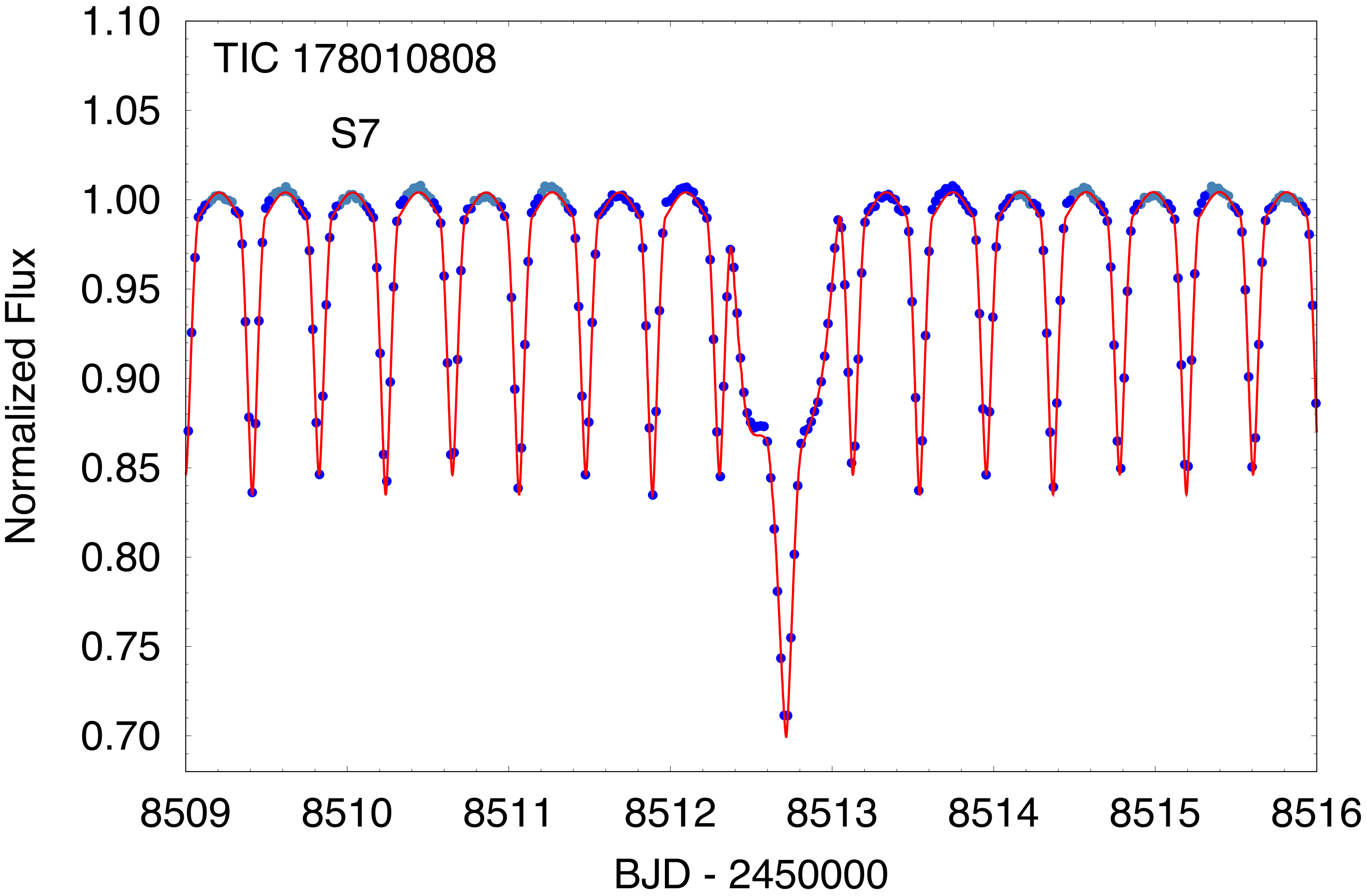} \vglue0.2cm
\includegraphics[width=0.45 \textwidth]{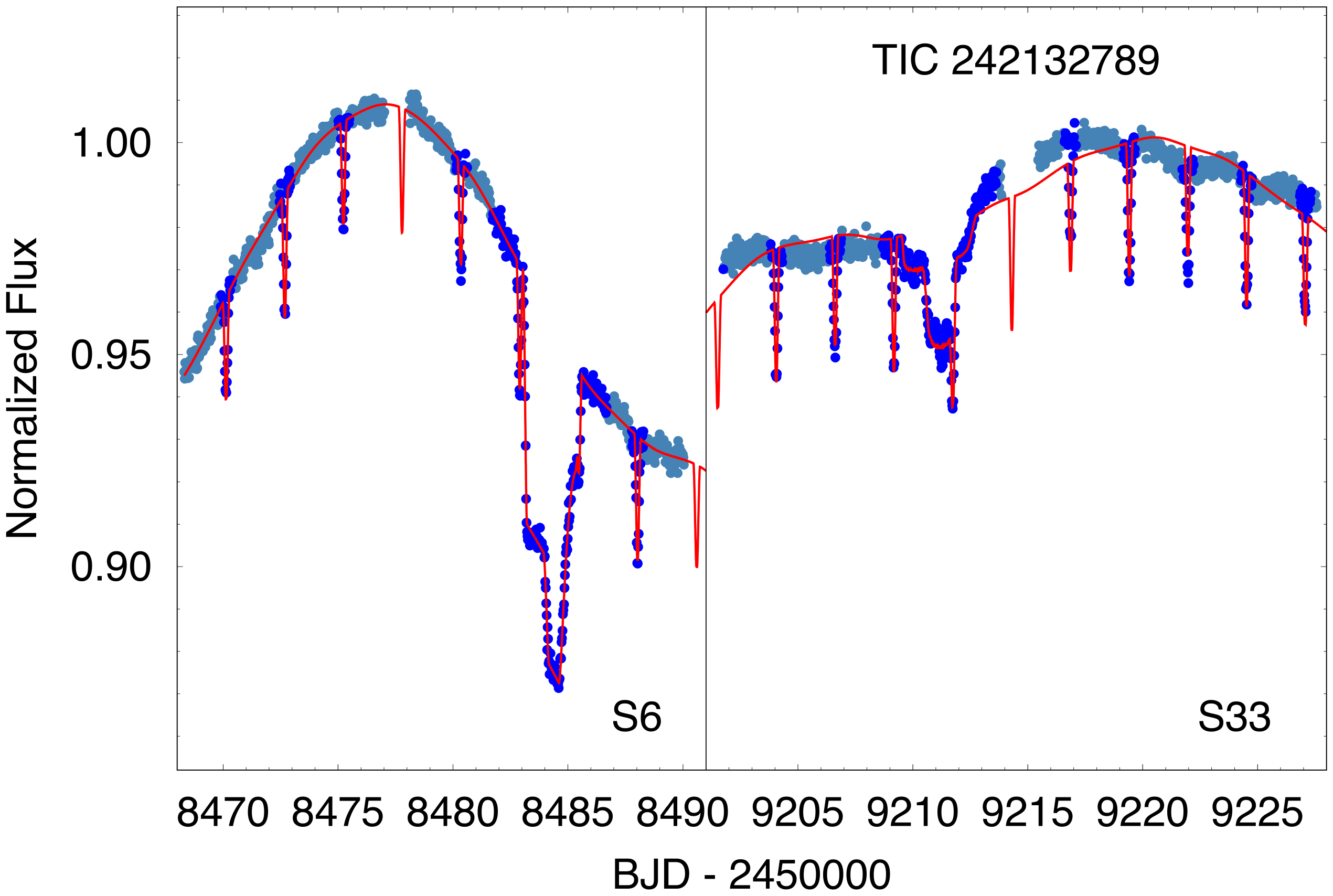} \hglue0.1cm  
\includegraphics[width=0.47 \textwidth]{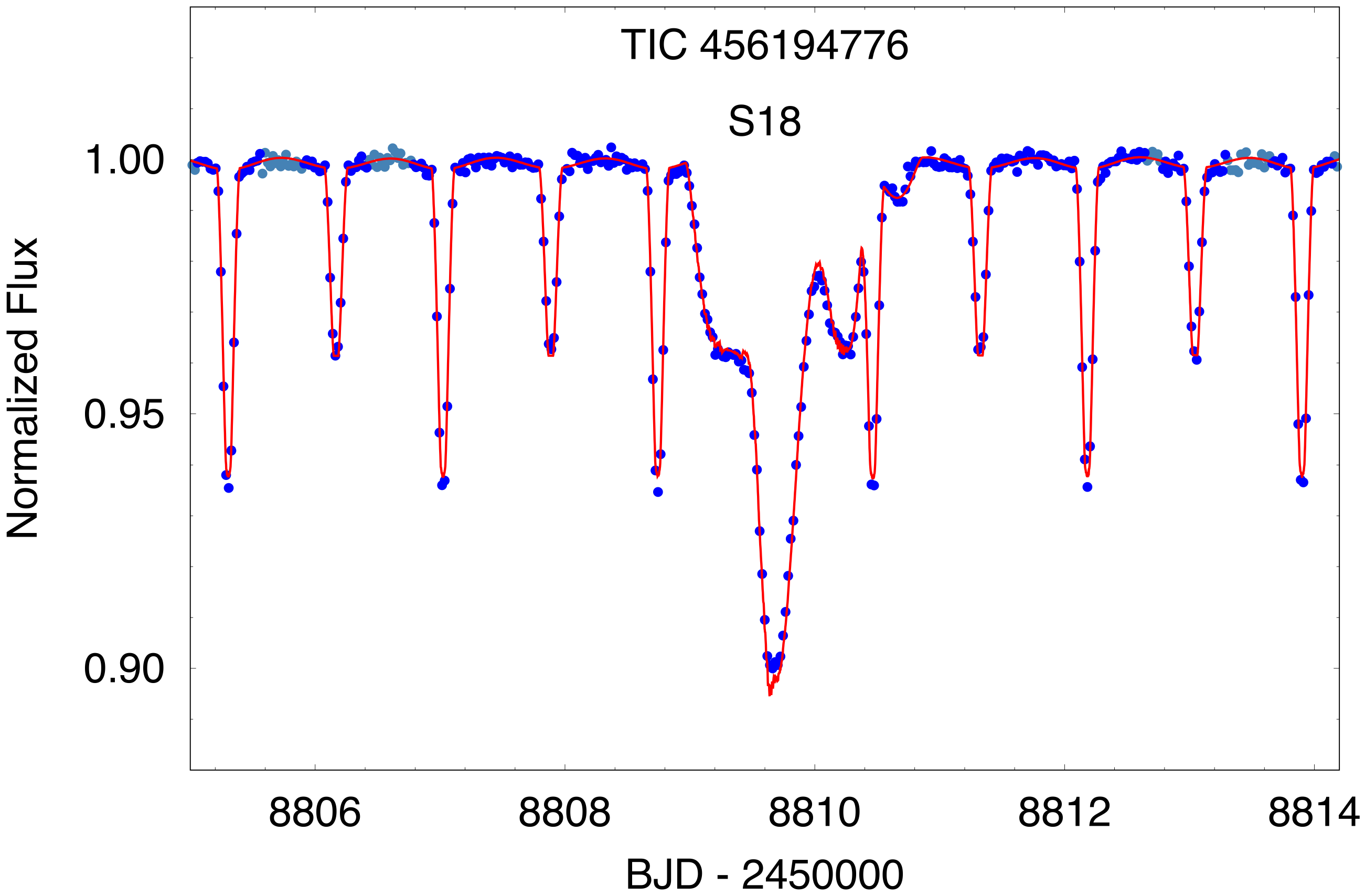} 

\caption{{\it TESS} third body lightcurves.  We present a portion of a sector's lightcurve for each source containing the third body event that led to their discoveries.  For two of the sources there is only a single third body event that was detected, while in the other four cases we show portions of two or three orbits which exhibited third body events.  The overplotted model lightcurves are discussed in Sect.~\ref{sec:photodynamical}. The lighter blue points in the out-of-eclipse region were omitted from the photodynamical fits to save computation time.}
\label{fig:triples}
\end{center}
\end{figure*}

All six targets were measured in full frame images from {\it TESS} with either 30-min or 10-min cadence.  A portion of the {\it TESS} lightcurves for all six sources are shown in Fig.\ref{fig:triples}. The sectors during which these sources were observed with {\it TESS} are summarized in Table \ref{tbl:sectors}. TICs 37743815, 42565581, 178010808, and 242132789 were observed during two widely separated sectors each, and a third body event was observed for each source in both sectors, except for TIC 178010808, for which only one third-body event was detected.  TIC 456194776 was observed in only one sector.  Finally, TIC 54060695 was observed during three sectors, with a third body event detected in each---two secondary and one primary outer eclipses.  

Given that the sectors are only approximately a month long and the outer orbital periods range from 42 to 123 days, it is somewhat fortuitous that we managed to observe third body events in nearly all the sectors.  However, most of these systems exhibit two eclipses per outer orbit, and we have selected these for presentation in this work precisely because the outer periods are relatively short and therefore easy to detect even in archival data sets.  In other words, there are several selection effects at work here.

\begin{figure*}
\begin{center}
\includegraphics[width=0.44 \textwidth]{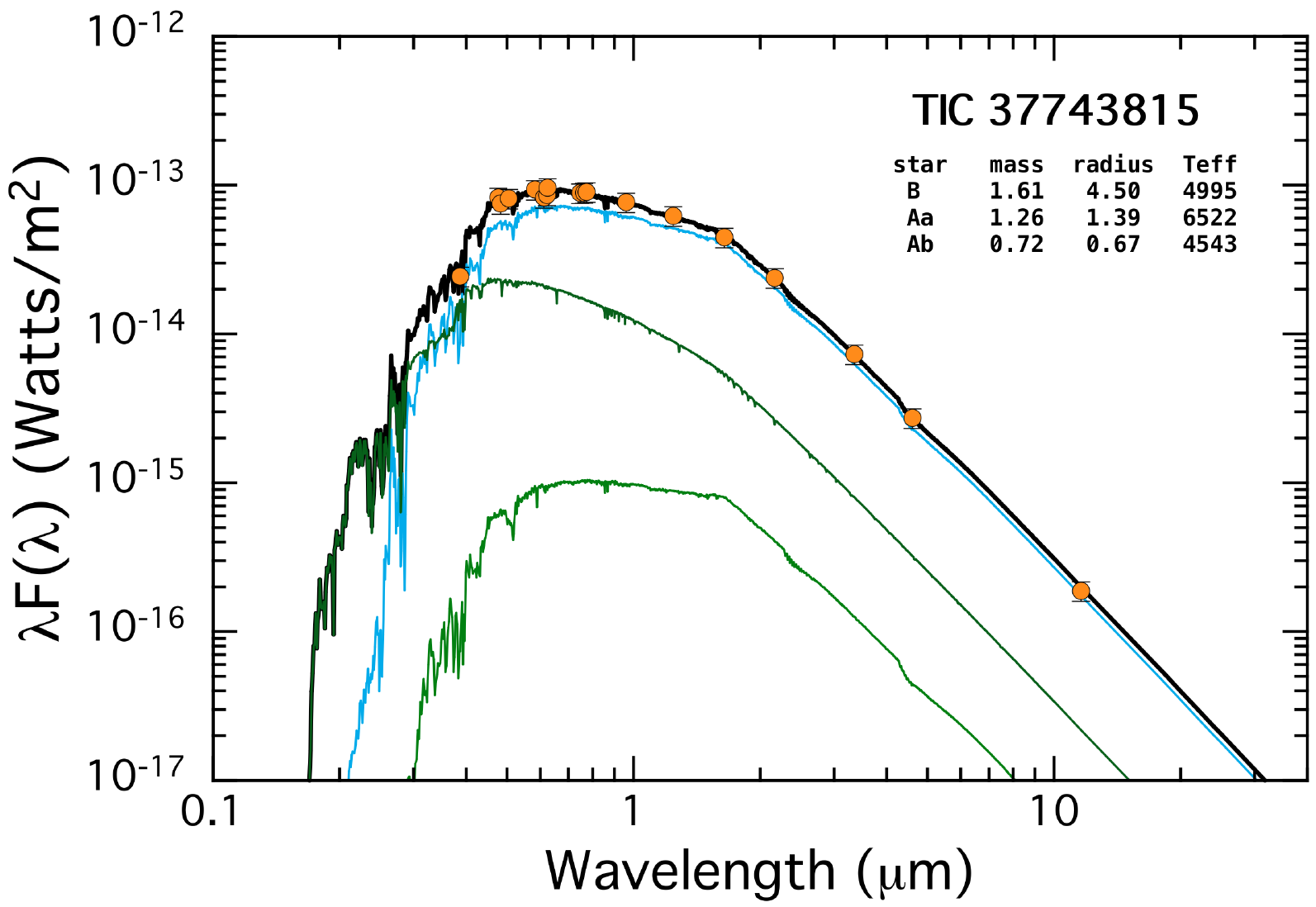} \hglue0.3cm
\includegraphics[width=0.45 \textwidth]{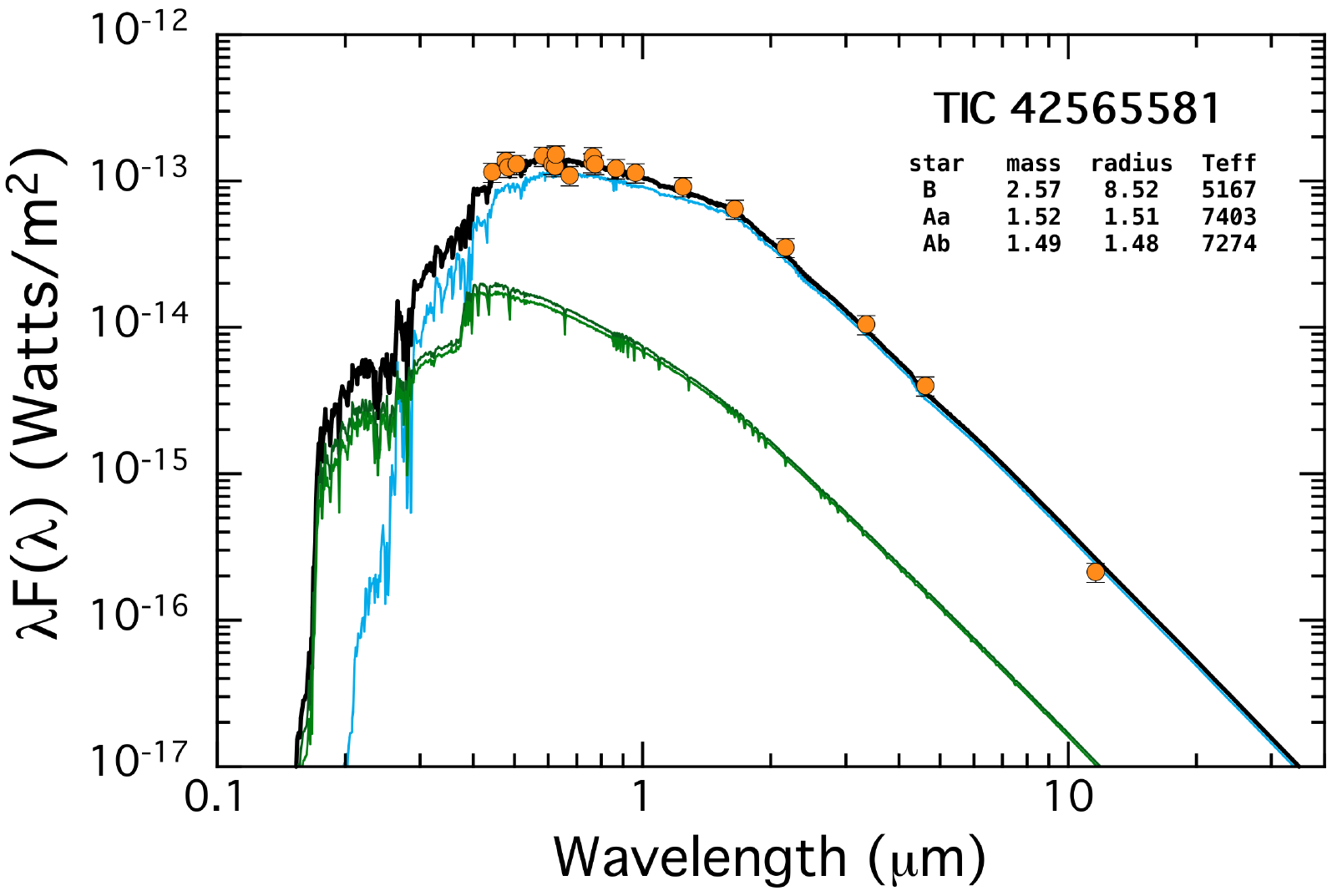} 
\includegraphics[width=0.45 \textwidth]{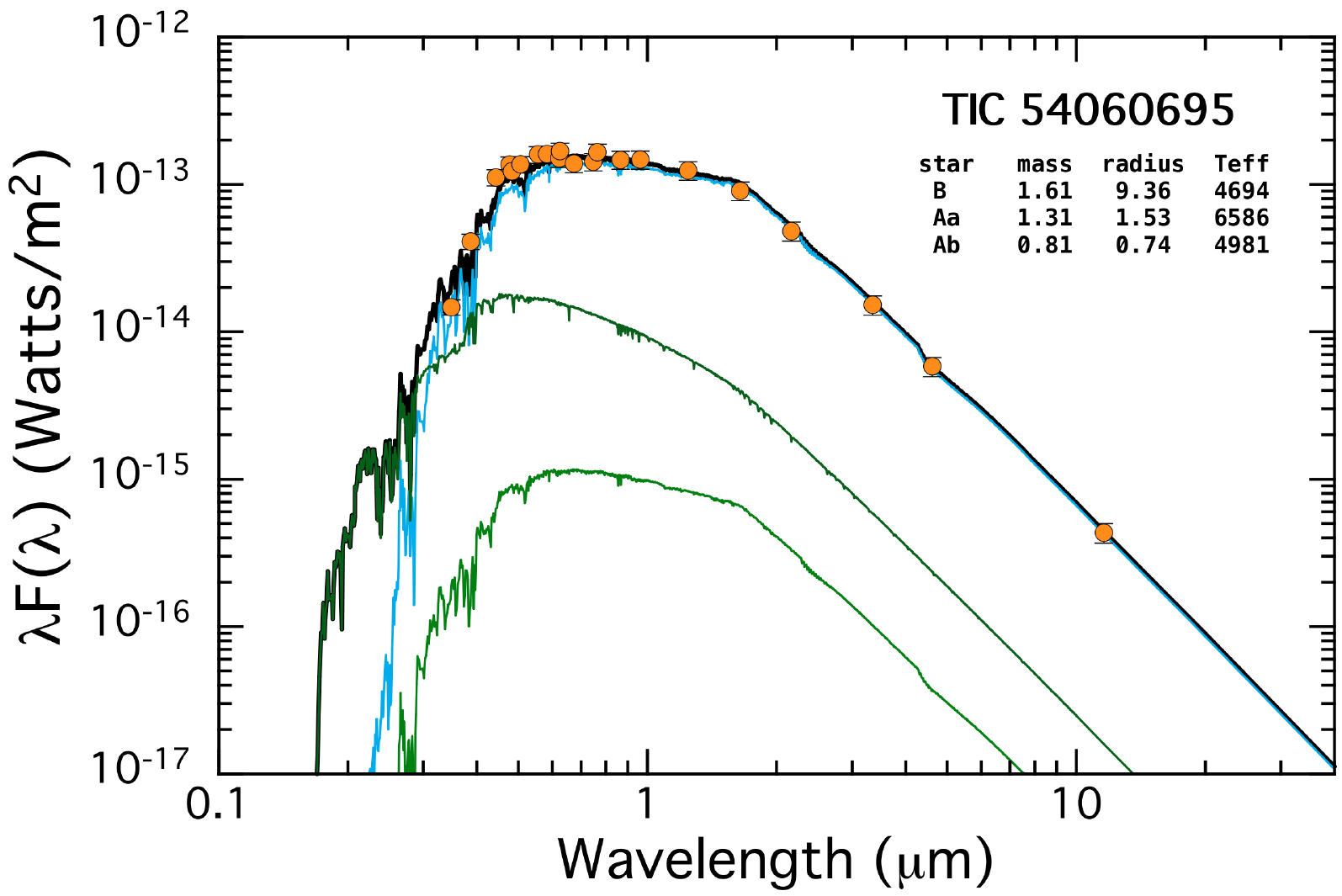} \hglue0.3cm
\includegraphics[width=0.45 \textwidth]{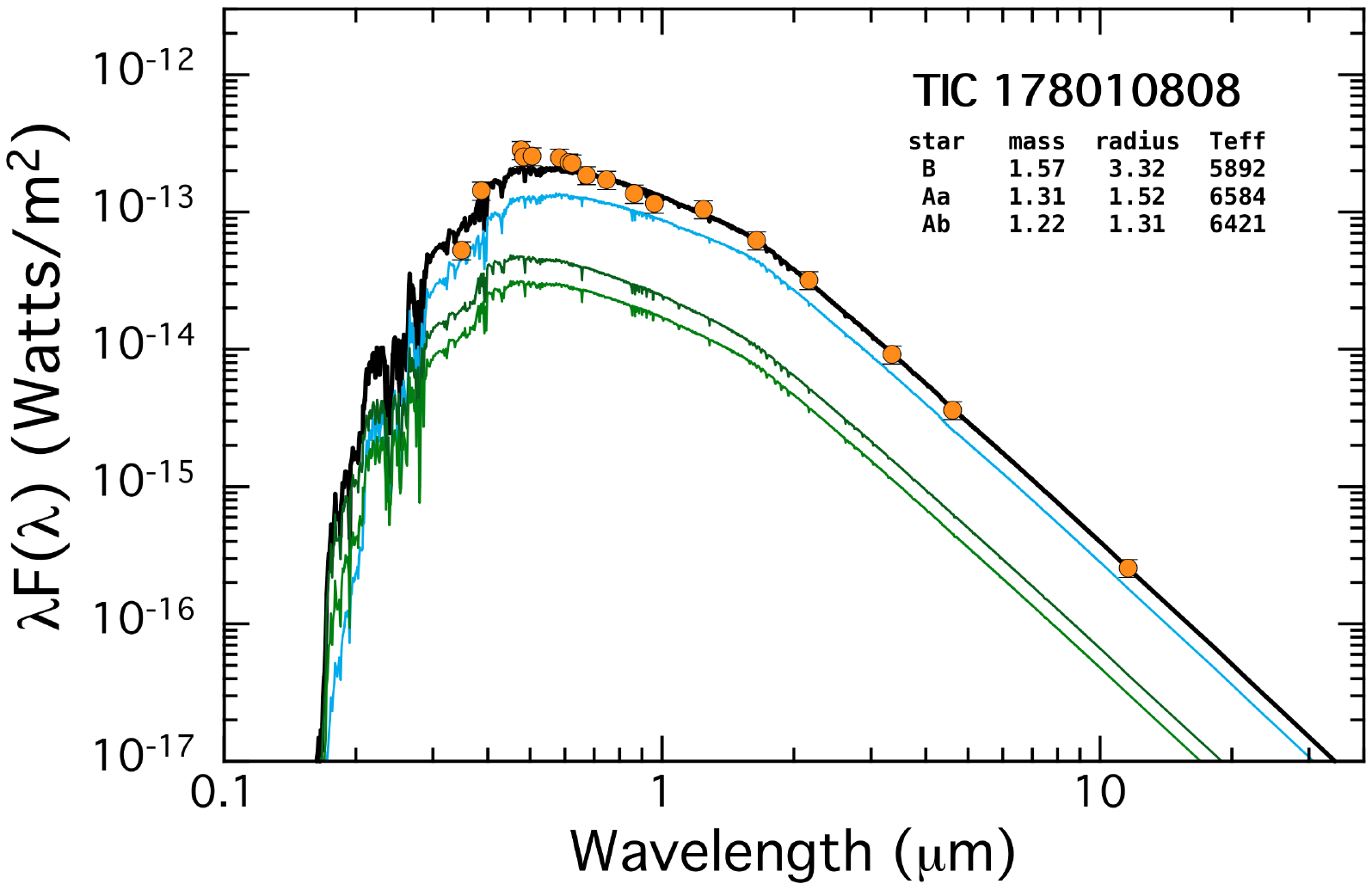}
\includegraphics[width=0.45 \textwidth]{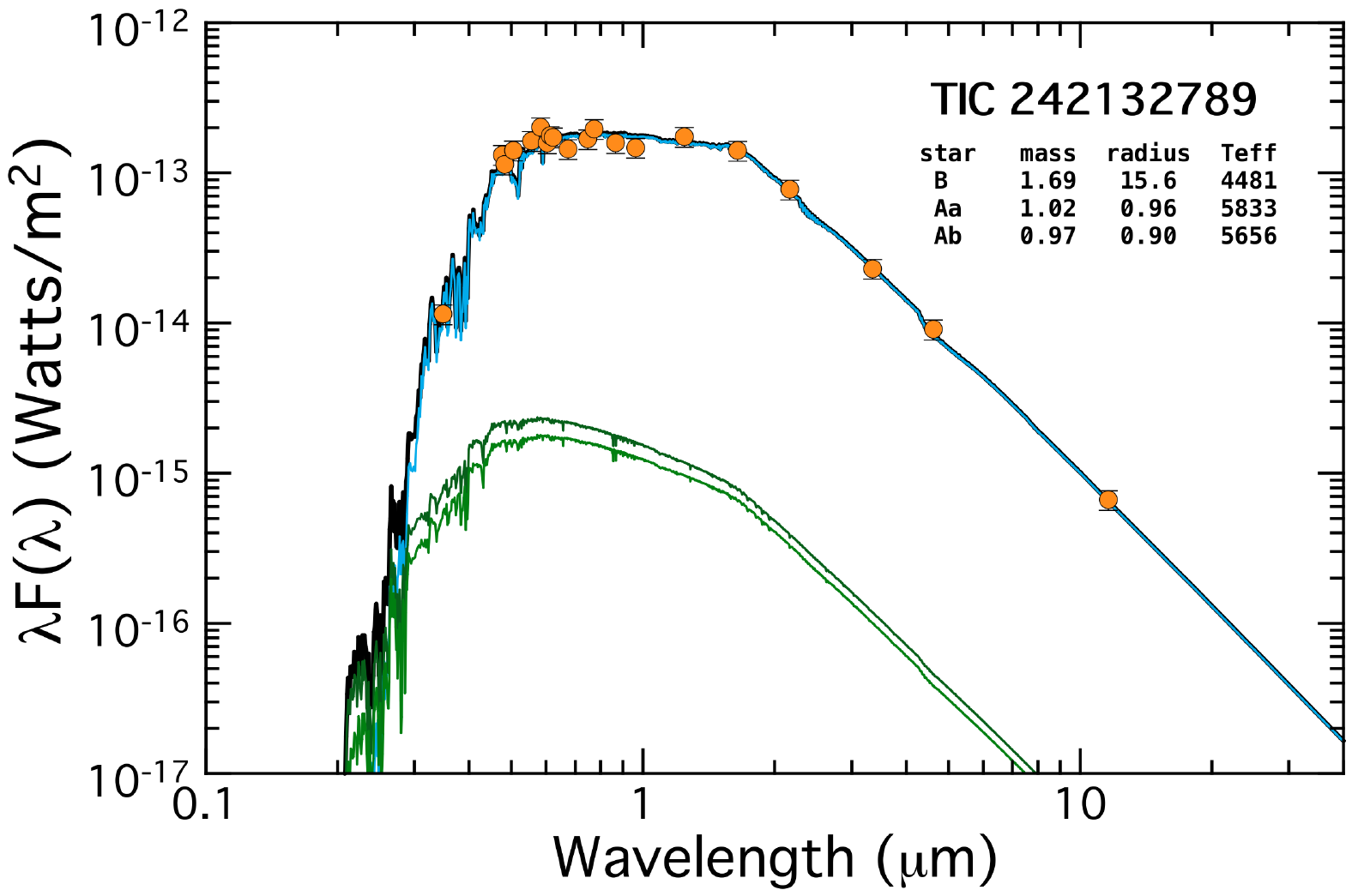} \hglue0.3cm
\includegraphics[width=0.45 \textwidth]{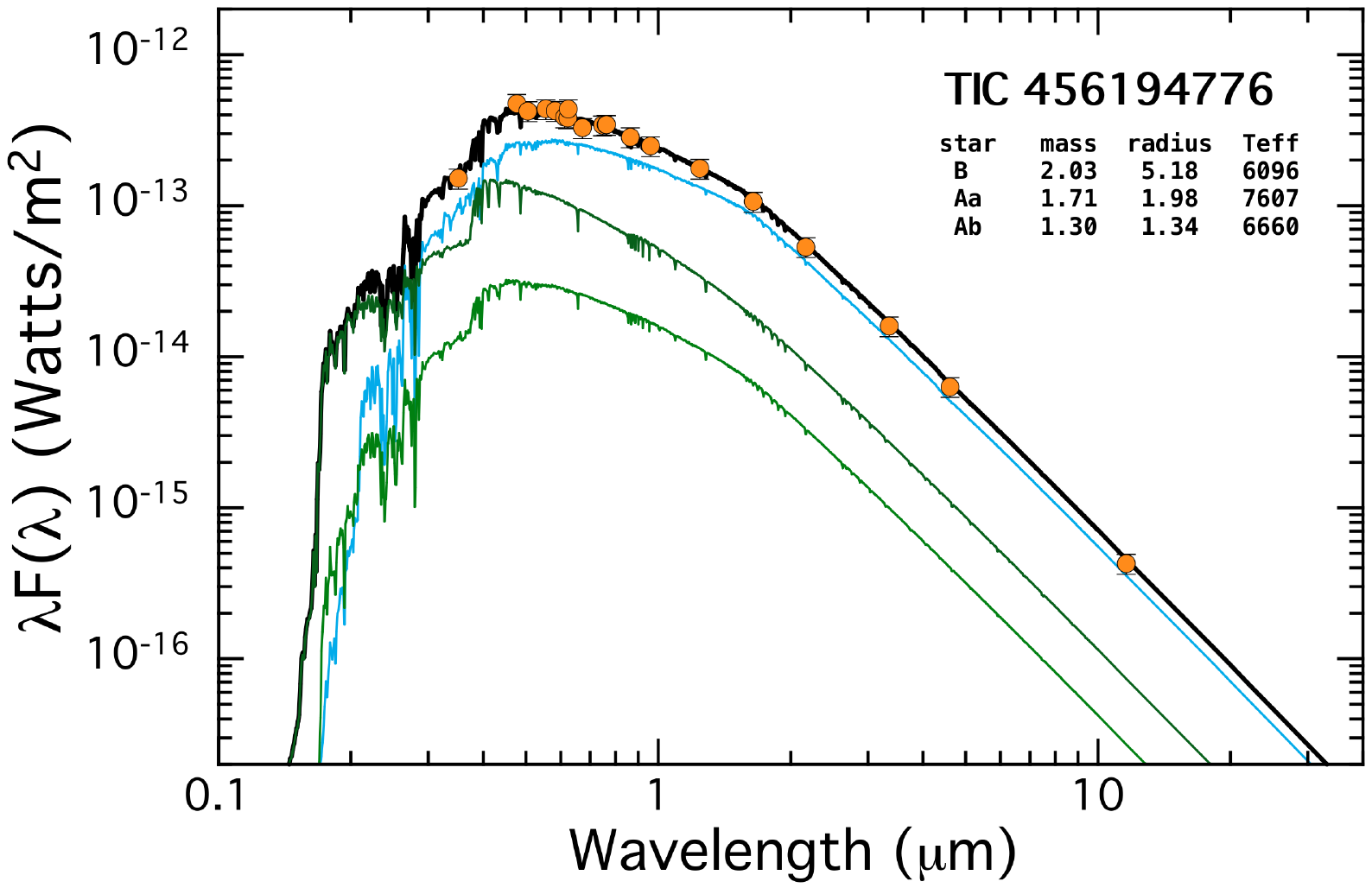}
\caption{SED fits for each of the six triply eclipsing triples discussed in this work. The cyan curve represents the model spectrum of the tertiary star (B) while the green curves represent the EB stars (Aa and Ab).  The black curve is the sum of the three model spectra. The fits for the three stellar masses, radii, and $T_{\rm eff}$'s were made using only the $\sim$20 measured SED points, a very loose constraint on the radius and $T_{\rm eff}$ for the tertiary star, and a temperature ratio for the inner EB based on eclipse depths (see Sect.~\ref{sec:SED} for details). We also explicitly make the assumption that the three stars are evolving in a coeval fashion without mass transfer. The units on the inset tables are M$_\odot$, R$_\odot$, and K.  Typical formal uncertainties on the masses are $\sim$10\%.}
\label{fig:SEDs}
\end{center}
\end{figure*} 

\begin{figure*}
\begin{center}
\includegraphics[width=0.45 \textwidth]{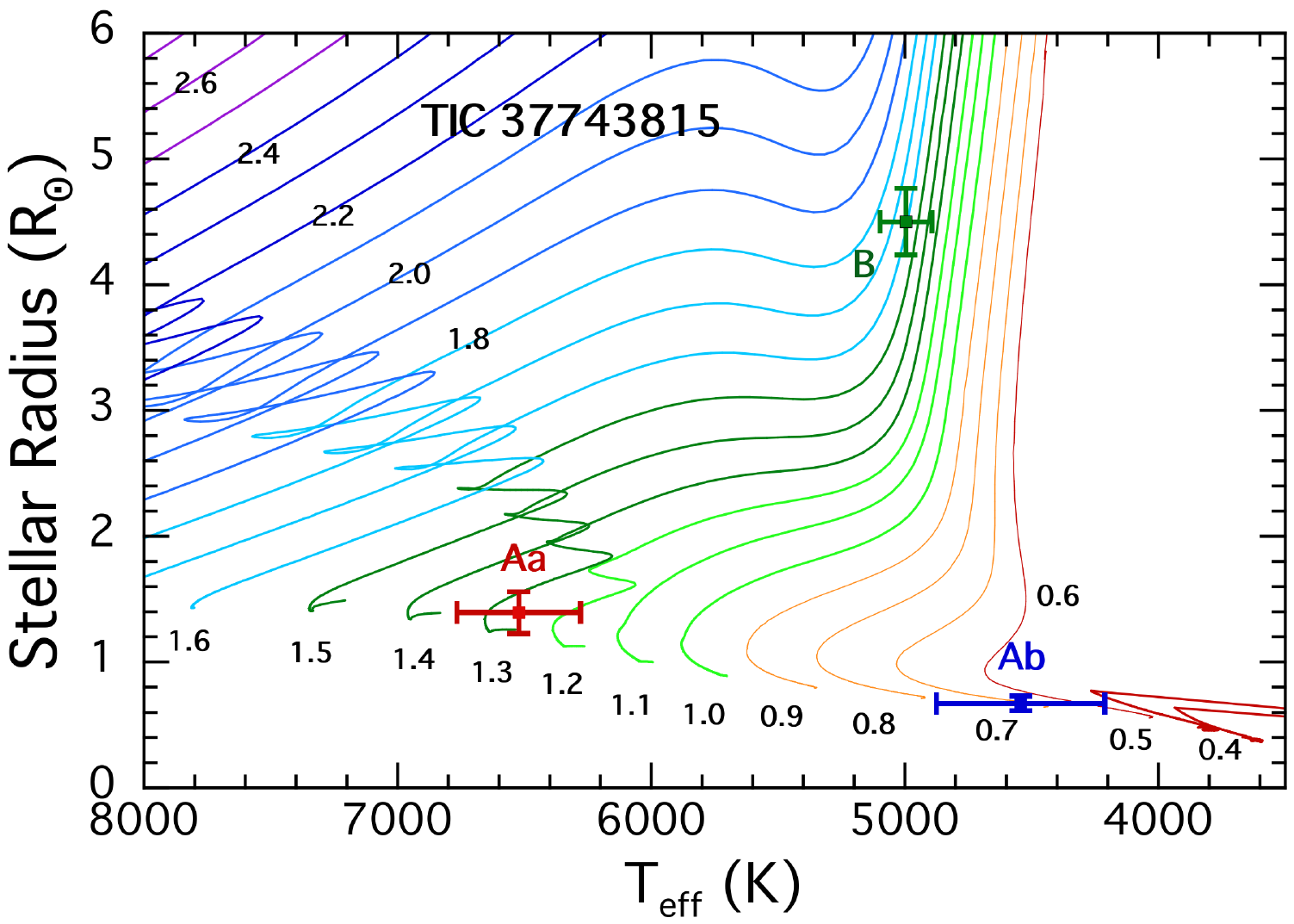} \hglue0.3cm
\includegraphics[width=0.45 \textwidth]{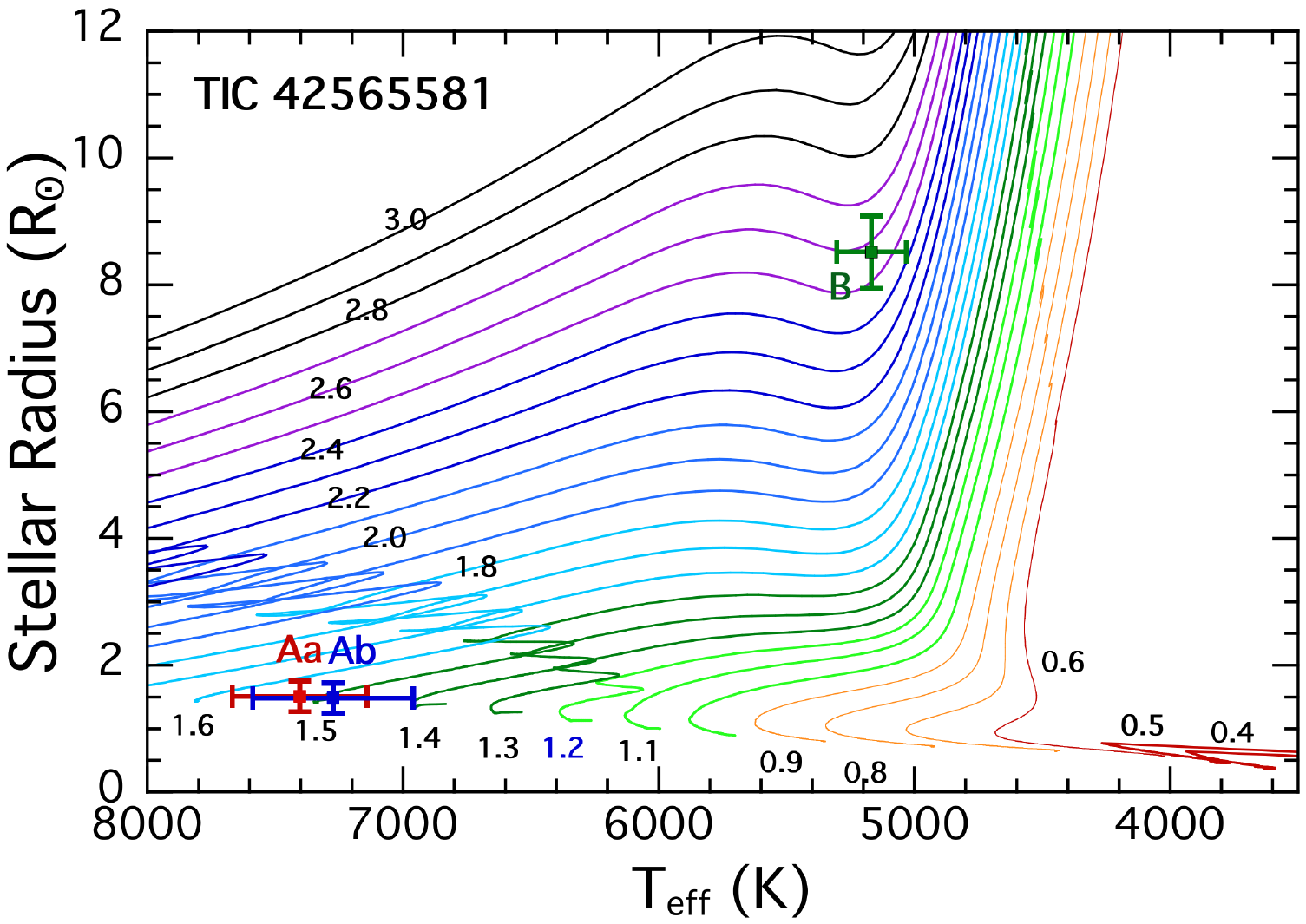}
\includegraphics[width=0.45 \textwidth]{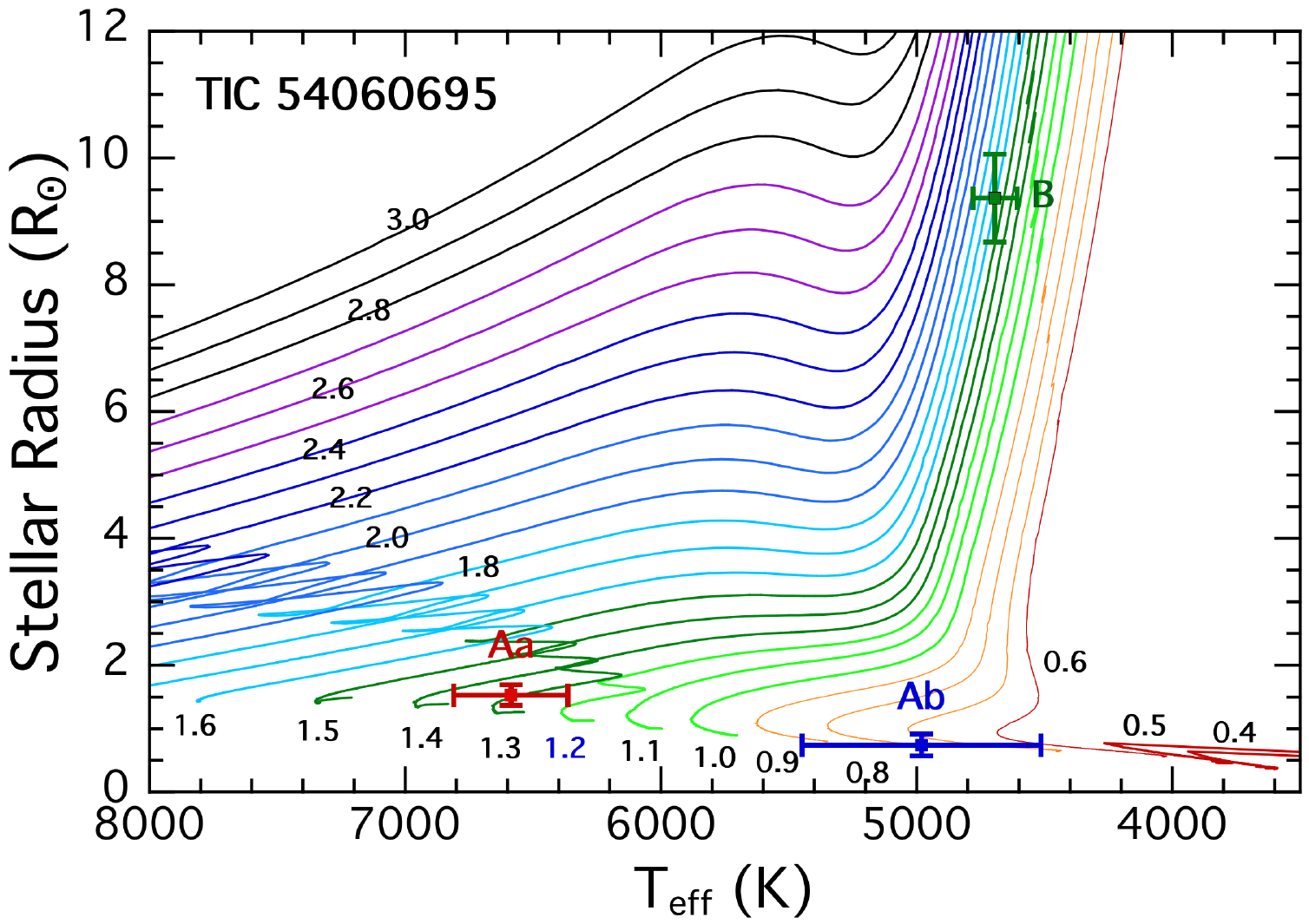} \hglue0.3cm
\includegraphics[width=0.453 \textwidth]{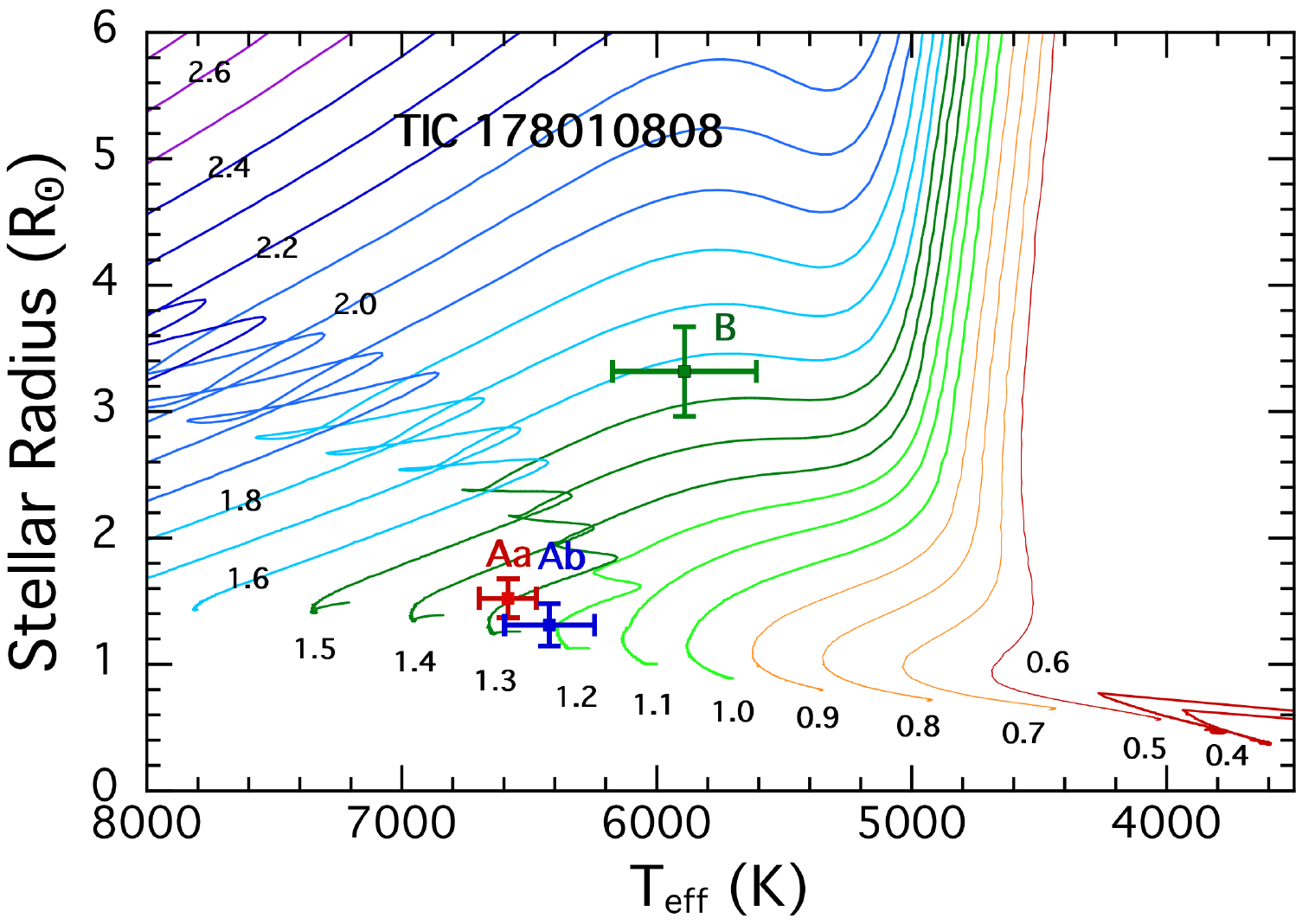} \hglue0.19cm
\includegraphics[width=0.45 \textwidth]{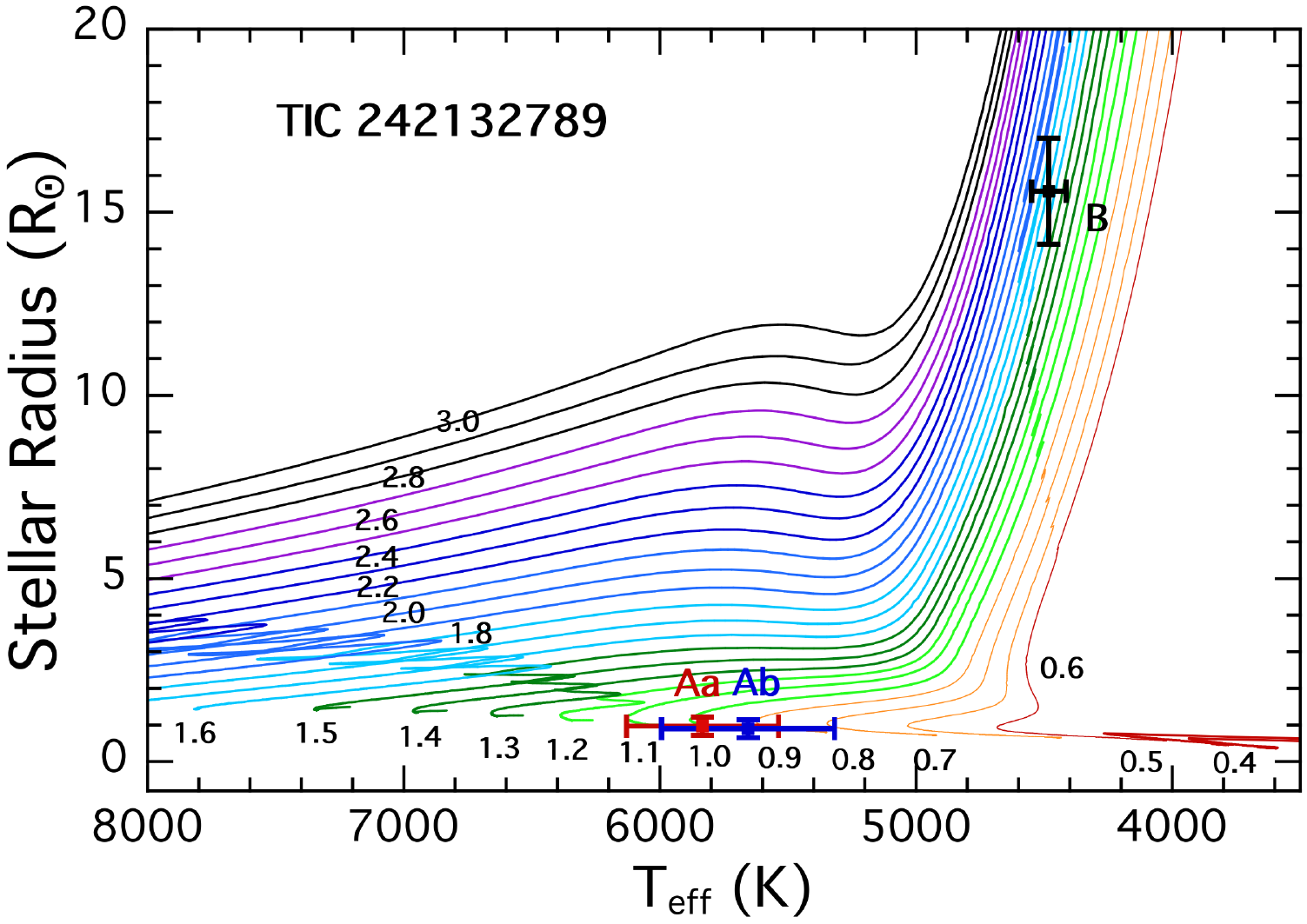} \hglue0.32cm
\includegraphics[width=0.453 \textwidth]{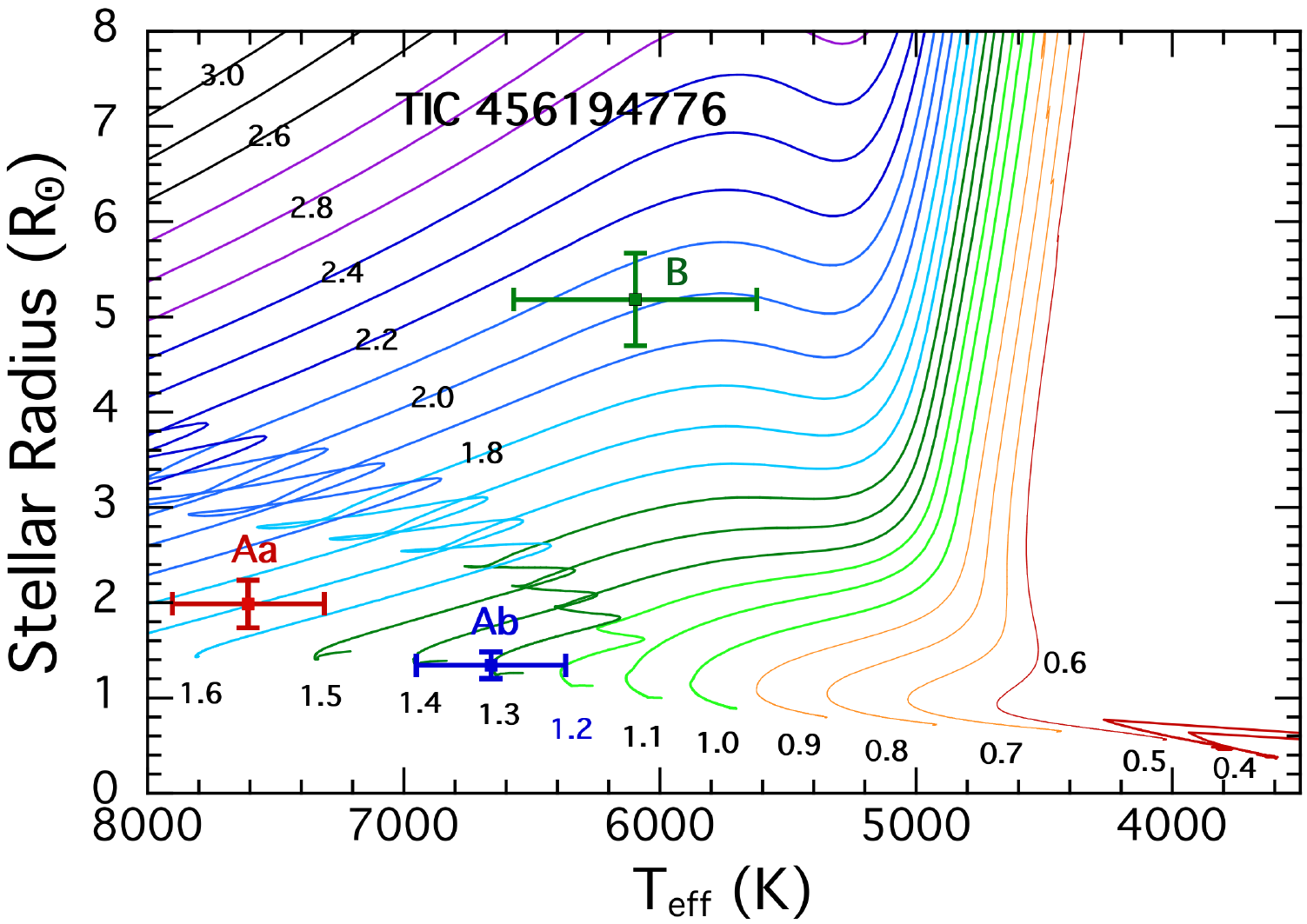}
\caption{The locations of the three stars in each of the six triple systems shown superposed on the {\tt MIST} stellar evolution tracks for stars of solar composition.  The numbers next to the tracks are the stellar masses in M$_\odot$. The locations of the stars in this diagram were taken from the SED fits shown in Fig.~\ref{fig:SEDs}. Somewhat more accurate stellar parameters are tabulated in Sect.~\ref{sec:photodynamical} based on the full photodynamical analyses.  However, the locations do not move appreciably (see Sect.~\ref{sec:individual}).  }
\label{fig:evolution}
\end{center}
\end{figure*} 

\begin{figure*}
\begin{center}
\includegraphics[width=0.435 \textwidth]{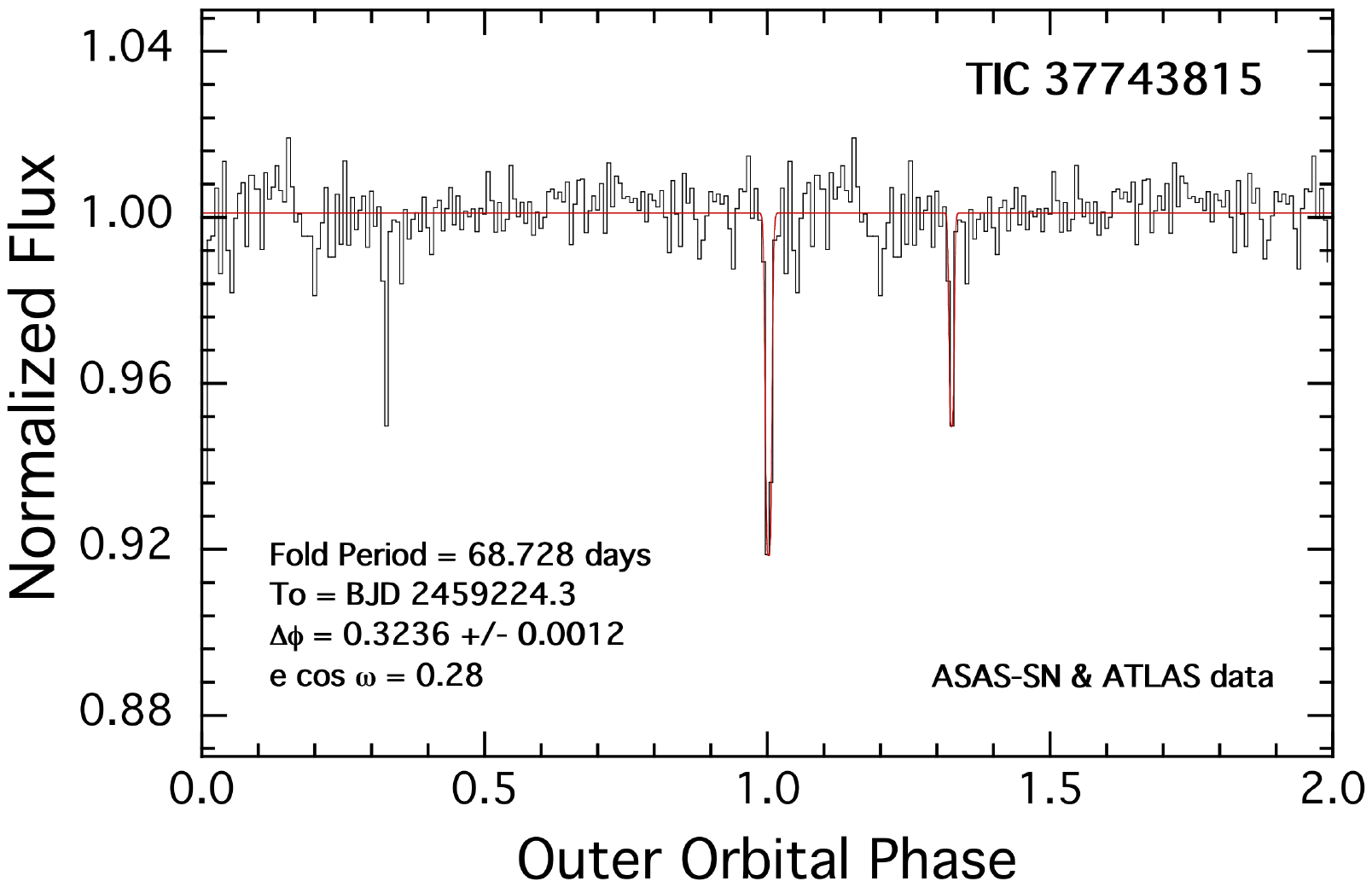} \hglue0.2cm
\includegraphics[width=0.445 \textwidth]{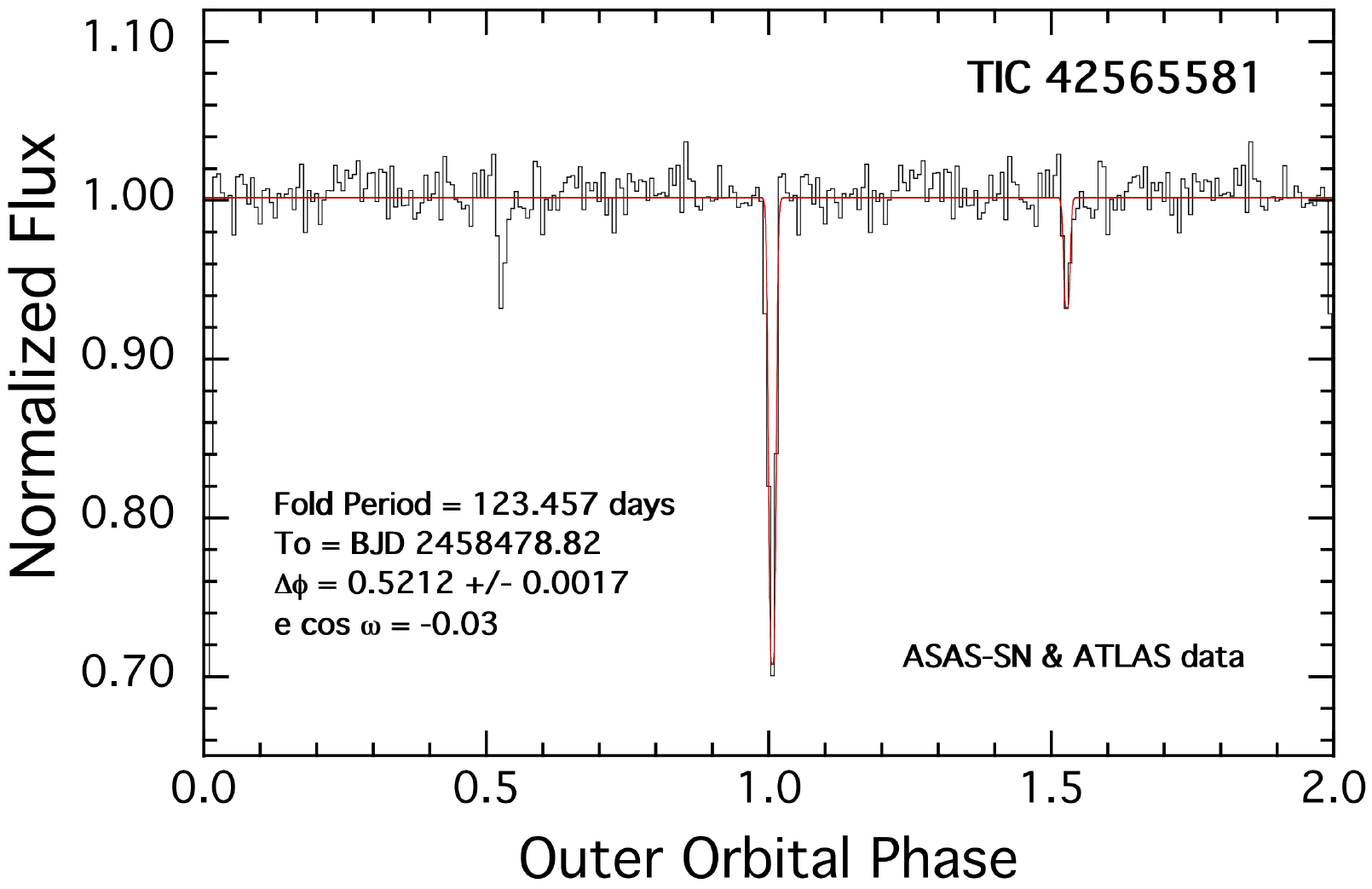}
\includegraphics[width=0.44 \textwidth]{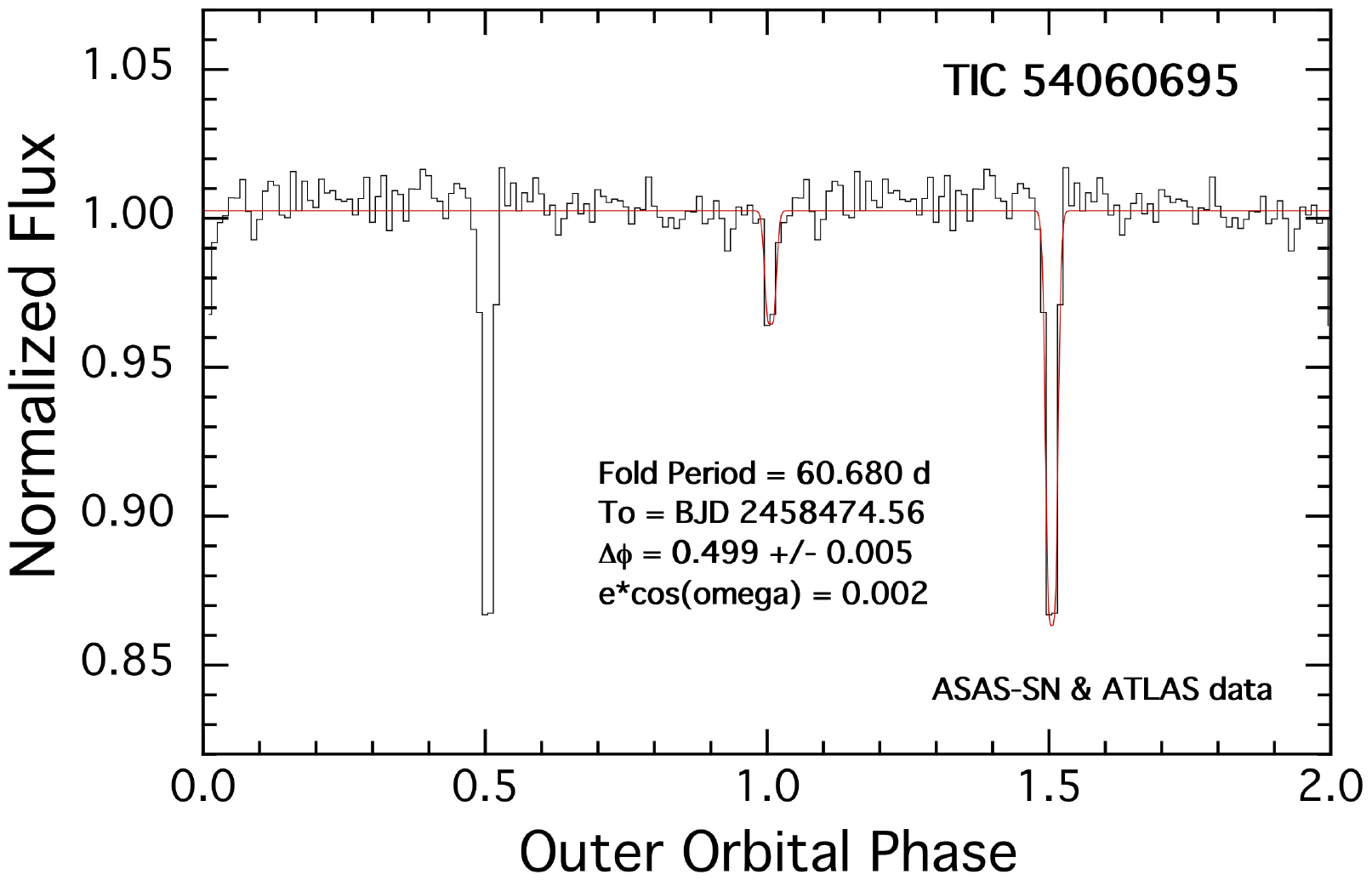}  \hglue0.2cm
\includegraphics[width=0.45 \textwidth]{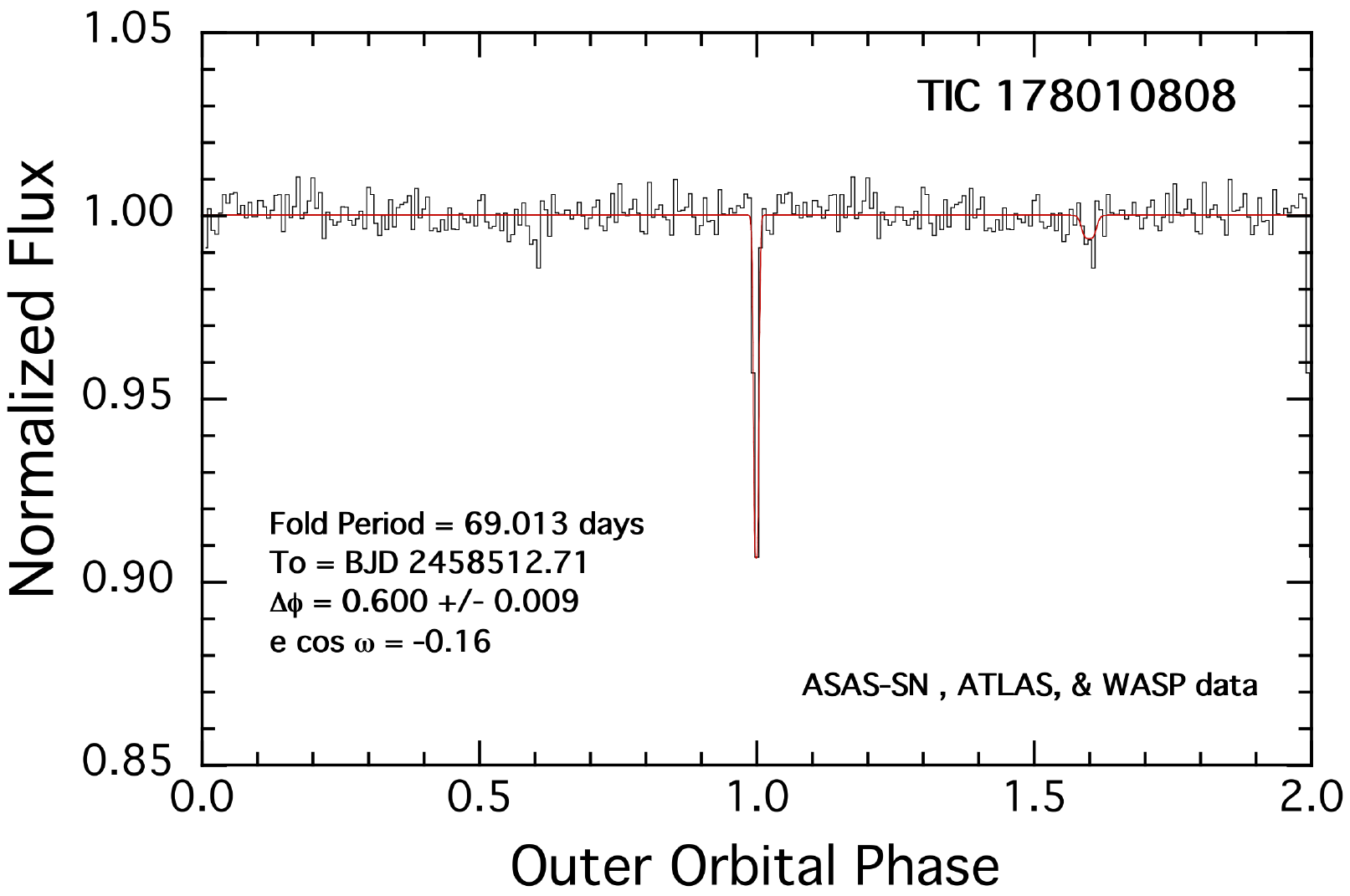}
\includegraphics[width=0.45 \textwidth]{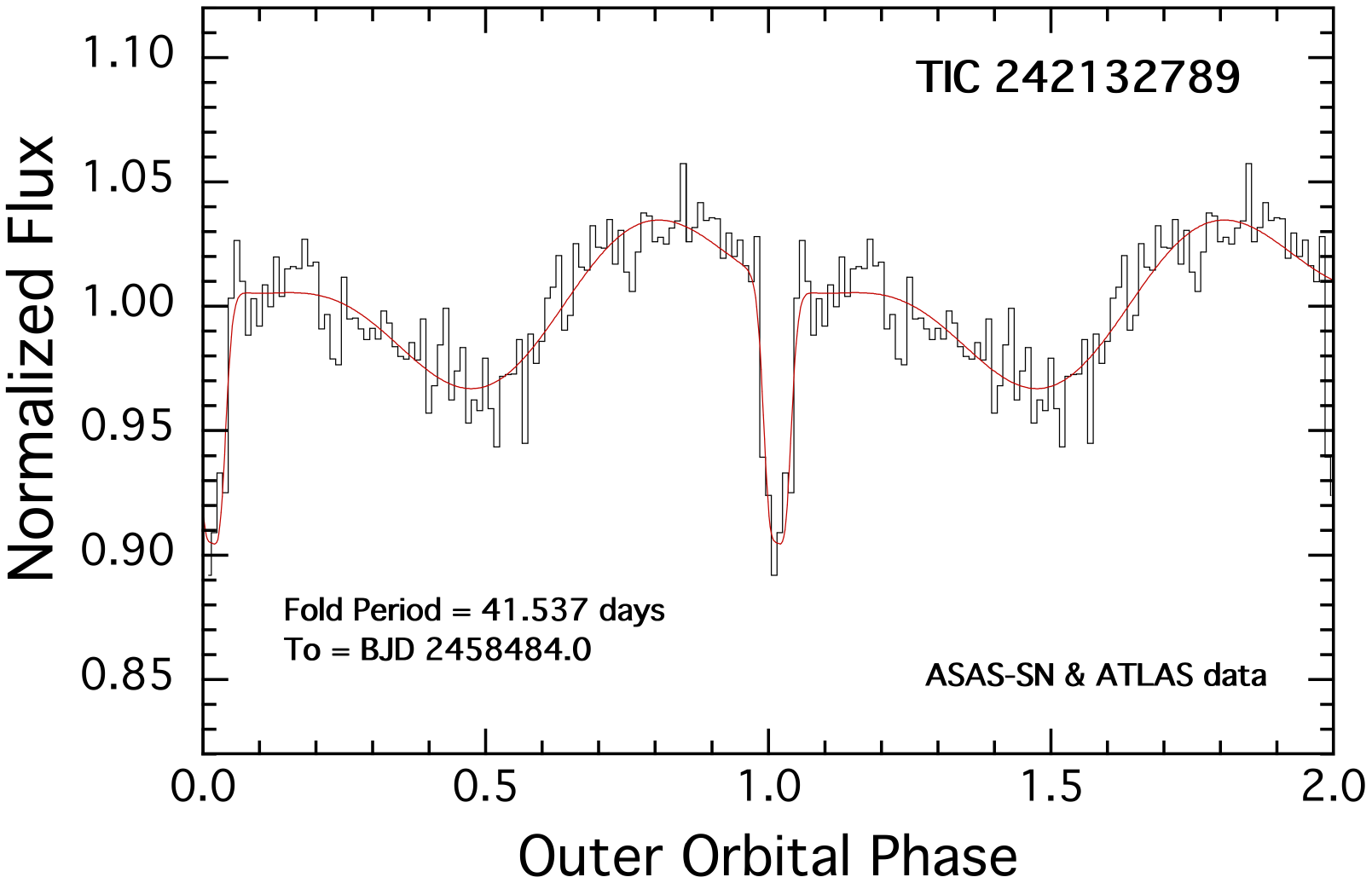} \hglue0.2cm
\includegraphics[width=0.45 \textwidth]{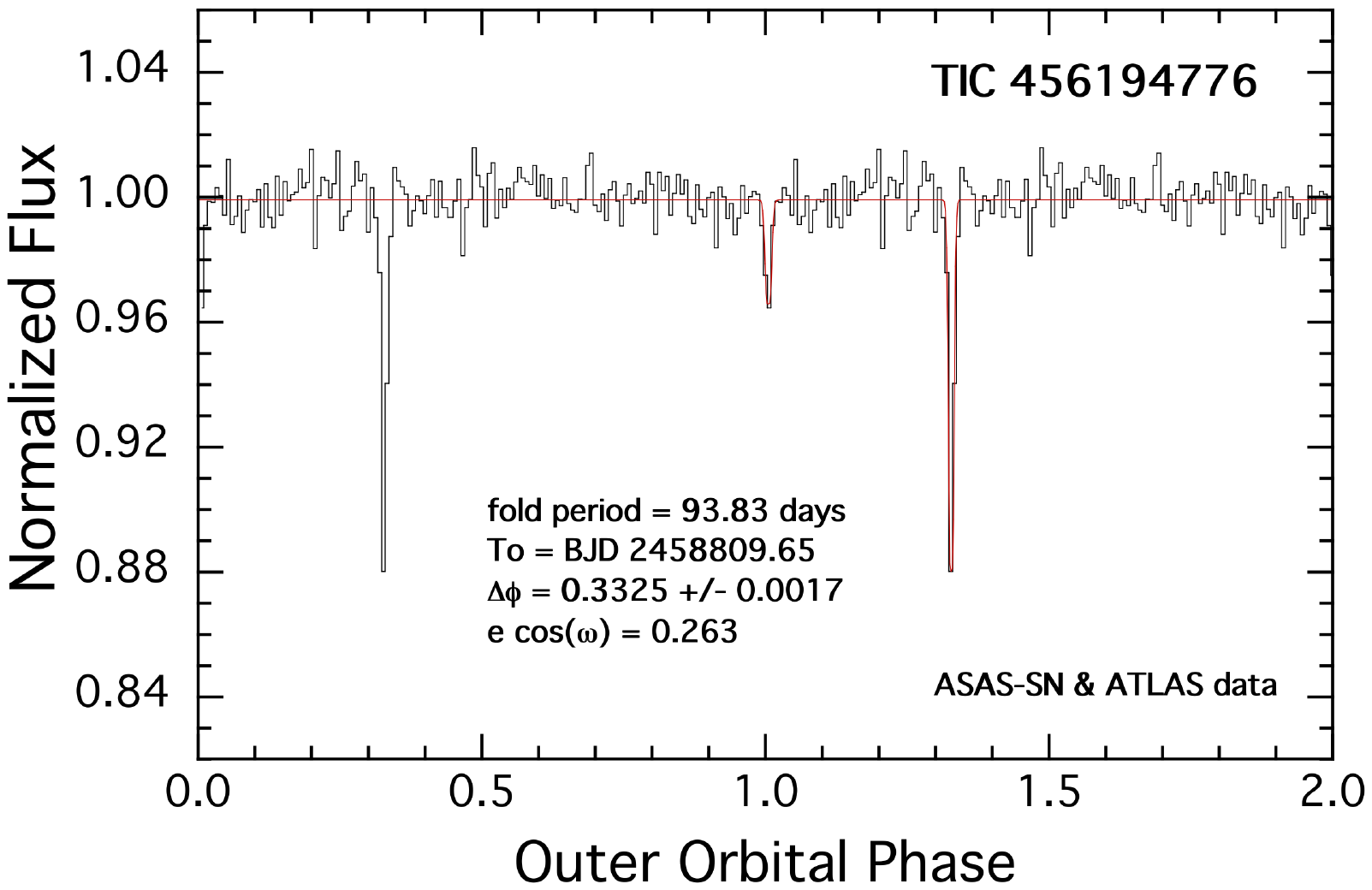}
\caption{Folded, binned, and averaged lightcurves for the outer orbits of the six triply eclipsing triple stars.  These are based on archival data from ATLAS, ASAS-SN, and WASP data (see Sect.~\ref{sec:outer_orbit} for details).  On each plot we write the fold period, the epoch reference time of phase zero, the orbital phase difference between the primary and secondary outer eclipses (if both are detected), and the inferred value of $e \cos \omega_{\rm out}$ based on the fold.  The red curve is a fit to the one or two outer eclipses during the second plotted orbital cycle only.  These are used to measure the orbital phase difference between the two eclipses and the widths (where possible and appropriate).}
\label{fig:outer_fold}
\end{center}
\end{figure*} 

\section{Preliminary SED Analysis}
\label{sec:SED}

Once we have discovered a triply eclipsing triple, we would like to develop some initial estimates of the nature of the three stars in the system.  To do this we make use of an analysis of the spectral energy distribution (SED).  We utilize the VizieR (\citealt{ochsenbein00}; A.-C.~Simon \& T.~Boch: \url{http://vizier.unistra.fr/vizier/sed/}) SED service which, in turn, utilizes systematic sky coverage of such surveys as Skymapper \citep{wolf18}, Pan-STARRS \citep{chambers16}, SDSS \citep{gunn98}, 2MASS \citep{2MASS}, WISE \citep{WISE}, and in some cases Galex \citep{bianchi17}. These typically provide $\sim$20 fluxes over the range 0.35 to 21 microns.  

Unless we have specific information to the contrary, we assume for our preliminary analysis that the three stars in the system have evolved in a coeval fashion since their birth as a triple system.  We further assume that there has been no mass transfer among the three stars, in particular between the binary components.  Under these assumptions, there are only four parameters that need to be fitted via a Markov chain Monte Carlo approach (see, e.g., \citealt{ford05}; \citealt{rappaport21}): $M_{\rm Aa}$, $M_{\rm Ab}$, $M_{\rm B}$, and the age of the system, where $Aa$ and $Ab$ refer to the stars in the inner binary, while $B$ is the tertiary star in the outer orbit.  We also make use of {\em MIST} stellar evolution tracks (\citealt{paxton11}; \citealt{paxton15}; \citealt{paxton19}; \citealt{dotter16}; \citealt{choi16}) for an assumed solar composition\footnote{Adopted for this preliminary analysis only. See Sect.~\ref{sec:photodynamical} for a description of the full, and more general, photodynamical analysis.}, as well as stellar atmosphere models from \citep{castelli03}.  If these four parameters can be determined, then the evolution tracks simultaneously determine the stellar radii and effective temperature of all three stars.  

In order to fit an SED, one typically requires an accurate distance to the source and the corresponding interstellar extinction, $A_V$.  Since Gaia \citep{GaiaEDR3} provides a secure distance with typically better than 5\% accuracy, and we can find information on the extinction from a variety of sources (e.g., Bayestar19; \citealt{green19}), in principle we do not need to fit for these parameters.  However, we usually add these two quantities to the fitted MCMC parameters, but with priors limited to just $\pm$ 4 times the listed uncertainties on them.

In spite of having some 20 SED data points to work with, and only 4-6 free parameters to fit for, this is most often quite insufficient for a decent solution.  The reasons are that (i) many of the SED points are sufficiently close to each other in wavelength so that they are not really independent, and (ii) a typical SED curve contains essentially only 4 or 5 defining characteristics, e.g., flux at any given wavelength on the Rayleigh-Jeans tail, wavelength of the peak in the SED curve, sharpness of the falloff at short wavelengths, etc. Therefore, we find it extremely helpful to add a few supplementary constraints in the MCMC fit.  These include estimates of the radius and $T_{\rm eff}$ of the tertiary and the ratio $T_{\rm eff,Aa}/T_{\rm eff,Ab}$.  Because all of the tertiary stars in this work are giants\footnote{The least evolved tertiary is the one in TIC 178010808, a 1.6 M$_\odot$ star of radius 2.9 R$_\odot$ whose luminosity exceeds that of the combined EB stars by nearly a factor of 2.}, they tend to dominate the light from the system.  Therefore, we make use of the Gaia DR2 and {\it TESS} Input Catalog (TIC v8.2) estimates of the `composite' radius and $T_{\rm eff}$ of the entire system measured as a single object (see Table \ref{tbl:mags}).  However, because those estimates are hardly a perfect representation of the tertiary, i.e., there are two other stars in the system, though of considerably lower luminosity, we take uncertainties on $T_{\rm eff}$ to be $\pm 300$ K and on $R_B$ to be $\pm 2 \,R_\odot$.  Finally, the temperature ratio of the two EB stars can be estimated approximately from the ratio of their eclipse depths.  

Using only this limited information, we fit the SED for all the properties of the stars in our six systems.  The results are shown in Fig.~\ref{fig:SEDs}.  In each case, the measured SED points (orange circles) have been corrected for interstellar extinction.  The continuous green curves represent the model fits for the stars in the EB, while the cyan curve is for the tertiary.  The heavier black curve is the total flux from all three stars.  One can get a very good sense from these plots just how much the tertiary dominates the system light.  We list on the plots only the nominal best-fitting values for $M$, $R$, and $T_{\rm eff}$ for each of the three stars.  We do not list the corresponding uncertainties on these parameters here because we employ a more comprehensive photodynamical analysis in Sect.~\ref{sec:photodynamical} which utilizes the EB and third-body eclipse contributions to the lightcurve, as well as the SED, to determine the final and more accurate stellar parameters.  Nonetheless, these first estimates of the stellar parameters provide a very quick estimate of what kinds of stars we are dealing with. The formal uncertainties on the masses are typically 10\%. These parameters can then be utilized as the initial input guesses to the photodynamical analysis.  

We utilize the system parameters for the stars found from this preliminary SED analysis to show in Fig.~\ref{fig:evolution} the locations of the stars superposed on the {\tt MIST} evolution tracks.  The tertiaries range from $\sim$3 to 15 R$_\odot$, with $T_{\rm eff}$ ranging from 6000 K to 4500 K, respectively.  By contrast, most of the binary stars are still on the main sequence, and, with only one exception, range in mass from 1.0-1.8\,M$_\odot$.  Within four of the systems the two EB stars tend to have similar masses.  

As a final note on the SED analysis, we point out that we have not considered pre-MS solutions.  These are considered and rejected in the photodynamical analysis.  

\section{Outer Orbital Period Determination with Archival Data}
\label{sec:outer_orbit}

Once we have discovered a triply eclipsing triple system with \textit{TESS}, the most important question to answer after determining the basic parameters of the constituent stars is the nature of its outer orbit, in particular the period and eccentricity.  For this purpose, in the absence of RV data, we most often make use of the ASAS-SN (\citealt{shappee14}; \citealt{kochanek17}) and ATLAS (\citealt{tonry18}; \citealt{smith20}) archival data sets. The ASAS-SN data sets typically have $\sim$1500-3000 photometric measurements of a given target, while the ATLAS archives often have approximately 1700 photometric measurements.  The ATLAS data have the advantage of going somewhat deeper than the ASAS-SN data, but the disadvantage of saturating on brighter stars where ASAS-SN may still perform well. When these two data sets are of comparable quality, we typically add them.  Naturally, we also check for KELT \citep{pepperetal07,pepperetal12}, WASP \citep{2006PASP..118.1407P}, HAT \citep{bakosetal02} and MASCARA data \citep{talens17} to see whether they are available for a particular source.  

For all the sources, with the exception of TIC 242132789, there were WASP archival data available.  We found these to be quite useful in helping to determine the long-term average EB periods.  But, it turns out that these data were generally too noisy to aid in the search for the outer third body eclipses, except in the case of TIC 178010808 where the WASP data nicely complemented the ATLAS and ASAS-SN data.  For TIC 242132789 there was a set of KELT data in addition to ATLAS and ASAS-SN data. The KELT data marginally detected the spot and ellipsoidal light modulations associated with the outer orbit, but not the outer eclipses.  

We do a blind search for the outer eclipses (either outer primary or outer secondary eclipse) using a Box Least Squares transform \citep{kovacs02}.  Before doing the BLS search we remove the lightcurve of the inner EB by Fourier means (as described in \citealt{powell21}).  In the process we remove between 5 and 100 orbital harmonics depending on the sharpness of the features in the EB lightcurve.  This cleaning process requires knowing the orbital period of the EB very accurately.  In turn, we determine the long-term average binary period from the {\it TESS} data or from the archival data, whichever yield a more precise result.  

We show the results of the above procedure for each of our six triply eclipsing triple stars in Fig.~\ref{fig:outer_fold}.  In each panel we show a folded, binned, and averaged lightcurve of the archival data about the period corresponding to the largest and most significant peak in the BLS transform. In all cases, the zero phase for the outer orbit is taken to be the time of one of the third-body eclipses observed in the {\it TESS} data.  In four of the six sources, both the primary and secondary outer eclipses are clearly detected. In those cases, the value of $e \cos \omega_{\rm out}$ is also accurately determined, in addition to the outer period of the triple.  In the fifth source, TIC 178010808, the secondary outer eclipse is only barely detected, if at all.  In the sixth source, TIC 242132789, the outer orbital lightcurve has a clear undulating structure superposed on the very clear primary eclipse.  Because the slow modulations have the same period as the outer eclipses, we attribute this to starspot(s) on a giant tertiary that is corotating with the binary orbit as well as to ellipsoidal light variations from the giant.  Note that the giant in this system has $R \simeq 15\,{\rm R}_\odot$.   Not only is this star large, but its outer orbital period of 42 days is the shortest among our sample, and one of the shortest period triples known.


\section{Photodynamical Analysis for the System Parameters}
\label{sec:photodynamical}

For all six of our triply eclipsing triples, we have carried out a photodynamical analysis with the software package {\sc Lightcurvefactory} \citep[see, e.g.][and references therein]{borkovitsetal19a,borkovitsetal20a}. As described in earlier work, the code contains (i) a built-in numerical integrator to calculate the three-body perturbed coordinates and velocities of the three stars in the system; (ii) emulators for the {\it TESS} light curve, the ETVs extracted therefrom, and radial velocity curve (if available), and (iii) an MCMC-based search routine for the system parameters. The latter utilizes an implementation of the generic Metropolis-Hastings algorithm \citep[see e.g.][]{ford05}. The use of this software package and the consecutive steps of the entire analysis process have been previously explained in detail as the code was applied to a wide range of multistellar systems \citep{borkovitsetal18,borkovitsetal19a,borkovitsetal19b,borkovitsetal20a,borkovitsetal20b,borkovitsetal21,mitnyanetal20}.  These included tight and wider triple systems (with and without outer eclipses),  as well as quadruple systems with either a 2+2 or 2+1+1 configuration. Here we discuss only a few specific points related to the current triples.

\begin{table*}
 \centering
\caption{Orbital and astrophysical parameters of TIC 37743815 and TIC 42565581 from the joint photodynamical \textit{TESS}, ETV, SED and \texttt{PARSEC} isochrone solution. Note that the orbital parameters are instantaneous, osculating orbital elements and are given for epoch $t_0$ (first row).  Therefore, the orbital periods, in particular, cannot be used for predicting the times of future eclipses; see Table \ref{tab:ephemerides} for the latter, and \citet{kostovetal21} for a detailed explanation.}
 \label{tab:syntheticfit_TIC37743815+42565581}
\begin{tabular}{@{}lllllll}
\hline
 & \multicolumn{3}{c}{TIC\,37743815} & \multicolumn{3}{c}{TIC\,42565581} \\
\hline
\multicolumn{7}{c}{orbital elements} \\
\hline
   & \multicolumn{3}{c}{subsystem} & \multicolumn{3}{c}{subsystem}  \\
   & \multicolumn{2}{c}{Aa--Ab} & A--B & \multicolumn{2}{c}{Aa--Ab} & A--B \\
  \hline
  $t_0$ [BJD - 2400000] & \multicolumn{3}{c}{58468.0} & \multicolumn{3}{c}{58468.0} \\
  $P$ [days] & \multicolumn{2}{c}{$0.906926_{-0.000009}^{+0.000008}$} & $68.7998_{-0.0025}^{+0.0029}$ & \multicolumn{2}{c}{$1.823537_{-0.000035}^{+0.000031}$} & $123.5467_{-0.0039}^{+0.0041}$  \\
  $a$ [R$_\odot$] & \multicolumn{2}{c}{$4.779_{-0.037}^{+0.026}$} & $105.8_{-0.6}^{+1.1}$ & \multicolumn{2}{c}{$9.61_{-0.16}^{+0.12}$} & $187.9_{-2.8}^{+3.3}$ \\
  $e$ & \multicolumn{2}{c}{$0.0093_{-0.0035}^{+0.0065}$} & $0.361_{-0.016}^{+0.018}$ & \multicolumn{2}{c}{$0.01227_{-0.00040}^{+0.00041}$} & $0.161_{-0.017}^{+0.107}$ \\
  $\omega$ [deg] & \multicolumn{2}{c}{$95.0_{-3.8}^{+3.6}$} & $40.5_{-3.6}^{+2.8}$ & \multicolumn{2}{c}{$163.5_{-3.4}^{+4.3}$} & $64_{-28}^{+16}$ \\ 
  $i$ [deg] & \multicolumn{2}{c}{$89.55_{-0.39}^{+0.53}$} & $89.60_{-0.56}^{+0.16}$ & \multicolumn{2}{c}{$90.26_{-0.53}^{+0.31}$} & $89.10_{-0.27}^{+1.35}$ \\
  $\mathcal{T}_0^\mathrm{inf}$ [BJD - 2400000]& \multicolumn{2}{c}{$58469.09761_{-0.00004}^{+0.00014}$} & $58468.2849_{-0.0096}^{+0.0111}$ & \multicolumn{2}{c}{$58468.9993_{-0.0002}^{+0.0002}$} & $58479.1476_{-0.0096}^{+0.0076}$ \\
  $\tau$ [BJD - 2400000]& \multicolumn{2}{c}{$58486.6584_{-0.0097}^{+0.0092}$} & $58417.24_{-0.61}^{+0.47}$ & \multicolumn{2}{c}{$58468.469_{-0.017}^{+0.022}$} & $58405.4_{-16.5}^{+7.4}$ \\
  $\Omega$ [deg] & \multicolumn{2}{c}{$0.0$} & $-1.08_{-2.17}^{+1.92}$ & \multicolumn{2}{c}{$0.0$} & $4.72_{-9.91}^{+1.59}$ \\
  $i_\mathrm{m}$ [deg] & \multicolumn{3}{c}{$1.84_{-0.82}^{+1.49}$} & \multicolumn{3}{c}{$5.53_{-1.69}^{+1.63}$} \\
  \hline
  mass ratio $[q=m_\mathrm{sec}/m_\mathrm{pri}]$ & \multicolumn{2}{c}{$0.642_{-0.010}^{+0.010}$} & $0.905_{-0.053}^{+0.017}$ & \multicolumn{2}{c}{$0.997_{-0.004}^{+0.004}$} & $0.638_{-0.014}^{+0.036}$ \\
 $K_\mathrm{pri}$ [km\,s$^{-1}$] & \multicolumn{2}{c}{$104.3_{-1.5}^{+1.1}$} & $39.9_{-2.2}^{+0.7}$ & \multicolumn{2}{c}{$132.6_{-1.0}^{+1.1}$} & $30.6_{-0.7}^{+1.3}$ \\ 
 $K_\mathrm{sec}$ [km\,s$^{-1}$] & \multicolumn{2}{c}{$162.3_{-2.5}^{+1.3}$} & $44.0_{-0.3}^{+0.4}$ & \multicolumn{2}{c}{$132.9_{-0.9}^{+1.2}$} & $47.5_{-1.0}^{+1.7}$ \\ 
  \hline  
\multicolumn{7}{c}{stellar parameters} \\
\hline
   & Aa & Ab &  B & Aa & Ab &  B \\
  \hline
 \multicolumn{7}{c}{Relative quantities} \\
  \hline
 fractional radius [$R/a$]  & $0.2201_{-0.0039}^{+0.0031}$ & $0.1376_{-0.0023}^{+0.0008}$ & $0.0377_{-0.0014}^{+0.0037}$ & $0.2382_{-0.0025}^{+0.0024}$ & $0.2358_{-0.0025}^{+0.0026}$ & $0.0451_{-0.0027}^{+0.0020}$ \\
 temperature relative to $(T_\mathrm{eff})_\mathrm{Aa}$ & $1$ & $0.7202_{-0.0094}^{+0.0107}$ & $0.9244_{-0.0362}^{+0.0163}$ & $1$ & $1.0001_{-0.0006}^{+0.0007}$ & $0.6959_{-0.0133}^{+0.0204}$ \\
 fractional flux [in \textit{TESS}-band] & $0.0804_{-0.0011}^{+0.0011}$ & $0.0081_{-0.0006}^{+0.0006}$ & $0.8957_{-0.0251}^{+0.0118}$ & $0.1417_{-0.0016}^{+0.0017}$ & $0.1393_{-0.0015}^{+0.0015}$ & $0.6093_{-0.0303}^{+0.0334}$ \\
 \hline
 \multicolumn{7}{c}{Physical Quantities} \\
  \hline 
 $m$ [M$_\odot$] & $1.082_{-0.021}^{+0.020}$ & $0.697_{-0.024}^{+0.012}$ & $1.605_{-0.085}^{+0.052}$ & $1.793_{-0.089}^{+0.067}$ & $1.787_{-0.087}^{+0.068}$ & $2.255_{-0.113}^{+0.233}$ \\
 $R$ [R$_\odot$] & $1.052_{-0.028}^{+0.020}$ & $0.659_{-0.017}^{+0.006}$ & $4.011_{-0.152}^{+0.362}$ & $2.288_{-0.047}^{+0.034}$ & $2.266_{-0.047}^{+0.035}$ & $8.410_{-0.485}^{+0.583}$ \\
 $T_\mathrm{eff}$ [K]& $5899_{-70}^{+94}$ & $4255_{-85}^{+112}$ & $5434_{-210}^{+140}$ & $7521_{-153}^{+117}$ & $7521_{-155}^{+119}$ & $5218_{-130}^{+168}$ \\
 $L_\mathrm{bol}$ [L$_\odot$] & $1.214_{-0.082}^{+0.075}$ & $0.126_{-0.009}^{+0.012}$ & $12.79_{-0.72}^{+0.60}$ & $15.14_{-1.61}^{+0.87}$ & $14.83_{-1.53}^{+0.92}$ & $47.74_{-3.09}^{+3.77}$ \\
 $M_\mathrm{bol}$ & $4.56_{-0.07}^{+0.08}$ & $7.02_{-0.10}^{+0.08}$ & $2.00_{-0.05}^{+0.06}$ & $1.82_{-0.06}^{+0.12}$ & $1.84_{-0.07}^{+0.12}$ & $0.57_{-0.08}^{+0.07}$ \\
 $M_V           $ & $4.59_{-0.07}^{+0.08}$ & $7.85_{-0.20}^{+0.18}$ & $2.16_{-0.06}^{+0.07}$ & $1.76_{-0.06}^{+0.14}$ & $1.78_{-0.07}^{+0.13}$ & $0.78_{-0.09}^{+0.08}$ \\
 $\log g$ [dex] & $4.427_{-0.009}^{+0.012}$ & $4.644_{-0.002}^{+0.006}$ & $3.438_{-0.090}^{+0.036}$ & $3.969_{-0.010}^{+0.012}$ & $3.976_{-0.010}^{+0.012}$ & $2.935_{-0.030}^{+0.057}$ \\
 \hline
\multicolumn{7}{c}{Global system parameters} \\
  \hline
$\log$(age) [dex] &\multicolumn{3}{c}{$9.384_{-0.021}^{+0.069}$} &\multicolumn{2}{c}{$9.044_{-0.059}^{+0.066}$} & $8.935_{-0.089}^{+0.027}$ \\
$[M/H]$  [dex]    &\multicolumn{3}{c}{$0.140_{-0.086}^{+0.122}$} &\multicolumn{3}{c}{$0.050_{-0.176}^{+0.145}$} \\
$E(B-V)$ [mag]    &\multicolumn{3}{c}{$0.158_{-0.054}^{+0.047}$} &\multicolumn{3}{c}{$0.392_{-0.039}^{+0.051}$} \\
extra light $\ell_4$ [in \textit{TESS}-band] & \multicolumn{3}{c}{$0.016_{-0.012}^{+0.025}$} & \multicolumn{3}{c}{$0.110_{-0.034}^{+0.031}$} \\
$(M_V)_\mathrm{tot}$  &\multicolumn{3}{c}{$2.04_{-0.06}^{+0.07}$} &\multicolumn{3}{c}{$0.13_{-0.06}^{+0.10}$} \\
distance [pc]           &\multicolumn{3}{c}{$1789_{-55}^{+112}$} &\multicolumn{3}{c}{$3150_{-117}^{+196}$} \\  
\hline
\end{tabular}
\end{table*}

\begin{table*}
 \centering
\caption{The same as in Table~\ref{tab:syntheticfit_TIC37743815+42565581} above, but for TIC 54060695 and TIC 178010808.}
 \label{tab:syntheticfit_TIC54060695+178010808}
\begin{tabular}{@{}lllllll}
\hline
 & \multicolumn{3}{c}{TIC\,54060695} & \multicolumn{3}{c}{TIC\,178010808} \\
\hline
\multicolumn{7}{c}{orbital elements} \\
\hline
   & \multicolumn{3}{c}{subsystem} & \multicolumn{3}{c}{subsystem}  \\
   & \multicolumn{2}{c}{Aa--Ab} & A--B & \multicolumn{2}{c}{Aa--Ab} & A--B \\
  \hline
  $t_0$ [BJD - 2400000] & \multicolumn{3}{c}{58468.0} & \multicolumn{3}{c}{58491.5} \\
  $P$ [days] & \multicolumn{2}{c}{$1.060801_{-0.000017}^{+0.000014}$} & $60.7759_{-0.0011}^{+0.0011}$ & \multicolumn{2}{c}{$0.826496_{-0.000014}^{+0.000014}$} & $69.083_{-0.030}^{+0.022}$  \\
  $a$ [R$_\odot$] & \multicolumn{2}{c}{$5.905_{-0.033}^{+0.031}$} & $107.8_{-0.5}^{+0.8}$ & \multicolumn{2}{c}{$5.100_{-0.047}^{+0.017}$} & $114.9_{-1.3}^{+0.5}$ \\
  $e$ & \multicolumn{2}{c}{$0.00211_{-0.00085}^{+0.00096}$} & $0.0154_{-0.0090}^{+0.0090}$ & \multicolumn{2}{c}{$0.00026_{-0.00012}^{+0.00014}$} & $0.289_{-0.007}^{+0.024}$ \\
  $\omega$ [deg] & \multicolumn{2}{c}{$129_{-12}^{+22}$} & $89.8_{-2.5}^{+1.8}$ & \multicolumn{2}{c}{$69_{-24}^{+57}$} & $67.9_{-1.0}^{+1.3}$ \\ 
  $i$ [deg] & \multicolumn{2}{c}{$89.11_{-0.46}^{+0.44}$} & $88.98_{-0.12}^{+0.17}$ & \multicolumn{2}{c}{$86.16_{-0.22}^{+0.27}$} & $88.485_{-0.046}^{+0.021}$ \\
  $\mathcal{T}_0^\mathrm{inf/sup}$ [BJD - 2400000]& \multicolumn{2}{c}{$58486.6459_{-0.0003}^{+0.0003}$} & ${58474.5313_{-0.0088}^{+0.0088}}^*$ & \multicolumn{2}{c}{$58492.0717_{-0.0001}^{+0.0001}$} & ${58512.6976_{-0.0024}^{+0.0023}}^*$ \\
  $\tau$ [BJD - 2400000]& \multicolumn{2}{c}{$58486.231_{-0.036}^{+0.064}$} & $58473.6_{-31.5}^{+29.0}$ & \multicolumn{2}{c}{$58491.607_{-0.059}^{+0.116}$} & $58510.45_{-0.13}^{+0.19}$ \\
  $\Omega$ [deg] & \multicolumn{2}{c}{$0.0$} & $-2.81_{-2.15}^{+3.52}$ & \multicolumn{2}{c}{$0.0$} & $-1.84_{-0.29}^{+0.36}$ \\
  $i_\mathrm{m}$ [deg] & \multicolumn{3}{c}{$3.20_{-2.36}^{+1.81}$} & \multicolumn{3}{c}{$2.95_{-0.26}^{+0.25}$} \\
  \hline
  mass ratio $[q=m_\mathrm{sec}/m_\mathrm{pri}]$ & \multicolumn{2}{c}{$0.621_{-0.006}^{+0.005}$} & $0.858_{-0.010}^{+0.011}$ & \multicolumn{2}{c}{$0.940_{-0.004}^{+0.003}$} & $0.635_{-0.005}^{+0.004}$ \\
$K_\mathrm{pri}$ [km\,s$^{-1}$] & \multicolumn{2}{c}{$107.8_{-0.7}^{+0.6}$} & $41.5_{-0.3}^{+0.6}$ & \multicolumn{2}{c}{$151.0_{-1.4}^{+0.6}$} & $34.1_{-0.2}^{+0.2}$ \\ 
$K_\mathrm{sec}$ [km\,s$^{-1}$] & \multicolumn{2}{c}{$173.8_{-1.2}^{+1.2}$} & $48.3_{-0.2}^{+0.3}$ & \multicolumn{2}{c}{$160.6_{-1.1}^{+0.5}$} & $53.8_{-0.1}^{+0.1}$ \\ 
  \hline  
\multicolumn{7}{c}{stellar parameters} \\
\hline
   & Aa & Ab &  B & Aa & Ab &  B \\
  \hline
 \multicolumn{7}{c}{Relative quantities} \\
  \hline
 fractional radius [$R/a$]  & $0.2744_{-0.0024}^{+0.0017}$ & $0.1426_{-0.0015}^{+0.0013}$ & $0.0774_{-0.0010}^{+0.0009}$ & $0.2994_{-0.0011}^{+0.0014}$ & $0.2642_{-0.0021}^{+0.0016}$ & $0.0249_{-0.0003}^{+0.0017}$ \\
 temperature relative to $(T_\mathrm{eff})_\mathrm{Aa}$ & $1$ & $0.7593_{-0.0064}^{+0.0057}$ & $0.6911_{-0.0056}^{+0.0050}$ & $1$ & $0.9819_{-0.0009}^{+0.0010}$ & $0.9543_{-0.0368}^{+0.0058}$ \\
 fractional flux [in \textit{TESS}-band] & $0.1070_{-0.0008}^{+0.0008}$ & $0.0122_{-0.0003}^{+0.0003}$ & $0.8396_{-0.0169}^{+0.0270}$ & $0.2008_{-0.0024}^{+0.0027}$ & $0.1482_{-0.0019}^{+0.0014}$ & $0.6340_{-0.0112}^{+0.0084}$ \\
 \hline
 \multicolumn{7}{c}{Physical Quantities} \\
  \hline 
 $m$ [M$_\odot$] & $1.513_{-0.027}^{+0.025}$ & $0.939_{-0.015}^{+0.012}$ & $2.099_{-0.032}^{+0.066}$ & $1.341_{-0.035}^{+0.012}$ & $1.261_{-0.035}^{+0.013}$ & $1.650_{-0.060}^{+0.028}$ \\
 $R$ [R$_\odot$] & $1.622_{-0.022}^{+0.015}$ & $0.842_{-0.013}^{+0.011}$ & $8.345_{-0.119}^{+0.112}$ & $1.526_{-0.008}^{+0.007}$ & $1.348_{-0.019}^{+0.012}$ & $2.859_{-0.036}^{+0.158}$ \\
 $T_\mathrm{eff}$ [K]& $7358_{-87}^{+151}$ & $5590_{-48}^{+61}$ & $5085_{-47}^{+68}$ & $6331_{-47}^{+198}$ & $6214_{-43}^{+200}$ & $6028_{-82}^{+52}$ \\
 $L_\mathrm{bol}$ [L$_\odot$] & $6.838_{-0.329}^{+0.723}$ & $0.618_{-0.029}^{+0.042}$ & $41.62_{-1.17}^{+2.16}$ & $3.369_{-0.093}^{+0.376}$ & $2.447_{-0.063}^{+0.229}$ & $9.772_{-0.261}^{+0.849}$ \\
 $M_\mathrm{bol}$ & $2.68_{-0.11}^{+0.05}$ & $5.29_{-0.07}^{+0.05}$ & $0.72_{-0.05}^{+0.03}$ & $3.45_{-0.11}^{+0.03}$ & $3.80_{-0.10}^{+0.03}$ & $2.29_{-0.09}^{+0.03}$ \\
 $M_V           $ & $2.64_{-0.11}^{+0.06}$ & $5.38_{-0.08}^{+0.06}$ & $0.95_{-0.06}^{+0.04}$ & $3.42_{-0.10}^{+0.03}$ & $3.78_{-0.09}^{+0.03}$ & $2.30_{-0.07}^{+0.03}$ \\
 $\log g$ [dex] & $4.199_{-0.006}^{+0.006}$ & $4.559_{-0.006}^{+0.007}$ & $2.919_{-0.015}^{+0.010}$ & $4.196_{-0.005}^{+0.003}$ & $4.278_{-0.003}^{+0.005}$ & $3.744_{-0.065}^{+0.012}$ \\
 \hline
\multicolumn{7}{c}{Global system parameters} \\
  \hline
$\log$(age) [dex] &\multicolumn{3}{c}{$9.017_{-0.037}^{+0.022}$} &\multicolumn{3}{c}{$9.335_{-0.018}^{+0.014}$} \\
$[M/H]$  [dex]    &\multicolumn{3}{c}{$-0.042_{-0.039}^{+0.048}$} &\multicolumn{3}{c}{$0.263_{-0.245}^{+0.085}$} \\
$E(B-V)$ [mag]    &\multicolumn{3}{c}{$0.115_{-0.018}^{+0.024}$} &\multicolumn{3}{c}{$0.058_{-0.010}^{+0.018}$} \\
extra light $\ell_4$ [in \textit{TESS}-band] & \multicolumn{3}{c}{$0.041_{-0.027}^{+0.017}$} & \multicolumn{3}{c}{$0.024_{-0.016}^{+0.019}$} \\
$(M_V)_\mathrm{tot}$  &\multicolumn{3}{c}{$0.73_{-0.07}^{+0.04}$} &\multicolumn{3}{c}{$1.78_{-0.08}^{+0.02}$} \\
distance [pc]           &\multicolumn{3}{c}{$2427_{-34}^{+33}$} &\multicolumn{3}{c}{$1415_{-14}^{+35}$} \\  
\hline
\end{tabular}

{\textit{Notes. }{$\mathcal{T}_0^\mathrm{inf/sup}$ denotes the moment of an inferior or superior conjunction of the secondary (Ab) and the tertiary (B) along their inner and outer orbits, respectively. Superior conjunctions are noted with $^*$.}}
\end{table*}

\begin{table*}
 \centering
\caption{The same as in Table~\ref{tab:syntheticfit_TIC37743815+42565581} above, but for TIC 242132789 and TIC 456194776.}
 \label{tab:syntheticfit_TIC242132789+456194776}
\begin{tabular}{@{}lllllll}
\hline
 & \multicolumn{3}{c}{TIC\,242132789} & \multicolumn{3}{c}{TIC\,456194776} \\
\hline
\multicolumn{7}{c}{orbital elements} \\
\hline
   & \multicolumn{3}{c}{subsystem} & \multicolumn{3}{c}{subsystem}  \\
   & \multicolumn{2}{c}{Aa--Ab} & A--B & \multicolumn{2}{c}{Aa--Ab} & A--B \\
  \hline
  $t_0$ [BJD - 2400000] & \multicolumn{3}{c}{58468.0} & \multicolumn{3}{c}{58790.0} \\
  $P$ [days] & \multicolumn{2}{c}{$5.1287_{-0.0013}^{+0.0013}$} & $42.0317_{-0.0085}^{+0.0091}$ & \multicolumn{2}{c}{$1.7192540_{-0.0000075}^{+0.0000071}$} & $93.915_{-0.038}^{+0.045}$   \\
  $a$ [R$_\odot$] & \multicolumn{2}{c}{$16.98_{-0.18}^{+0.13}$} & $81.02_{-0.91}^{+0.68}$ & \multicolumn{2}{c}{$8.287_{-0.030}^{+0.019}$} & $143.8_{-1.1}^{+0.4}$ \\
  $e$ & \multicolumn{2}{c}{$0.01644_{-0.00042}^{+0.00041}$} & $0.0055_{-0.0030}^{+0.0037}$ & \multicolumn{2}{c}{$0.00293_{-0.00043}^{+0.00060}$} & $0.288_{-0.043}^{+0.040}$ \\
  $\omega$ [deg] & \multicolumn{2}{c}{$311.4_{-2.3}^{+2.4}$} & $171_{-51}^{+106}$ & \multicolumn{2}{c}{$204_{-9}^{+15}$} & $198.9_{-1.8}^{+2.0}$ \\ 
  $i$ [deg] & \multicolumn{2}{c}{$88.08_{-0.40}^{+0.47}$} & $89.47_{-0.15}^{+0.14}$ & \multicolumn{2}{c}{$89.50_{-0.85}^{+0.39}$} & $88.578_{-0.035}^{+0.035}$ \\
  $\mathcal{T}_0^\mathrm{inf/sup}$ [BJD - 2400000] & \multicolumn{2}{c}{${58470.1198_{-0.0011}^{+0.0011}}^*$} & $58484.4801_{-0.0053}^{+0.0054}$ & \multicolumn{2}{c}{$58791.5538_{-0.0003}^{+0.0002}$} & ${58809.8646_{-0.0046}^{+0.0046}}^*$ \\
  $\tau$ [BJD - 2400000] & \multicolumn{2}{c}{$58468.163_{-0.033}^{+0.035}$} & $58471.0_{-18.9}^{+5.0}$ & \multicolumn{2}{c}{$58791.237_{-0.041}^{+0.069}$} & $58735.9_{-1.3}^{+1.8}$ \\
  $\Omega$ [deg] & \multicolumn{2}{c}{$0.0$} & $-0.85_{-0.93}^{+2.66}$ & \multicolumn{2}{c}{$0.0$} & $-1.06_{-0.46}^{+0.71}$ \\
  $i_\mathrm{m}$ [deg] & \multicolumn{3}{c}{$2.00_{-0.64}^{+0.88}$} & \multicolumn{3}{c}{$1.52_{-0.76}^{+0.40}$} \\
  \hline
  mass ratio $[q=m_\mathrm{sec}/m_\mathrm{pri}]$ & \multicolumn{2}{c}{$0.852_{-0.008}^{+0.008}$} & $0.618_{-0.005}^{+0.005}$ & \multicolumn{2}{c}{$0.762_{-0.005}^{+0.005}$} & $0.750_{-0.020}^{+0.011}$ \\
$K_\mathrm{pri}$ [km\,s$^{-1}$] & \multicolumn{2}{c}{$77.0_{-0.7}^{+0.7}$} & $37.3_{-0.5}^{+0.4}$ & \multicolumn{2}{c}{$105.4_{-0.6}^{+0.5}$} & $34.5_{-0.5}^{+0.4}$  \\ 
$K_\mathrm{sec}$ [km\,s$^{-1}$] & \multicolumn{2}{c}{$90.4_{-1.1}^{+0.8}$} & $60.3_{-0.6}^{+0.5}$ & \multicolumn{2}{c}{$138.4_{-0.5}^{+0.4}$} & $46.1_{-0.6}^{+0.6}$ \\ 
  \hline  
\multicolumn{7}{c}{stellar parameters} \\
\hline
   & Aa & Ab &  B & Aa & Ab &  B \\
  \hline
 \multicolumn{7}{c}{Relative quantities} \\
  \hline
 fractional radius [$R/a$] & $0.1027_{-0.0018}^{+0.0018}$ & $0.0712_{-0.0012}^{+0.0013}$ & $0.1509_{-0.0010}^{+0.0010}$ & $0.1995_{-0.0017}^{+0.0019}$ & $0.1274_{-0.0013}^{+0.0013}$  & $0.0344_{-0.0005}^{+0.0005}$ \\
 temperature relative to $(T_\mathrm{eff})_\mathrm{Aa}$ & $1$ & $0.9658_{-0.0090}^{+0.0085}$ & $0.7170_{-0.0074}^{+0.0102}$ & $1$ & $0.8808_{-0.0050}^{+0.0038}$ & $0.8787_{-0.0082}^{+0.0065}$ \\
 fractional flux [in \textit{TESS}-band] & $0.0597_{-0.0013}^{+0.0013}$ & $0.0256_{-0.0010}^{+0.0013}$ & $0.8779_{-0.0279}^{+0.0269}$ & $0.1354_{-0.0028}^{+0.0028}$ & $0.0368_{-0.0005}^{+0.0006}$ & $0.8153_{-0.0230}^{+0.0087}$ \\
 fractional flux [in $R_C$-band] & $-$ & $-$ & $-$ & $0.1399_{-0.0049}^{+0.0058}$ & $0.0362_{-0.0014}^{+0.0012}$ & $0.7944_{-0.0307}^{+0.0181}$ \\
 \hline
 \multicolumn{7}{c}{Physical Quantities} \\
  \hline 
 $m$ [M$_\odot$] & $1.346_{-0.045}^{+0.031}$ & $1.146_{-0.034}^{+0.027}$ & $1.539_{-0.060}^{+0.046}$ & $1.464_{-0.015}^{+0.010}$ & $1.115_{-0.014}^{+0.010}$ & $1.939_{-0.056}^{+0.035}$ \\
 $R$ [R$_\odot$] & $1.741_{-0.030}^{+0.031}$ & $1.207_{-0.027}^{+0.029}$ & $12.22_{-0.13}^{+0.12}$ & $1.653_{-0.017}^{+0.017}$ & $1.055_{-0.013}^{+0.012}$ & $4.940_{-0.084}^{+0.066}$ \\
 $T_\mathrm{eff}$ [K] & $6568_{-43}^{+101}$ & $6367_{-66}^{+33}$ & $4734_{-63}^{+30}$ & $6709_{-138}^{+263}$ & $5924_{-114}^{+176}$ & $5920_{-120}^{+142}$ \\
 $L_\mathrm{bol}$ [L$_\odot$] & $5.118_{-0.262}^{+0.266}$ & $2.135_{-0.135}^{+0.154}$ & $67.27_{-3.34}^{+1.79}$ & $5.059_{-0.505}^{+0.708}$ & $1.245_{-0.117}^{+0.131}$ & $27.12_{-2.54}^{+2.54}$ \\
 $M_\mathrm{bol}$ & $3.00_{-0.06}^{+0.06}$ & $3.95_{-0.08}^{+0.07}$ & $0.20_{-0.03}^{+0.06}$ & $3.01_{-0.14}^{+0.11}$ & $4.53_{-0.11}^{+0.11}$ & $1.19_{-0.10}^{+0.11}$ \\
 $M_V           $ & $2.99_{-0.06}^{+0.06}$ & $3.95_{-0.08}^{+0.07}$ & $0.58_{-0.04}^{+0.10}$ & $2.97_{-0.13}^{+0.11}$ & $4.55_{-0.12}^{+0.12}$ & $1.26_{-0.11}^{+0.10}$ \\
 $\log g$ [dex] & $4.083_{-0.016}^{+0.017}$ & $4.331_{-0.013}^{+0.012}$ & $2.450_{-0.008}^{+0.007}$ & $4.166_{-0.008}^{+0.007}$ & $4.437_{-0.007}^{+0.008}$ & $3.336_{-0.015}^{+0.013}$ \\
 \hline
\multicolumn{7}{c}{Global system parameters} \\
  \hline
$\log$(age) [dex] & \multicolumn{2}{c}{$9.393_{-0.053}^{+0.040}$} & $9.401_{-0.023}^{+0.023}$ & \multicolumn{3}{c}{$9.144_{-0.019}^{+0.021}$} \\
$[M/H]$  [dex]    & \multicolumn{3}{c}{$-0.087_{-0.080}^{+0.072}$} & \multicolumn{3}{c}{$0.221_{-0.203}^{+0.089}$} \\
$E(B-V)$ [mag]    & \multicolumn{3}{c}{$0.548_{-0.028}^{+0.013}$} & \multicolumn{3}{c}{$0.136_{-0.038}^{+0.051}$} \\
extra light $\ell_4$ [in \textit{TESS}-band] & \multicolumn{3}{c}{$0.036_{-0.026}^{+0.029}/0.025_{-0.019}^{+0.019}$} & \multicolumn{3}{c}{$0.013_{-0.009}^{+0.024}$} \\
extra light $\ell_4$ [in $R_C$-band] & \multicolumn{3}{c}{$-$} & \multicolumn{3}{c}{$0.028_{-0.019}^{+0.040}$} \\
$(M_V)_\mathrm{tot}$  & \multicolumn{3}{c}{$0.42_{-0.04}^{+0.09}$} & \multicolumn{3}{c}{$1.01_{-0.11}^{+0.10}$} \\
distance [pc]         & \multicolumn{3}{c}{$2667_{-28}^{+28}$} & \multicolumn{3}{c}{$1609_{-24}^{+23}$} \\  
\hline
\end{tabular}

\textit{Notes. }{$\mathcal{T}_0^\mathrm{inf/sup}$ denotes the moment of an inferior or superior conjunction of the secondary (Ab) and the tertiary (B) along their inner and outer orbits, respectively. Superior conjunctions are noted with $^*$.}
\end{table*}

In relatively close systems, as we are studying here, perturbations to the EB orbit and the detailed profiles of the third body eclipses carry important information about the system parameters, including constraints on the masses and orbital elements.  However, with only one exception in this current work, we have no RV measurements to help constrain the system parameters.  Therefore, we adopt a somewhat different strategy.  In the analysis we utilize some a priori knowledge of stellar astrophysics and evolution with the use of \texttt{PARSEC} isochrones and evolutionary tracks \citep{PARSEC}. We make use of tabulated three-dimensional grids in triplets of \{age, metallicity, initial stellar mass\} of \texttt{PARSEC} isochrones that have stellar temperatures, radii, surface gravities, luminosities, and magnitudes in different passbands of several photometric systems.  Then, we allow the three parameters \{age, metallicity, initial stellar mass\} to vary as adjustable MCMC variables. The stellar temperature, radius, and actual passband magnitude are calculated through trilinear interpolations from the grid points and these values are used to generate synthetic lightcurves and an SED that can be compared to their observational counterparts.  This process, which is also built into {\sc Lightcurvefactory}, is described in detail in \citet{borkovitsetal20a}. In our prior work, we have termed these solutions `MDN' (model-dependent-no-RV solutions) which is what we implement here.  

Regarding the technical details, in the case of the stellar evolution model dependent runs, the freely adjusted (i.e., trial) parameters were as follows:
\begin{itemize}
\item[(i)] Stars: Three stellar masses and the `extra light' contamination, $\ell_4$, from a possible fourth star (or any other contaminating sources in the \textit{TESS} aperture). Additionally, the metallicity of the system ([$M/H$]), the (logarithmic) age of the three stars ($\log\tau$), the interstellar reddening $E(B-V)$ toward the given triple, and its distance, were also varied.
\item[(ii)] Orbits: Three of six orbital-element related parameters of the inner, and six parameters of the outer orbits, i.e., the components of the eccentricity vectors of the two orbits $(e\sin\omega)_\mathrm{in,out}$, $(e\cos\omega)_\mathrm{in,out}$, the inclinations relative to the plane of the sky ($i_\mathrm{in,out}$), and moreover, three other parameters for the outer orbit, including the period ($P_\mathrm{out}$), time of the first (inferior or superior) conjunction of the tertiary star observed in the \textit{TESS} data ($\mathcal{T}_\mathrm{out}^\mathrm{inf,sup})$ and, finally, the longitude of the node relative to the inner binary's node ($\Omega_\mathrm{out}$).
\end{itemize} 

Here we add some additional notes about the `age' and the `distance'.  First, regarding the `age' parameter, our previous experience has led us to believe that, in some cases, it is better to allow the ages of the three stars to be adjusted individually instead of requiring strict coeval evolution. This issue was briefly discussed in \citet{rowdenetal20} and \citet{borkovitsetal21}, and we discuss it below in the case of some individual sources. Regarding the distance of the triple system, one can argue that the accurate trigonometric distances obtained with Gaia \citep{bailer-jonesetal21} should be used as Gaussian priors to penalize the model solutions. But, given that neither DR2 nor the recently released eDR3 Gaia parallaxes have been corrected for the multistellar nature of the objects, we consider the published parallaxes and corresponding distances to be not necessarily accurate for our systems. Therefore, we decided not to utilize the Gaia distances. Instead, we constrained the distance by minimizing the $\chi^2_\mathrm{SED}$ value a posteriori, at the end of each trial step (for results see Sect.~\ref{sec:individual}).

A couple of other parameters were {\it constrained} instead of being adjusted or held constant during our analyses. Specifically, the orbital period of the inner binary ($P_\mathrm{in}$) and its orbital phase (through the time of an arbitrary primary eclipse or, more strictly, the time of the inferior conjunction of the secondary star -- $\mathcal{T}^\mathrm{inf}_\mathrm{in}$) are in this category.  They were constrained internally through the ETV curves.

Regarding the atmospheric parameters of the stars, we handled them in a similar manner as in our previous photodynamical studies. We utilized a logarithmic limb-darkening law \citep{klinglesmithsobieski70} for which the passband-dependent linear and non-linear coefficients were interpolated in each trial step via the tables from the original version of the {\tt Phoebe} software \citep{Phoebe}. We set the gravity darkening exponents for all late type stars to $\beta=0.32$ in accordance with the classic model of \citet{lucy67} valid for convective stars and hold them constant. For several of our systems, however, the analysis of the net SED has revealed that the EBs contain hotter stars, having radiative envelopes. For these stars, we set $\beta=1.0$.  The choice of this parameter, however, has only minor consequences, as the stars under the present investigation are close to spheroids.

For the photodynamical analysis we utilized a more sophisticated processing for the {\it TESS} photometric data by employing the convolution-based differential image analysis tasks of the FITSH package \citet{pal12}. Furthermore, we note that in preparing the observational data for analysis, to save computational time we dropped out the out-of-eclipse sections of the 30-min cadence \textit{TESS} lightcurves, retaining only the $\pm0\fp15$ phase-domain regions around the binary eclipses themselves.  However, during sections of the data containing the third-body (i.e., `outer') eclipses, we kept the data for an entire binary period both before and after the first and last contacts of the given third-body eclipse. 

Moreover, the mid-eclipse times of the inner binaries, used to define the ETV curves, were calculated in a manner that was described in detail in \cite{borkovitsetal16}. The ETVs were one of the inputs to the photodynamical analysis.  We tabulate all the eclipse times in Appendix~\ref{app:ToMs} as online only tables.

Finally, note also some system specific departures from the standard procedures described above in the case of two of our triples. These are as follows: 

(1) In the case of TIC\,242132789, the lightcurve of both \textit{TESS} sectors display non-linear, somewhat erratic variations in the mean out-of-eclipse flux levels. This can partly be attributed to the ellipsoidal light variations (ELV) of the red giant tertiary, which is handled internally by our lightcurve emulator and, therefore, nothing special has to be done to model the ELVs. However, an additional, more erratic contribution to these variations might arise from either time-varying spot activity on the surface of the red giant or any stray light in the aperture (or both). In any case, independent of the origin(s) of this time-varying, irregular contaminating flux, we modelled it by fitting an eighth order polynomial, separately for the two sectors, simultaneously with the triple star lightcurve modelling, during each MCMC step.

(2) The other triple that was handled somewhat differently is TIC 456194776. This system was originally the fourth target of the ground-based photometric follow-up campaign that was described in detail in Sect. 2.3 of \citet{borkovitsetal22}.  In contrast to the other three triply eclipsing triples, of which the detailed analyses were published in \citet{borkovitsetal22}, unfortunately, we were unable to catch any further third-body eclipses during our observing runs.  This is the reason why this system was not included the above-mentioned study, but rather appears in this work.  On the other hand, however, we were able to observe 9 regular inner eclipses from the ground between 12 August 2020 and 12 December 2021. Thus, we decided to include these eclipsing lightcurves (in Cousins $R_C$-band) and also the extracted eclipse times into our analysis.  In addition, near the final stages of our analysis we also acquired RV data for this target. Therefore, we carried out a second kind of analysis for this target which we have termed `MDR' (model-dependent-with-RV) with the inclusion of these RV points. We discuss this latter, MDR solution and compare it with the MDN results in Appendix~\ref{app:T456194776RV}.  This second type of analysis might also be called a `spectro-photodynamical solution'.

\section{Stellar and Orbital Parameters for the Six Triples}
\label{sec:results}

\subsection{Individual Triples}
\label{sec:individual}

In this section we briefly discuss the results for each of the six triply eclipsing triples.  When we refer to parameters, they are the ones taken from Tables \ref{tab:syntheticfit_TIC37743815+42565581} to \ref{tab:ephemerides} based on the photodynamical fits (unless otherwise specifically indicated).   

\subsubsection{TIC 37743815}

The outer period for this system is 68.80 days from the photodynamical solution and 68.72\footnote{In this section we cite the eclipsing periods, taken from Table \ref{tab:ephemerides}.} from the BLS transform of the ASAS-SN and ATLAS data sets.  The EB period is 0.9071 d.  The quantity $e_{\rm out} \cos \omega_{\rm out}$ from a fold of the latter data is 0.28.  The photodyamical fit yields $e_{\rm out} = 0.361$ and $\omega_{\rm out} = 40.5^\circ$, which combine to give $e_{\rm out} \cos \omega_{\rm out} = 0.27$, in excellent agreement with the findings of the archival data.  The mutual inclination angle between the inner EB and outer orbit is $1.8^\circ \pm 1^\circ$.  The overall system is incredibly flat with $i_{\rm in}$ and $i_{\rm out}$ both within 1/2$^\circ$ of 90$^\circ$.

The mass and radius of the tertiary are 1.6 M$_\odot$ and 4.0 R$_\odot$, respectively.  Both EB stars are considerably lower in mass at 1.1 M$_\odot$ and 0.7 M$_\odot$.  The results from the SED fit (Fig.~\ref{fig:SEDs}) give the same mass for the tertiary star and a radius that is 12\% larger than the photodyamical solution. The primary in the EB has a 16\% higher mass from the SED fit than given by the photodynamical fit, while the secondary EB mass is in excellent agreement. In fact, it is instructive to compare the results of the SED fits with those of the more detailed photodyamical results as a type of calibration of the former approach.  We mention these comparisons for all the systems.

The distance from the photodynamical fit of 1790 pc is within the uncertainties of 1857 pc given in Table \ref{tbl:mags}.  $E(B-V)$ from the photodynamical fit is a bit higher, but significantly so, than the value given in Table \ref{tbl:mags}.  We estimate the age of the system as 2.4 Gyr.

\subsubsection{TIC 42565581}

The outer orbital period of this triple is the longest among our sample of six systems at 123.5 days.  This is in excellent agreement with what we found from the archival photometric data.  The EB period is 1.8231 days.  The system is not exceptionally flat with $i_{\rm mut} = 5.5^\circ \pm 1.6^\circ$ with the inclination angles of the inner and outer orbits both within a few tenths of a degree of edge on.  The eccentricity of the outer orbit is $e_{\rm out} =0.16$ with $\omega_{\rm out} \simeq 61^{+16}_{-28}$ deg.  This combination would be in disagreement with  the value of $e_{\rm out} \cos \omega_{\rm out} = 0.03$ found from the archival photometric data unless the former value of $\omega_{\rm out}$ is near its 1-$\sigma$ upper limit of $\sim$77$^\circ$.

The tertiary star is the most massive of our sample at 2.2 M$_\odot$, and is substantially evolved off the main sequence with R = 8.4\,R$_\odot$.  The EB stars are near twins with masses of 1.79 M$_\odot$, which are much hotter at 7500 K than the tertiary at 5200 K.  These are all in satisfactory agreement with the results of our simpler SED fit, except that the mass of the tertiary in the latter fit was about 14\% higher, while the EB stars were $\sim$20\% lower in mass.  The radius of the tertiary was the same in both fits.

We find a photometric distance to this source of 3150 pc (with an uncertainty of $\sim$150 pc), which agrees well with the Gaia distance of $3180 \pm 160$ pc.  The fitted $E(B-V)$ for this source is $0.39 \pm 0.04$ compared with the literature tabulated value of only $0.21 \pm 0.02$.  Finally, we find an age $0.89 \pm 0.12$ Gyr for the tertiary vs. $1.11 \pm 0.15$ Gyr for the EB when the two are allowed to have independent ages.

\subsubsection{TIC 54060695}

The outer period for this system is 60.72 days, in good agreement with the long term mean period of 60.68 days found from the combined archival data of ASAS-SN and ATLAS. The two outer eclipses seen in the outer orbit fold of the archival data yielded an $e \cos \omega_{\rm out} \simeq 0.002$, while the photodynamical fit yielded separate parameter values of $e_{\rm out} = 0.015$ and $\omega_{\rm out} = 89^\circ$.  The mutual inclination angle, $i_{\rm mut}$, between the plane of the inner EB (with period 1.0605 d) and outer orbit is $3^\circ \pm 2^\circ$, while the inclination angles of the individual orbital planes are both close to $89^\circ$.  These are all consistent with the presence of both a primary and a secondary outer eclipse.

The tertiary star has a mass of 2.1\,M$_\odot$ and is substantially evolved off the MS with a radius of 8.3\,R$_\odot$.  Its $T_{\rm eff}$ is only 5100 K. The two EB stars are close to 1.5 and 1\,M$_\odot$, with the primary being considerably hotter at 7350 K.  The giant still dominates the system's light with 85\% of the luminosity.  These values are in decent agreement with those found from the SED fit only (see Fig.~\ref{fig:SEDs}), but the masses are consistently lower by $\sim$16\% in the SED fit. 

The photometric distance of $2427 \pm 33$ pc which is formally 3.4 $\sigma$ farther than the Gaia distance of $2221 \pm 50$ pc. The fitted value of $E(B-V)$ compares well to the one listed in Table \ref{tbl:mags}. The system has an inferred age of 1.04 Gyr.

\begin{figure*}
\begin{center}
\includegraphics[width=0.32 \textwidth]{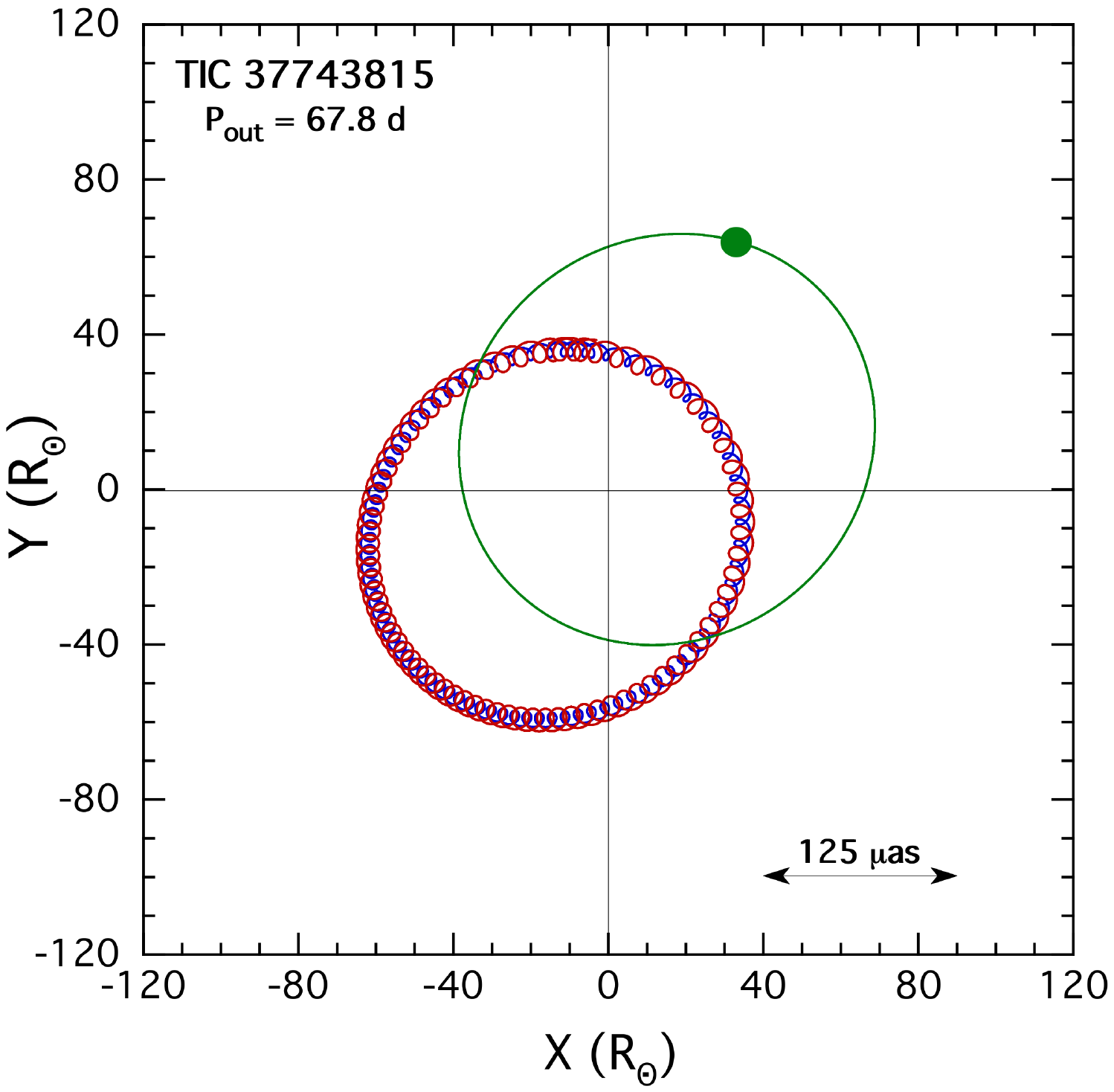}
\includegraphics[width=0.32 \textwidth]{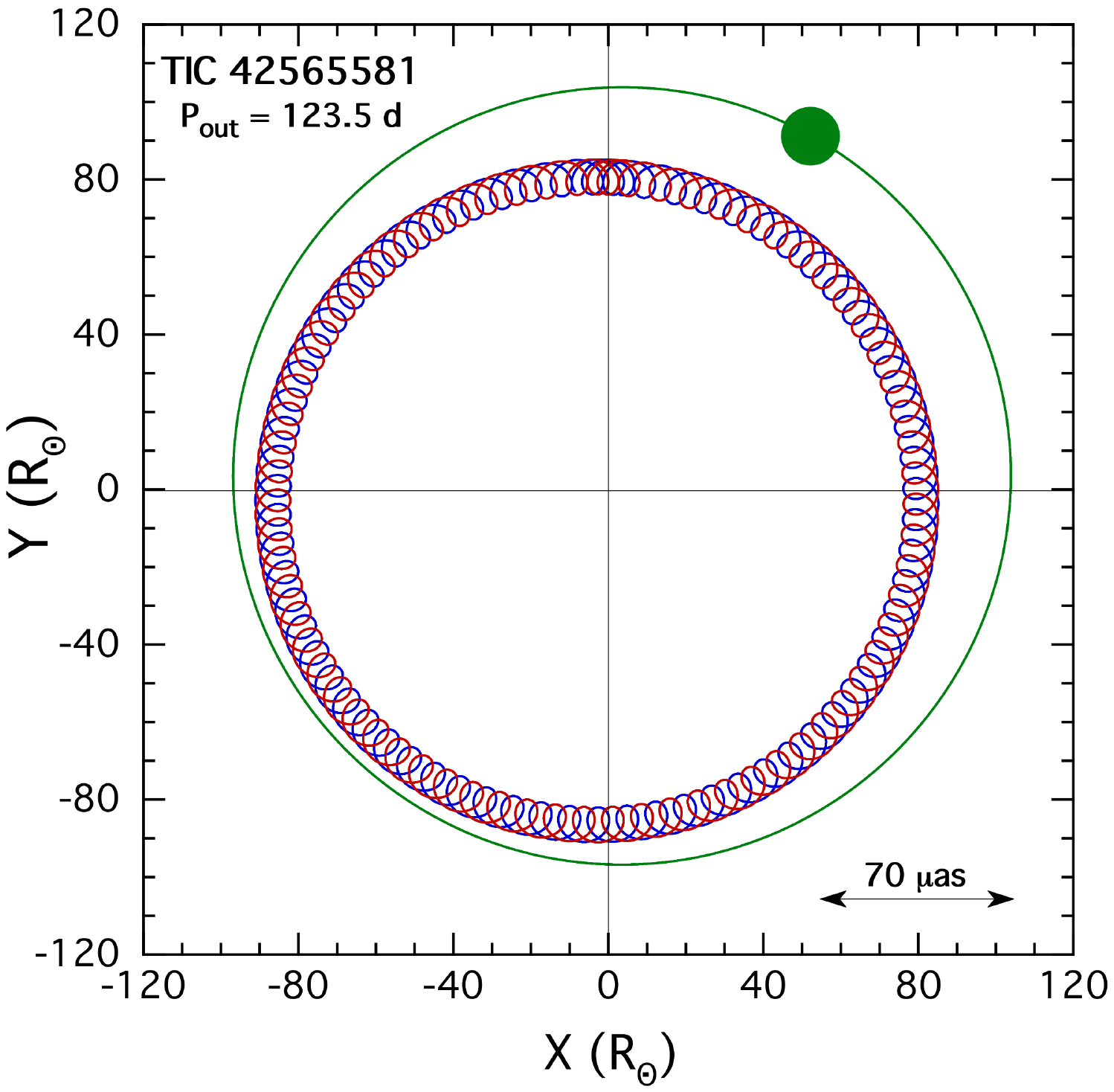}
\includegraphics[width=0.32 \textwidth]{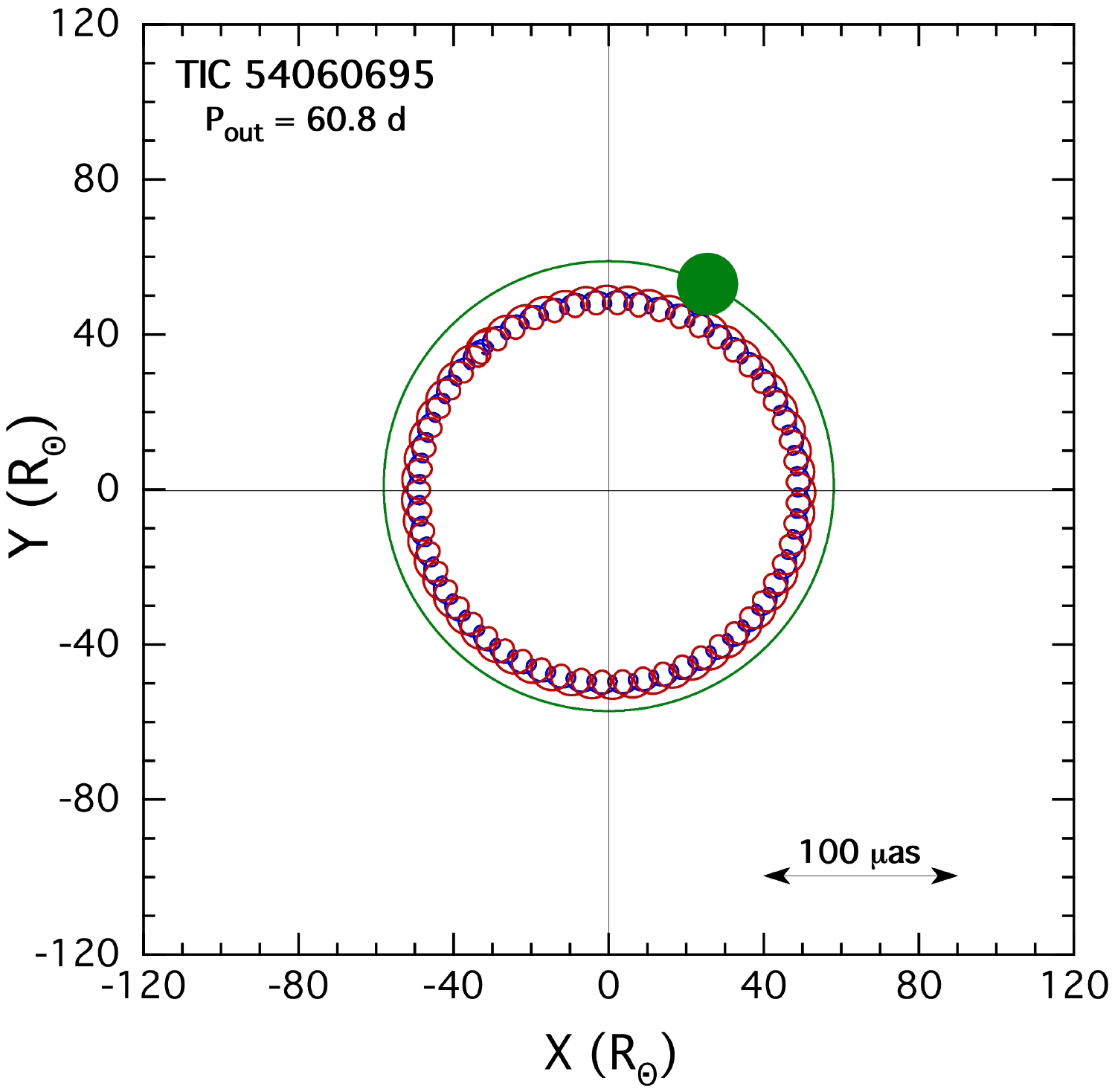}
\includegraphics[width=0.32 \textwidth]{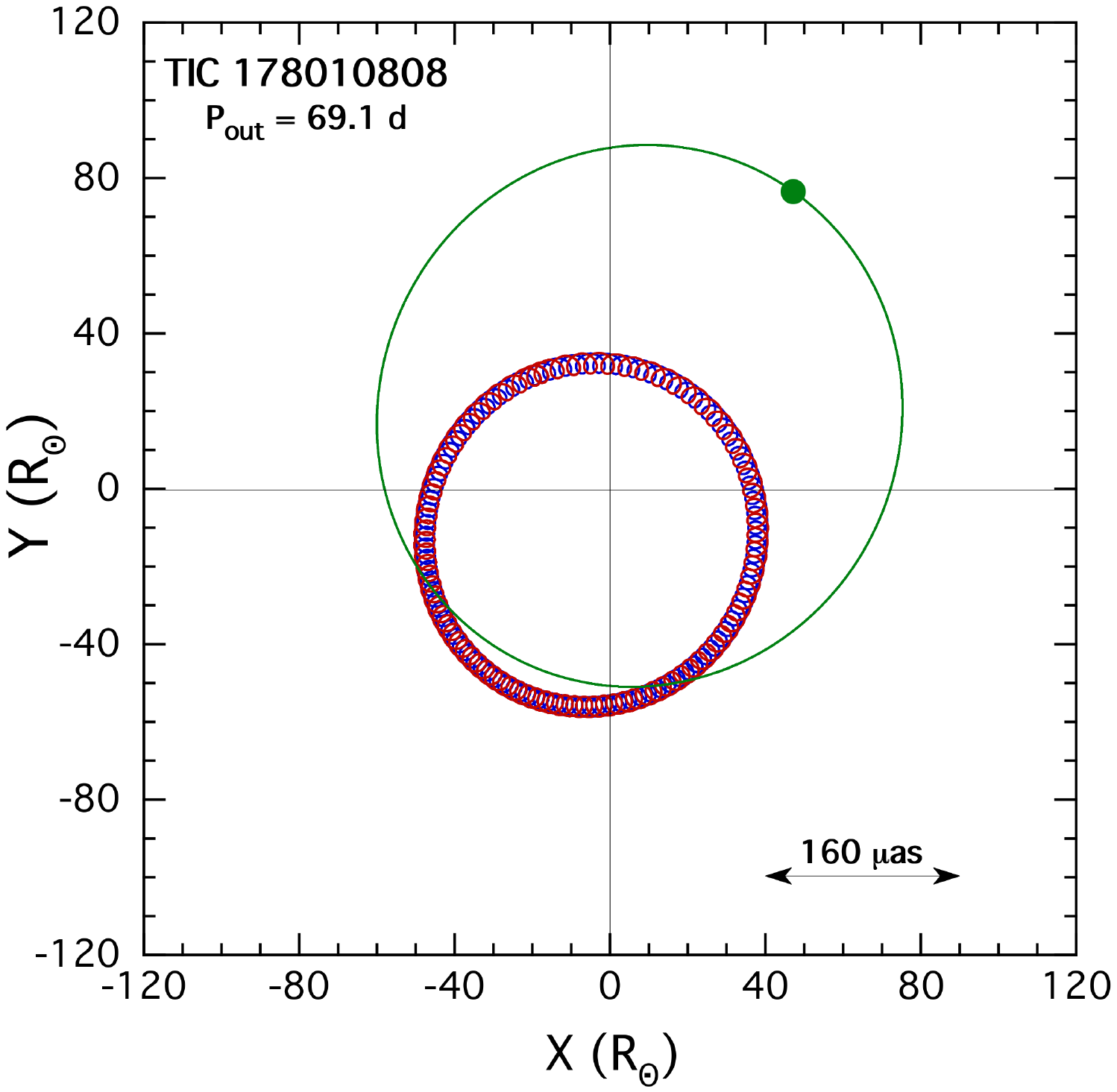}
\includegraphics[width=0.32 \textwidth]{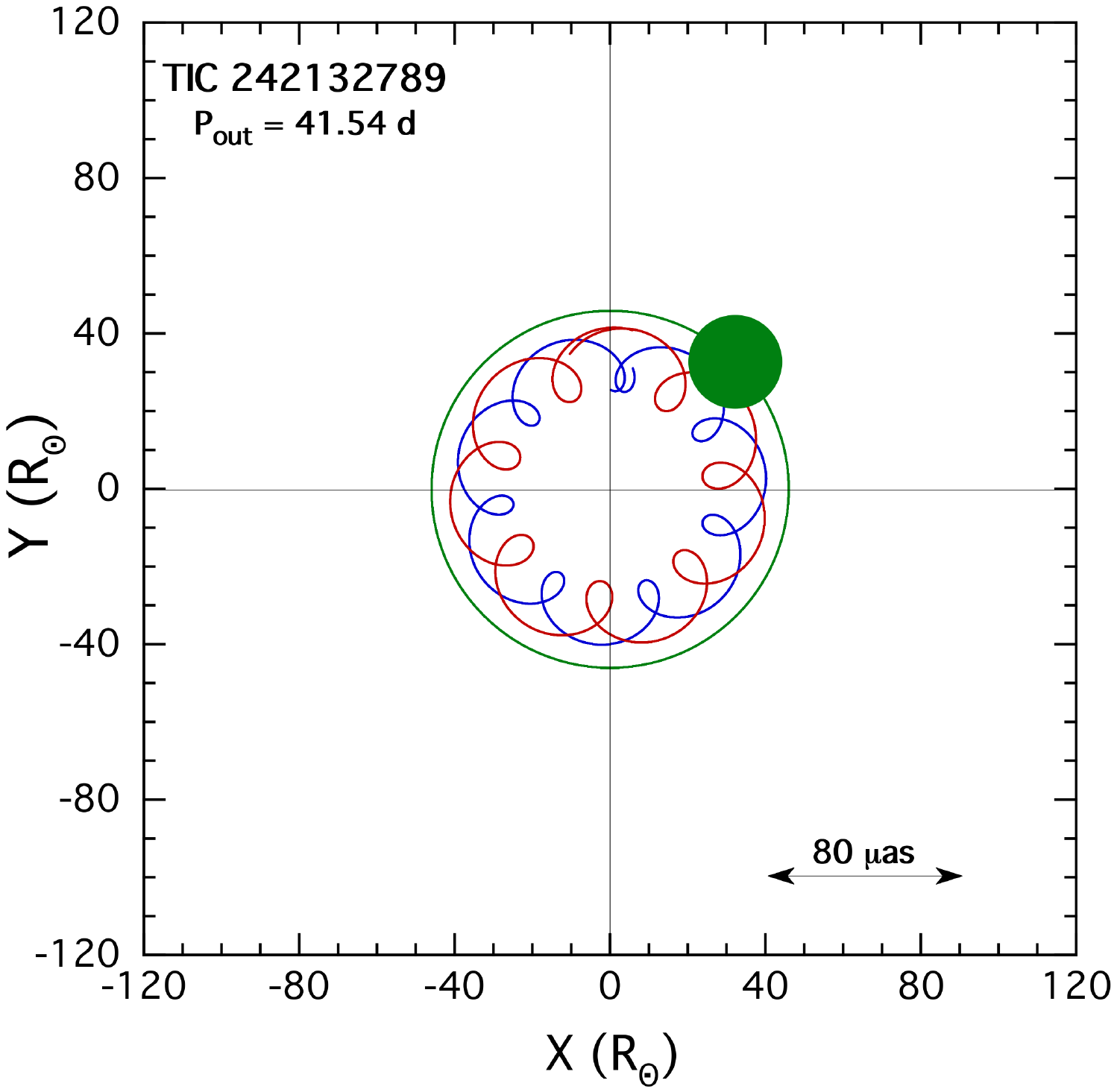}   
\includegraphics[width=0.32 \textwidth]{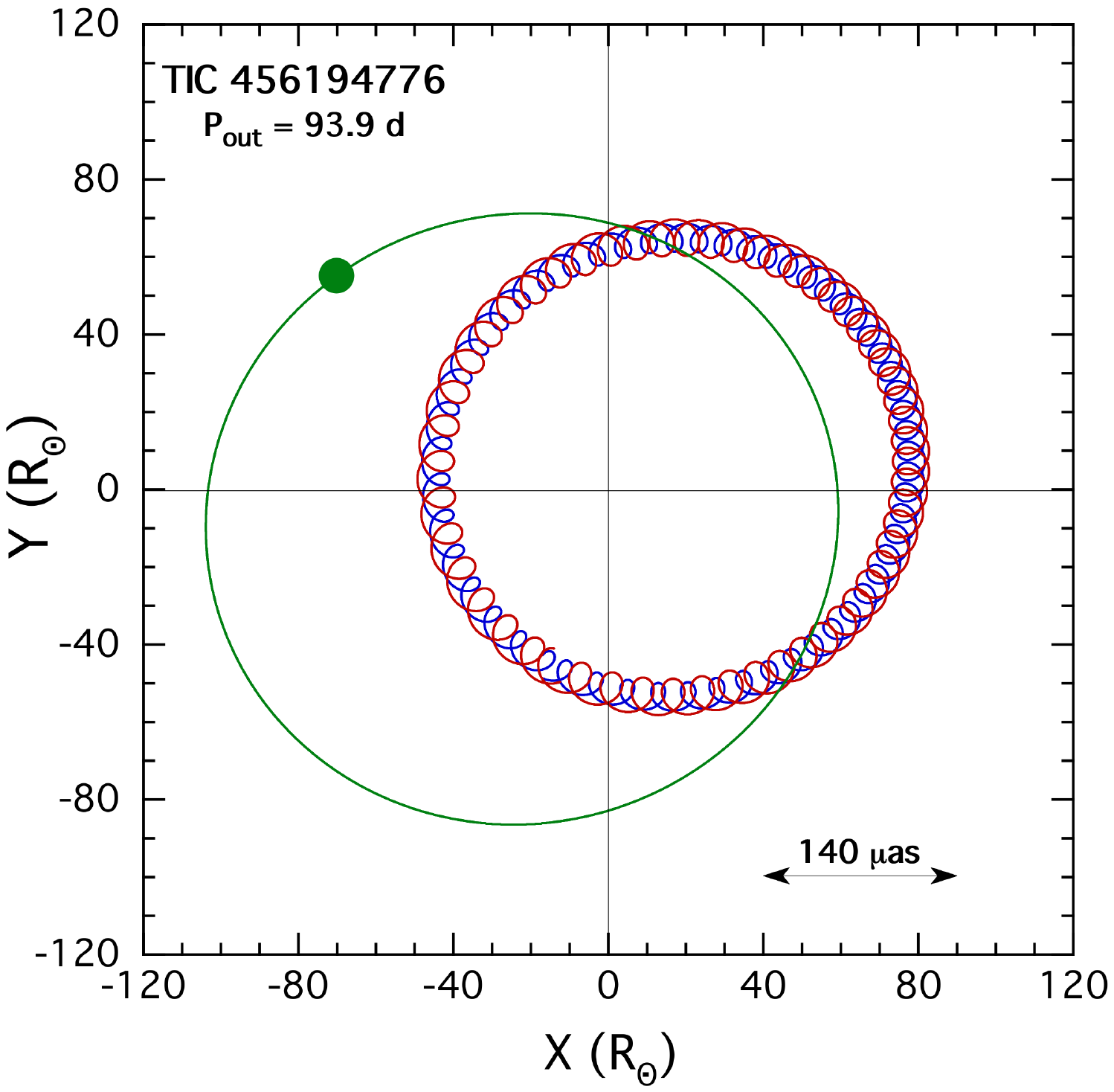}   
\caption{The outer orbits of the six triply eclipsing systems seen from above the orbital plane.  The stars are all moving counter-clockwise.  The observer is at $y \rightarrow \infty$.  Red and blue tracks are for the primary and secondary stars in the EB, respectively, while the green track is that of the tertiary star.  The heavy filled green circle represents the size of the tertiary to scale on the plot.  Each panel has an angular scale in units of micro-arcseconds.  A discussion about how the stellar motions may affect the Gaia distance determinations is given in Appendix \ref{sec:lightcentroid}.}
\label{fig:orbits}
\end{center}
\end{figure*} 

\subsubsection{TIC 178010808}

The outer orbit of TIC 178010808 has a period of 69.02 days, and is also rather flat with $i_{\rm mut} = 3^\circ \pm 0.3^\circ$, and relatively close to edge on with $i_{\rm in} = 86^\circ$, and $i_{\rm out} = 88.5^\circ$. The EB period is 0.8257 days. The eccentricity is characterized by $e_{\rm out} = 0.29$ with $\omega_{\rm out} = 68^\circ$.  The outer orbit fold using the ASAS-SN,  ATLAS, and WASP archival data (see Fig.~\ref{fig:outer_fold}) shows a clear detection of the primary outer eclipse, but there is merely a suggestion of a secondary outer eclipse which is not statistically significant with $e_{\rm out} \cos \omega_{\rm out} \simeq 0.16$. This is in substantial disagreement with $e_{\rm out} \cos \omega_{\rm out} \simeq 0.1$ found in the photodynamical fit.  

All three stars in this system have comparable masses (within $\sim$20\% of each other).  The tertiary has the largest mass at 1.65\,M$_\odot$ with a slightly evolved radius of 2.9\,R$_\odot$.  All three stars have $T_{\rm eff} \simeq 6200 K$.  All the stellar parameters are in good agreement with the results of the simple SED fit which employs only minimal constraints, but the SED-fitted radius for the tertiary is about 13\% larger at 3.3\,R$_\odot$.  In either case, this is the least evolved tertiary among our set of six triples.  

The photodynamic distance of $1415 \pm 30$ pc is in essentially perfect agreement with the Gaia distance (see Table \ref{tbl:mags}).  The fitted value of $E(B-V)$ is small at 0.058, but in agreement with the value listed in Table \ref{tbl:mags}.  Finally, the inferred age of the system is 2.2 Gyr.

\subsubsection{TIC 242132789}

The outer orbit of TIC 242132789 is the shortest among our set of six triply eclipsing triples at 41.5 days.  In that sense it is the most compact of our triples and the fourth shortest period triple system known.  It is also, by far the tightest of our systems with a period ratio of $P_{\rm out}/P_{\rm in}$ of only 8.2\footnote{We define the `tightness' of a binary as $P_{\rm out}/P_{\rm in}$, where the smaller the ratio, the tighter the binary.  In this case, as a technical matter, we use the ratio of periods from Table \ref{tab:syntheticfit_TIC37743815+42565581}.}, the second smallest ratio (after KIC 76668648) among all triple stellar systems where both the inner and outer periods are known with sufficient precision.  TIC 242132789 is now the tightest known triple with the tertiary being the most massive component. The tertiary in the system is also the largest among our six, with R = $12.2 \pm 0.1$\,R$_\odot$. This implies that $R_B/a \simeq 0.15$ so perhaps it is not surprising that the outer orbit has nearly circularized with $e_{\rm out} =0.006$. The mutual inclination angle is marginally significantly different from zero at $i_{\rm mut} = 2.0^{+0.9}_{-0.6}$ degrees.  The two orbital inclination angles are $i_{\rm in} = 88.1^\circ$ and $i_{\rm out} = 89.5^\circ$.

All three stellar components are F stars with the tertiary just higher enough in mass (1.54 M$_\odot$) to have evolved well off the main sequence while the EB primary, at 1.35 M$_\odot$ is only slightly evolved.  The SED fit is in agreement with these basic facts, but it yields $R_B = 15.6 \pm 1.6$\,R$_\odot$, just barely consistent with the photodynamical analysis, and the primary EB mass is lower by 25\% compared to the photodynamical results.  Much of this discrepancy may be attributed to the fact that the SED fit yields a distance of 3250 pc, which in excellent agree with Gaia, while the photodynamical solution prefers a much closer distance.

In fact, the photodynamic distance is $2667 \pm 28$ pc.  This is to be compared to the Gaia distance of $3258 \pm 165$ pc.    This is formally a 3.6 $\sigma$ discrepancy, and may partially be accounted for if the Gaia measurement is affected by the EB stars which contribute 10\% of the system light.  The fitted $E(B-V)$ is close to 0.55, while the value from Table \ref{tbl:mags} is lower at 0.34.  The age of the system is found to be 2.5 Gyr.

This system exhibits the largest ETVs among our sample of triples.  The amplitude is 0.01 days with a period of $P_{\rm out}/2 \simeq 21$ days.  The ETV curve for this source is explored further in Appendix \ref{app:242132789}.

\subsubsection{TIC 456194776}

This system has an outer orbital period of 93.83 days found from the ASAS-SN and ATLAS archival data, which, in turn, is in excellent agreement with the period found from the photodynamical modeling (93.90 days).  The EB period is 1.7193 days. The mutual inclination angle between the inner binary and the outer orbit is 1.5$^\circ$, while $i_{\rm in}$ and $i_{\rm out}$ are 89.5$^\circ$ and 88.6$^\circ$, respectively.  Again, this is a flat and edge on system.  The parameter $e \cos \omega_{\rm out}$ based on the fold of the outer orbit using ASAS-SN and ATLAS data is 0.26.  By comparison the individual components from the photodynamical fits are: $e = 0.29$ and $\omega_{\rm out} = 199^\circ$, leading to $|e \cos \omega_{\rm out}|=0.27$, which is in excellent agreement.

The mass of the tertiary is 1.9 M$_\odot$ and its evolved radius is 4.9 R$_\odot$.  The EB stars are near the main sequence with lower masses of 1.5 and 1.1 R$_\odot$.  The EB primary is considerably hotter at 6700 K than the evolved tertiary at 5900 K.  The SED fit alone yields somewhat hotter EB stars with 16\% higher mass, and good agreement with the photodynamical fit for the radius of the tertiary.  

The photometric distance of 1609 pc is in excellent agreement with the Gaia value of 1590 pc, given that both uncertainties are of order 30 pc.  The fitted photodynamical and  MAST values for $E(B-V)$ of 0.14 vs.~0.17, respectively, are in quite reasonable agreement.  The system age is 1.4 Gyr.

	This was the one triple system in this work for which we also obtained radial velocity data.  We explore in Appendix \ref{app:T456194776RV} how the RVs added to the photodynamical solution.

\subsection{Common Properties}
\label{sec:common}

What all of these six triply eclipsing triple systems have in common is that they are all remarkably flat with $i_{\rm mut}$ within a couple of degrees of zero.  In fact, only one of these systems has $i_{\rm mut} \gtrsim 3^\circ$ and that is TIC 42565581 where $i_{\rm mut} = 5.5^\circ \pm 1.6^\circ$.  The inclination angles of the EB plane and the outer orbit are all within a few degrees of 90$^\circ$, and, of course, that is a strong selection effect since we are searching for outer eclipses as a definitive signature of this type of system.

The orbital periods of the six systems are, in order, 41.5, 60.7, 68.7, 69.0, 93.9, and 123.5 days. As a measure of how compact these systems are, we note that there are only 8 other triple systems known with comparably short periods (see, e.g., \citealt{borkovitsetal22}).  By contrast, only one of our systems is considered `tight', i.e., with a small value of $P_{\rm out}/P_{\rm in}$.  Specifically, five of the six systems have $P_{\rm out}/P_{\rm in}$ in the range of 55-83, while TIC 242132789 is unique in this group with a quite small value of $P_{\rm out}/P_{\rm in} = 8.2$.  This is the second tightest triple known.  

Two of the outer orbits of the six triples have notably small eccentricities ($ e \lesssim 0.02$).  It is interesting that in these two cases, the value of $R_B/a_{\rm out}$ is 0.077 for TIC 54060695 and 0.151 for TIC 242132789.  Therefore, it seems likely that tidal circularization may have played a role in the circularization of these systems where the tertiary is both large and substantially convective.  For the other four systems, $R_B/a_{\rm out}$ ranges from 0.025 to 0.040. 

The tertiary masses in these systems range from 1.6 to 2.2 M$_\odot$, and are all evolved off the main sequence.  They have radii between 2.9 and 12 R$_\odot$.  Their evolutionary ages all range from 1.0-2.5 Gyr.  The large radii of the tertiary stars are also something of a selection effect.  The probability of outer eclipses in these systems is roughly proportional to $R_B/a_{\rm out}$.  Thus, for similar orbital separations as in these systems (50-100 R$_\odot$), the outer eclipse probability can be enhanced by nearly an order of magnitude for the size of the tertiaries we have found as compared to when they were on the main sequence.

\section{Summary and Discussion}
\label{sec:summary}

In this paper we report the discovery and analyses of six triply eclipsing triple systems found from observations with the \textit{TESS} space telescope. They were observed during one to three \textit{TESS} sectors each, yielding precise space-borne photometric data trains.  In four cases the source was observed during two sectors, but those sectors were separated by two years.  However, when combined with archival ground-based survey photometric measurements (see Fig.~\ref{fig:outer_fold}), we were able to obtain reasonably accurate orbital and stellar parameters for all six triple systems via detailed photodynamical analyses.

This is part of an ongoing program to find and characterize compact triple systems via their signature third-body eclipses.  In all we have found 52 such systems during the first three {\it TESS} cycles.  For 20 of these there is sufficient archival data (typically from ASAS-SN and ATLAS, but also including WASP, KELT, and MASCARA) to have determined unambiguously the outer orbit via the long-term detection of third-body eclipses.  Generally, with {\it TESS} we see only one or two third body events because of the sparse coverage, and even when there are two such events, they are sufficiently far apart (i.e., two years), that the outer period is at best ambiguous.  There is of course, the additional possibility that even if two eclipses are seen with {\it TESS}, they are not of the same type, i.e., primary vs.~secondary. Of the 20 triply eclipsing triples where we now know the outer period, we have chosen six more of them from this extensive set to report here\footnote{Four of the sources from this collection have been reported previously in \citet{borkovitsetal20a} and \citet{borkovitsetal22}.}.  This choice of six sources was to strike a balance between being able to discuss each one in some detail, without making the paper too lengthy.  

The vast majority of the triply eclipsing systems that we have discovered in the {\it TESS} data have been found via visual surveys of the lightcurves by the VSG group \citep{kristiansen22}. We employ both a machine learning (ML) approach \citep{powell21} and a direct visual search in looking for multistellar systems.  Empirically, we have found that the ML approach is superior for finding large numbers of EBs and quadruples within the millions of {\it TESS} lightcurves, but that the visual approach is substantially more efficient at finding non-repeating, and odd shaped third body eclipses.  In all, the VSG has surveyed some 10 million lightcurves while finding 52 with a triply eclipsing triples signature as well as many other interesting and unique phenomena (see \citealt{kristiansen22}).  We anticipate that our list of triply eclipsing triples will grow roughly linearly with time as more of the {\it TESS} lightcurves are inspected.

Most of the definitive determinations of the outer orbital periods were made using BLS transforms of archival ground-based photometric data sets, after the existence of third-body events was established from the {\it TESS} data. This raises the question of whether there is a way to find the third body eclipses directly in these archival data sets without first knowing that they are present in a particular source. We are fairly certain that most triple star systems exhibiting eclipses of the tertiary will also contain an eclipsing binary. In that case the dominant source of `noise' in a BLS is from the presence of the much higher duty cycle orbital modulations.  In order to detect the third body events in the archival data, it is generally necessary to first subtract out the EB lightcurve.  While it is possible to automate such a search and removal operation, it seems more efficient at this point to spot the existence of the third body events first in the precision photometry of the {\it TESS} data set.

Now that these triply eclipsing triples are known, and their basic parameters determined, more focused follow-up ground-based photometry, especially with small amateur telescopes, would be welcome. All six objects have G magnitudes in the range of 12.2 to 13.5.  The ordinary primary eclipse depths range from 2-15\%, while the third body eclipses range from a few to 25\% deep. The ETV data from {\it TESS} itself was typically instrumental in determining some of the parameters found from the photodynamical analyses. Thus, future timing observations of the ordinary EB eclipses in these systems would be quite helpful in improving the parameter determinations.  The dynamical delays in these systems range from 0.1 to 14 minutes, while the LTTE delays are typically $\sim$2 minutes, so readily within the realm of amateur observations.  Searches for additional third-body events are difficult without advance approximate predictions since they occur relatively infrequently. In order to facilitate ground-based follow up observations of future third-body events, we provide ephemerides for such observations in Table~\ref{tab:ephemerides}. In some cases, however, these ephemerides are somewhat uncertain, and therefore, we recommend dedicated observations within a wider time domain around each forecasted mid-third-body-eclipse time.  Because of the flatness of all these systems, we predict that there will be no eclipse-depth variations either to search for or to cause long `outages' of eclipses.

The six systems discussed in this work are relatively old, of order 1 to a few Gyr.  They are manifestly dynamically stable and will last until the tertiary overflows its Roche lobe.  In principle, the EB stars could also evolve to mass exchange, but the tertiaries in these systems are sufficiently more massive, and already evolved so that they will fill their Roche lobes, which range in size from 36 to 51 R$_\odot$, before the EB stars grow by even 10\% in size.  Generally, once a star of $\sim$1.5-2.5 M$_\odot$ has grown to 3-12 times its original radius, we can expect it to fully ascend the giant branch within a small fraction of its total lifetime.  These lifetimes are  illustrated in Table \ref{tbl:lifetime}. They also illustrate the fraction of time that systems like these have tertiary stars that are substantially evolved, and hence easier in which to detect third-body eclipses. For example, a 2 M$_\odot$ tertiary that has $R \gtrsim 8\,R_\odot$ spends only 37/1350 = 2.7\% of its total lifetime in this state.

As mentioned above the VSG group \citep{kristiansen22} has visually examined the lightcurves of some 9 million anonymous {\it TESS} lightcurves as well as 1 million {\it TESS} lightcurves of preselected EBs (see \citealt{powell21}; E. Kruse, 2022 in preparation).  In all, they have found 52 triply eclipsing triples with periods in the range of $\sim$42-300 days.  Thirty two of these were found among the EB lightcurves, and only 20 from among the much more numerous anonymous lightcurves.  If each target was observed, on average, during two {\it TESS} sectors spanning about 50 days, then the probability of finding a third body event in at least one of those sectors is of order 50\%, especially when we consider that a fair fraction of the systems exhibit both types of outer eclipses (i.e., primary and secondary).  Here we focus on the 32 systems found from among the preselected EB lightcurves.  This suggests a `success rate' of $3 \times 10^{-5}$ per EB.  In order to assess the actual fraction of EBs that contain a third body in a compact outer orbit that is coaligned with the inner binary, we use a simple Monte Carlo approach.  This takes into account: (i) the probability of seeing a third body eclipse in 50 days of observing; (ii) the fraction of triply eclipsing systems missed because even if the three stars are perfectly aligned it is possible to detect only EB eclipses; and (iii) the detection enhancement because some of the tertiary stars are evolved, i.e., larger than their MS radius.  We find that a fraction equal to $2 \times 10^{-4}$ of all close binaries (period = 0.5 -- 20 days) host a third star in a compact 2+1 triple configuration that is flat.  Thus, while these are relatively fairly rare systems, there are probably several hundred thousand of them in the Galaxy.

Again, as we have suggested throughout the paper, selection effects favor (i) compact systems, (ii) at least partly flat architecture, and (iii) somewhat evolved tertiaries to enhance the outer eclipse probability.  Thus, it is not surprising that these are the systems we have predominantly spotted while surveying the lightcurves.

We have found that the masses of all the stars in the triples we studied are the same to within small factors of order 2. Otherwise the EB might not have been detected in the glare of a much more luminous tertiary.  Comparable masses and short outer periods imply an accretion-driven migration formation scenario (see, e.g., \citealt{tokovinin21} and references therein). In the latter case, the flatness of the systems might demonstrate that such migration is accompanied by orbit alignment.  Comparable masses also imply accretion from a common gas source. This mechanism predicts that the outer mass ratio must not exceed unity, and this is indeed the case for the systems presented here.  Moderate outer eccentricities as we observe in 4 of the 6 systems can be largely primordial, although circularization by tides in giants also works in some cases (e.g., for TIC 54060695 and TIC 242132789 as we mentioned earlier). 

There is also a possible link between our triple systems and 2+2 compact quadruples. Some giants in these triple systems could have originally been
second binaries that have merged. This can be revealed by apparent difference of ages, as hinted at in one of our systems: TIC 42565581, though the evidence there is only marginal. These 6 triples with giant tertiaries, favored by observational selection, are also favorable candidates for a triple common envelope phase in the future.  In this latter regard, see the studies by (\citealt{toonen13}; \citealt{hamers22}), with emphasis on triple common envelope evolution \citep{glanz21} and its end products. 

Finally, we note that in terms of stellar evolution theory, the tertiary in these systems fixes their age.  Thus, we obtain a triplet of reasonably accurate masses and radii with a known age. 

\begin{table*}
\centering 
\caption{Derived ephemerides for the six triple systems to be used for planning future observations.}
 \label{tab:ephemerides}
 \begin{tabular}{lllllll}
 \hline 
TIC ID               & 37743815       & 42565581  & 54060695 & 178010808       & 242132789  & 456194776 \\
\hline
&\multicolumn{6}{c}{Inner binary} \\
\hline
$P$  & 0.90707 &  1.823071   & 1.06049087 & 0.8257362 &  5.11458   & 1.719349 \\
$\mathcal{T}_0$  & 8\,469.101 & 8\,469.003 & 8\,486.635 & 8\,492.0725 & 8\,472.69286 & 8\,791.552 \\
$\mathcal{A}_\mathrm{ETV}$  & 0.001 & 0.002 & 0.001 & 0.001 & 0.011 & 0.007 \\
$D$  & 0.105 & 0.293 & 0.145 & 0.155 & 0.282 & 0.176 \\
\hline
&\multicolumn{6}{c}{Wide binary (third body eclipses)} \\
\hline
$P$ & 68.720 & 123.452 & 60.72 & 69.02: & 41.531 & 93.90: \\
$\mathcal{T}_0^\mathrm{inf}$  &  9\,224.3 &  8\,479.1 & 8\,504.9 &  8\,554.2: &  8\,484.3 & 8\,841.2: \\
$D^\mathrm{inf}$ & 1.65 & 3.23 &  2.20 & 0.85: & 3.35 &  1.70: \\
$\mathcal{T}_0^\mathrm{sup}$ &  9\,246.7: &  8\,543.4: &  8\,474.51 &  8\,512.75 &  8\,505.3 &  8\,809.90 \\
$D^\mathrm{sup}$ & 1.05: & 2.30: &  1.95 & 1.00 & 3.35 &  1.90 \\
\hline
\end{tabular}

\textit{Notes.} (a) For the inner pairs: $P$, $\mathcal{T}_0$, $\mathcal{A}_\mathrm{ETV}$, $D$ are the period, reference time of a primary minimum, half-amplitude of the ETV curve, and the full duration of an eclipse, respectively. $\mathcal{T}_0$ is given in BJD -- 2\,450\,000, while the other quantities are in days. As all the inner eccentricities are very small and, hence, the shifts of the secondary eclipses relative to phase 0.5 are negligible (quantitatively, they are much smaller than the full durations of the individual eclipses), the same reference times and periods can be used to predict the times of the secondary eclipses. (b) For the outer orbits we give separate reference times for the third body eclipses around the inferior and superior conjunctions of the tertiary component. The eclipse durations, $D$, of the third-body eclipses do not give the extent of any specific third body events.  Rather $D$ represents the time difference corresponding to the very first and last moments around a given third-body conjunction when the first/last contact of a third-body event may occur). Double dots (:) (1) at the outer periods of TICs~178010808 and 456194776 refer to larger uncertainties arising from the fact that in these two triples only one third-body eclipse was observed with {\it TESS}; (2) at superior/inferior conjunction times refer to those kinds of third-body events (i.e., primary vs.~secondary outer eclipses) that were not observed with \textit{TESS}. For these events, the ephemerides are based on the archival data folds (see Fig.~\ref{fig:outer_fold}), and they might be less certain.

\end{table*} 

\begin{table}
\centering
\caption{Evolution Times in Myr of the Tertiary Stars}
\begin{tabular}{lccc}
\hline
\hline
Phase & 1.5\,M$_\odot$ & 2.0\,M$_\odot$ & 2.5\,M$_\odot$  \\
\hline
3.0\,R$_\odot \rightarrow$ tip of RGB & 247  & 61 & 13 \\
8.0\,R$_\odot \rightarrow$ tip of RGB & 86 &37  & 8 \\ 
15\,R$_\odot \rightarrow$ tip of RGB & 28 & 26 & 4 \\
giant $\rightarrow$ AGB & 126 & 300 & 200 \\
total evolution & 2900 & 1350 & 780 \\
\hline
\label{tbl:lifetime}  
\end{tabular}

\textit{Notes.}  All times were computed with the {\tt MESA} stellar evolution code. 2.5 M$_\odot$ stars will not attain sufficiently large radii to fill their Roche lobes during the ascent of the RGB in the systems under discussion. 
\end{table}

\section*{Data availability}

The \textit{TESS} data underlying this article were accessed from MAST (Barbara A. Mikulski Archive for Space Telescopes) Portal (\url{https://mast.stsci.edu/portal/Mashup/Clients/Mast/Portal.html}). The ASAS-SN archival photometric data were accessed from \url{https://asas-sn.osu.edu/}. The ATLAS archival photometric data were accessed from \url{https://fallingstar-data.com/forcedphot/queue/}. A part of the data were derived from sources in the public domain as given in the respective footnotes. The derived data generated in this research and the code used for the photodynamical analysis will be shared upon a reasonable request to the corresponding author.

\section*{Acknowledgments}

We are grateful to Andrei Tokovinin for going through the manuscript and making some very helpful suggestions for how to interpret our findings.

We thank Allan R. Schmitt and Troy Winarski for making their light curve examining software tools LcTools and AKO-TPF freely available.

VBK is thankful for support from NASA grants 80NSSC21K0351. 

TB and ZG acknowledge the support of the Hungarian National Research, Development and Innovation Office (NKFIH) grant K-125015, a PRODEX Experiment Agreement No. 4000137122 between the ELTE E\"otv\"os Lor\'and University and the European Space Agency (ESA-D/SCI-LE-2021-0025), and the support of the city of Szombathely. ZG acknowledges the Lend\"ulet LP2018-7/2021 grant of the Hungarian Academy of Science, the VEGA grant of the Slovak Academy of
Sciences No. 2/0031/22, the Slovak Research and Development Agency contract No. APVV-20-0148.

AP acknowledges the financial support of the Hungarian National Research, Development and Innovation Office -- NKFIH Grant K-138962.

We thank David Latham for facilitating the TRES observations of TIC 456194776.

The operation of the BRC80 robotic telescope of Baja Astronomical Observatory has been supported by the project ``Transient Astrophysical Objects'' GINOP 2.3.2-15-2016-00033 of the National Research, Development and Innovation Office (NKFIH), Hungary, funded by the European Union.

This paper includes data collected by the \textit{TESS} mission. Funding for the \textit{TESS} mission is provided by the NASA Science Mission directorate. Some of the data presented in this paper were obtained from the Mikulski Archive for Space Telescopes (MAST). STScI is operated by the Association of Universities for Research in Astronomy, Inc., under NASA contract NAS5-26555. Support for MAST for non-HST data is provided by the NASA Office of Space Science via grant NNX09AF08G and by other grants and contracts.

We have made extensive use of the All-Sky Automated Survey for Supernovae archival photometric data.  See \citet{shappee14} and \citet{kochanek17} for details of the ASAS-SN survey.

We also acknowledge use of the photometric archival data from the Asteroid Terrestrial-impact Last Alert System (ATLAS) project.  See \citet{tonry18} and \citet{heinze18} for specifics of the ATLAS survey.

This work has made use  of data  from the European  Space Agency (ESA)  mission {\it Gaia}\footnote{\url{https://www.cosmos.esa.int/gaia}},  processed  by  the {\it   Gaia}   Data   Processing   and  Analysis   Consortium   (DPAC)\footnote{\url{https://www.cosmos.esa.int/web/gaia/dpac/consortium}}.  Funding for the DPAC  has been provided  by national  institutions, in  particular the institutions participating in the {\it Gaia} Multilateral Agreement.

This publication makes use of data products from the Wide-field Infrared Survey Explorer, which is a joint project of the University of California, Los Angeles, and the Jet Propulsion Laboratory/California Institute of Technology, funded by the National Aeronautics and Space Administration. 

This work also makes use of data products from the Two Micron All Sky Survey, which is a joint project of the University of Massachusetts and the Infrared Processing and Analysis Center/California Institute of Technology, funded by the National Aeronautics and Space Administration and the National Science Foundation.

We  used the  Simbad  service  operated by  the  Centre des  Donn\'ees Stellaires (Strasbourg,  France) and the ESO  Science Archive Facility services (data  obtained under request number 396301).



\begin{thebibliography}{99}

\bibitem[Alonso et al.(2015)]{alonso15} 
 Alonso R., Deeg H. J., Hoyer S., Lodieu N., Palle E., Sanchis-Ojeda R., 2015, \aap, 584, L8

\bibitem [\protect\citeauthoryear{Bailer-Jones et al.}{2021}]  {bailer-jonesetal21}
 Bailer-Jones, C. A. L., Rybizki, J., Fouesneau, M., Demleitner., M., Andrae, R., 2021, \aj, 161, 147

\bibitem [\protect\citeauthoryear{Bakos et al.}{2002}]{bakosetal02} 
 Bakos, G., \'A., L\'az\'ar, J., Papp, I., S\'ari, P., Green, E. M., 2002, \pasp, 114, 974

\bibitem[Bianchi et al.(2017)]{bianchi17} Bianchi, L., Shiao, B., \& Thilker, D. 2017, ApJS, 230, 24

\bibitem [\protect\citeauthoryear{Borkovits et al.}{2003}]  {borkovitsetal03}
Borkovits, T., \'Erdi, B., Forg\'acs-Dajka, E., Kov\'acs, T., 2003, \aap, 398, 1091

\bibitem [\protect\citeauthoryear{Borkovits et al.}{2013}]  {borkovitsetal13}
Borkovits, T., Derekas, A., Kiss, L. L., et al., 2013, \mnras, 428, 1656

\bibitem [\protect\citeauthoryear{Borkovits et al.}{2015}]  {borkovitsetal15}
Borkovits, T., Rappaport, S., Hajdu, T., Sztakovics, J., 2015, \mnras, 448, 946

\bibitem [\protect\citeauthoryear{Borkovits et al.}{2016}]  {borkovitsetal16}
Borkovits, T., Hajdu, T., Sztakovics, J., Rappaport, S., Levine, A., B\'\i r\'o, I. B., Klagyivik, P., 2016, \mnras, 455, 4136

\bibitem [\protect\citeauthoryear{Borkovits et al.}{2018}]  {borkovitsetal18}
Borkovits, T., Albrecht, S., Rappaport, S., et al. 2018, \mnras, 478, 513

\bibitem [\protect\citeauthoryear{Borkovits et al.}{2019a}]  {borkovitsetal19a}
Borkovits, T., Rappaport, S., Kaye, T., et al. 2019a, \mnras, 483, 1934

\bibitem [\protect\citeauthoryear{Borkovits et al.}{2019b}]  {borkovitsetal19b}
Borkovits, T.,  Sperauskas, J., Tokovinin, A., Latham, D. W., Cs\'anyi, I., Hajdu, T., Moln\'ar, L., 2019b, \mnras, 487, 4631

\bibitem [\protect\citeauthoryear{Borkovits et al.}{2020a}]  {borkovitsetal20a}
Borkovits, T., Rappaport, S., Hajdu, T., et al. 2020a, \mnras, 493, 5005

\bibitem[Borkovits et al.(2020b)]{borkovitsetal20b}
Borkovits, T., Rappaport, S., Tan, T.G., et al.  2020b, \mnras, 496, 4624

\bibitem [\protect\citeauthoryear{Borkovits et al.}{2021}]  {borkovitsetal21}
Borkovits, T., Rappaport, S., Maxted, P. F. L., et al. 2021, \mnras, 503, 3759

\bibitem[Borkovits et al.(2022)]{borkovitsetal22} 
Borkovits, T., Mitnyan, T., Rappaport, S., et al. 2022, \mnras, 510, 135

\bibitem[Borkovits (2022)]{borkovits22} Borkovits, T. 2022, `Eclipsing Binaries in Dynamically Interacting Close, Multiple
Systems', Galaxies, 10, 9 (arXiv:2201.01243)

\bibitem[Borucki et al.(2010)]{borucki10} Borucki W. J. et al., 2010, Science, 327, 977

\bibitem[Bouma et al.(2019)]{bouma19} Bouma, L., Hartman, J., Bhatti, W. 2019, \apjs, 245, 13.

\bibitem[\protect\citeauthoryear{Bressan et al.}{2012}]{PARSEC}
Bressan, A., Marigo, P., Girardi, L. et al. 2012, \mnras, 427, 127

\bibitem[Buchhave et al.(2010)]{buchhave10} Buchhave, L.A., Bakos, G. A., Hartman, J.D., et al. 2010,
\apj, 720, 1118, doi: 10.1088/0004-637X/720/2/1118

\bibitem[Carter et al.(2011)]{carter11} 
Carter, J.A., et al., 2011, Science, 331, 562

\bibitem[Castelli \& Kurucz (2003)]{castelli03} Castelli F., Kurucz R. L., 2003, in Piskunov, N.,Weiss, W.W., Gray D. F., eds,
Proc. IAU Symp.~210, Modelling of Stellar Atmospheres. p.A20, preprint (astro-ph/0405087)

\bibitem[Chambers et al.(2016)]{chambers16} Chambers, K.C., Magnier, E.A., Metcalfe, N., et al. 2016, arXiv:1612.05560

\bibitem[Choi et al.(2016)]{choi16} Choi J., Dotter A., Conroy C., Cantiello M., Paxton B., \& Johnson B. D., 2016, \apj, 823, 102

\bibitem [\protect\citeauthoryear{Cutri et al.}{2013}]{WISE} 
 Cutri, R.M., Wright, E.L., Conrow, T., et al.~2013, wise.rept, 1C.
 
\bibitem[Donati et al.(1997)]{donati97} Donati, J.-F., Semel, M., Carter, B. D., Rees, D. E., \&
Collier Cameron, A. 1997, \mnras, 291, 658, doi: 10.1093/mnras/291.4.658
 
 \bibitem[Dotter (2016)]{dotter16} Dotter A., 2016, \apjs, 222, 8

\bibitem[Fausnaugh et al.(2020)]{fausnaugh20} Fausnaugh, M.M., Burke, C.J., Ricker, G.R., \& Vanderspek, R. 2020, Research Notes of the AAS, 4, 251

\bibitem [\protect\citeauthoryear{Ford}{2005}] {ford05}
Ford, E. B., 2005, \aj, 129, 1706

\bibitem [\protect\citeauthoryear{Gaia collaboration}{2021}]  {GaiaEDR3}
Gaia Collaboration, Brown, A. G. A., Vallenari, A., Prusti, T. et al. 2021, \aap, 649, A1

\bibitem[Furesz (2008)]{furesz08} Furesz, G. 2008, PhD thesis, University of Szeged

\bibitem[Glanz \& Perets (2021)]{glanz21} Glanz, H., \& Perets, H.B. 2021, \mnras, 500, 1921

\bibitem[Green et al.(2019)]{green19} Green, G.M., Schlafly, E.F., Zucker, C., Speagle, J.S., \& Finkbeiner, D.P. 2019, arXiv:1905.02734

\bibitem[Gunn et al.(1998)]{gunn98} Gunn, J.E., Carr, M., Rockosi, C., et al. 1998, \aj, 116, 3040

\bibitem[Hamers et al.(2022)]{hamers22} Hammers, A.S., Perets, H.B., Thompson, T.A., \& Neunteufel, P. 2022, \apj, 925, 178

\bibitem[Heinze et al.(2018)]{heinze18} Heinze, A.N., Tonry, J.L., Denneau, L., et al. 2018, \aj, 156, 241

\bibitem [\protect\citeauthoryear{Henden et al.}{2015}] {APASS}
 Henden, A. A., Levine, S., Terrell, D., Welch, D. 2015, American Astronomical Society, AAS Meeting \#225, id.336.16
 
 \bibitem[Howell et al.(2014)]{howell14} Howell, S.B., Sobeck, C., Hass, M., et al. 2014, \pasp, 126, 398
 
\bibitem[Huang et al.(2020)]{huang20} 
Huang, C.X., Vanderburg, A., P\'{a}l, A., et al., 2020, RNAAS, 4, 206

\bibitem[Jenkins et al.(2016)]{jenkins16} Jenkins, J.M., Twicken, J.D., McCauliff, S., et al. 2016, in: Software and Cyberinfrastructure for Astronomy IV,
volume 9913. International Society for Optics and Photonics, p.~99133E

\bibitem[Klinglesmith \& Sobieski(1970)]{klinglesmithsobieski70} 
 Klinglesmith, D. A., Sobieski, S., 1970, \aj, 75, 175

\bibitem[Kochanek et al.(2017)]{kochanek17} 
Kochanek, C. S., Shappee, B. J., Stanek, K. Z., et al., 2017, \pasp, 129, 104502

\bibitem[\protect\citeauthoryear{Kostov et al.}{2021}]{kostovetal21} 
Kostov et al., 2021, \apj, 917, 93

\bibitem[Kostov et al.(2022)]{kostov22} 
 Kostov, V.B., Powell, B.P., Rappaport, S.A, et al. 2022, ApJS in press, arXiv:2202.05790

\bibitem[Kov\'{a}cs et al.(2002)]{kovacs02} 
Kov\'{a}cs, G., Zucker, S., Mazeh, T., 2002, \aap, 391, 369

\bibitem[Kozai(1962)]{kozai962}
 Kozai, Y. 1962, \aj, 67, 591

\bibitem[Kristiansen et al.(2022)]{kristiansen22} 
 Kristiansen, M.H., Rappaport, S., Vanderburg, A., et al. 2022, submitted to PASP.
 
\bibitem[Latham et al.(2002)]{Latham:2002} Latham, D.\ W., Stefanik, R.\ P., Torres, G., et al.\ 2002, \aj, 124, 1144

\bibitem[Lidov(1962)]{lidov962} Lidov, M. L., 1962, Planetary and Space Science, 9, 719

\bibitem[Lucy (1967)]{lucy67} Lucy, L.B. 1967, Zeitschrift f\"ur Astrophysik, 65, 89

\bibitem [\protect\citeauthoryear{Masuda et al.}{2015}] {masudaetal15}
Masuda, K., Uehara, S., Kawahara, H, 2015, \apj, 806, L37

\bibitem [\protect\citeauthoryear{Mitnyan et al.}{2020}] {mitnyanetal20}
Mitnyan, T., Borkovits, T., Rappaport, S., P\'al, A., Maxted, P. F. L., 2020, \mnras, 498, 6034

\bibitem[Nardiello et al.(2019)]{nardiello19} Nardiello, D., Borsato, L,, Piotto, G,, et al. 2019, NNRAS, 490, 3806

\bibitem[Nordstr\"om et al.(1994)]{Nordstrom:1994} Nordstr\"om, B., Latham, D.\ W., Morse, J.\ A., et al.\ 1994, \aap, 287, 338

\bibitem[Ochsenbein et al.(2000)]{ochsenbein00} Ochsenbein, F., Bauer, P., \& Marcout, J. 2000. A\&AS, 143, 23

\bibitem[Oelkers \& Stassun (2018)]{oelkers18} Oelkers, R.J., \& Stassun, K.G. 2018, \aj,156, 132

\bibitem [\protect\citeauthoryear{Orosz}{2015}] {orosz15}
Orosz, J., 2015, ASPC, 496, 55

\bibitem [\protect\citeauthoryear{Paegert et al.}{2021}] {TIC8}
 Paegert, M. et al, 2021, arXiv:2108.04778

\bibitem[P\'al (2012)] {pal12} P\'al, A., 2012, \mnras, 421, 1825

\bibitem[Paxton et al.(2011)]{paxton11} Paxton, B., Bildsten L., Dotter A., Herwig F., Lesaffre P., \& Timmes F., 2011, ApJS, 192, 3

\bibitem[Paxton et al.(2015)]{paxton15} Paxton, B., et al., 2015, ApJS, 220, 15

\bibitem[Paxton et al.(2019)]{paxton19} Paxton, B., et al., 2019, ApJS, 243, 10

\bibitem [\protect\citeauthoryear{Pepper et al.}{2007}]{pepperetal07} 
 Pepper, J., Pogge, R.W., DePoy, D.L., et al., 2007, \pasp, 119, 923
 
 \bibitem [\protect\citeauthoryear{Pepper et al.}{2012}]{pepperetal12} 
 Pepper, J., Kuhn, R.B., Siverd, R., et al., 2012, \pasp, 124, 230

\bibitem [\protect\citeauthoryear{Pollacco et al.}{2006}]{2006PASP..118.1407P} 
 Pollacco, D. L., Skillen, I., Collier Cameron, A., et al. 2006, \pasp, 118, 1407

\bibitem[Powell et al.(2021)]{powell21} 
Powell, B.P., Kostov, V.B., Rappaport, S., et al. 2021, \aj, 161, 162

\bibitem [\protect\citeauthoryear{Pr\v{s}a \& Zwitter}{2005}]{Phoebe} 
Pr\v{s}a, A., \& Zwitter, T., 2005, \apj, 628, 426

\bibitem[Rappaport et al.(2021)]{rappaport21} Rappaport, S., Kurtz, D, Handler, G., et al. 2021, MNRAS, 503, 254
 
\bibitem[Ricker et al.(2015)]{ricker15} Ricker, G.R., Winn, J.N., Vanderspek, R., et al. 2015, JATIS, 1, 014003

\bibitem [\protect\citeauthoryear{Rowden et al.}{2020}] {rowdenetal20}
 Rowden, P., et al., 2020, \aj, 160, 76

\bibitem[Schmitt et al.(2019)]{schmitt19} 
Schmitt, A.R., Hartman, J.D., \& Kipping, D.M. 2019, arXiv:1910.08034

\bibitem[Shappee et al.(2014)]{shappee14} 
Shappee, B. J., Prieto, J. L., Grupe, D., et al. 2014, \apj, 788, 48

\bibitem[Skrutskie et al.(2006)]{2MASS} 
 Skrutskie, M.F., Cutri, R.M., Stiening, R., et al. 2006, \aj, 131, 1163

\bibitem[Smith et al.(2020)]{smith20}  
Smith, K.W, Smartt, S.J., Young, D.R., Tonry, J.L., et al. 2020, \pasp, 132, 5002. 
 
 \bibitem[Talens et al.(2017)]{talens17} Talens, G.J.J., Spronck, J.F.P., Lesage, A.-L., Otten, G.P.P.L., Stuik, R., Pollacco, D. \& Snellen, I.A.G. 2017, \aap, 601, A11


\bibitem[\protect\citeauthoryear{Tokovinin}{2021}]{tokovinin21} Tokovinin, A.,  2021, Universe, 7, 352

\bibitem[Toonen \& Nelemans (2013)]{toonen13} Toonen, S., \& Nelemans, G. 2013, \aap, 557, 87

\bibitem[Tonry et al.(2018)]{tonry18} 
Tonry, J.L., Denneau, L., Heinze, A.N., et al. 2018, \pasp, 130, 4505

\bibitem[von Zeipel(1910)]{vonzeipel910} von Zeipel, 1910, AN, 183, 345

\bibitem[Wolf et al.(2018)]{wolf18} Wolf, C., Onken, C.A., Luvaul, L.C., et al. 2018, PASA, 35, 10 

\end{thebibliography}




\appendix

\section{Supplementary Material -- RV and ETV fits}
\label{app:242132789}

\subsection{Interpretation of the large amplitude ETV of TIC 242132789}

\begin{figure}
\begin{center}
\includegraphics[width=0.99 \columnwidth]{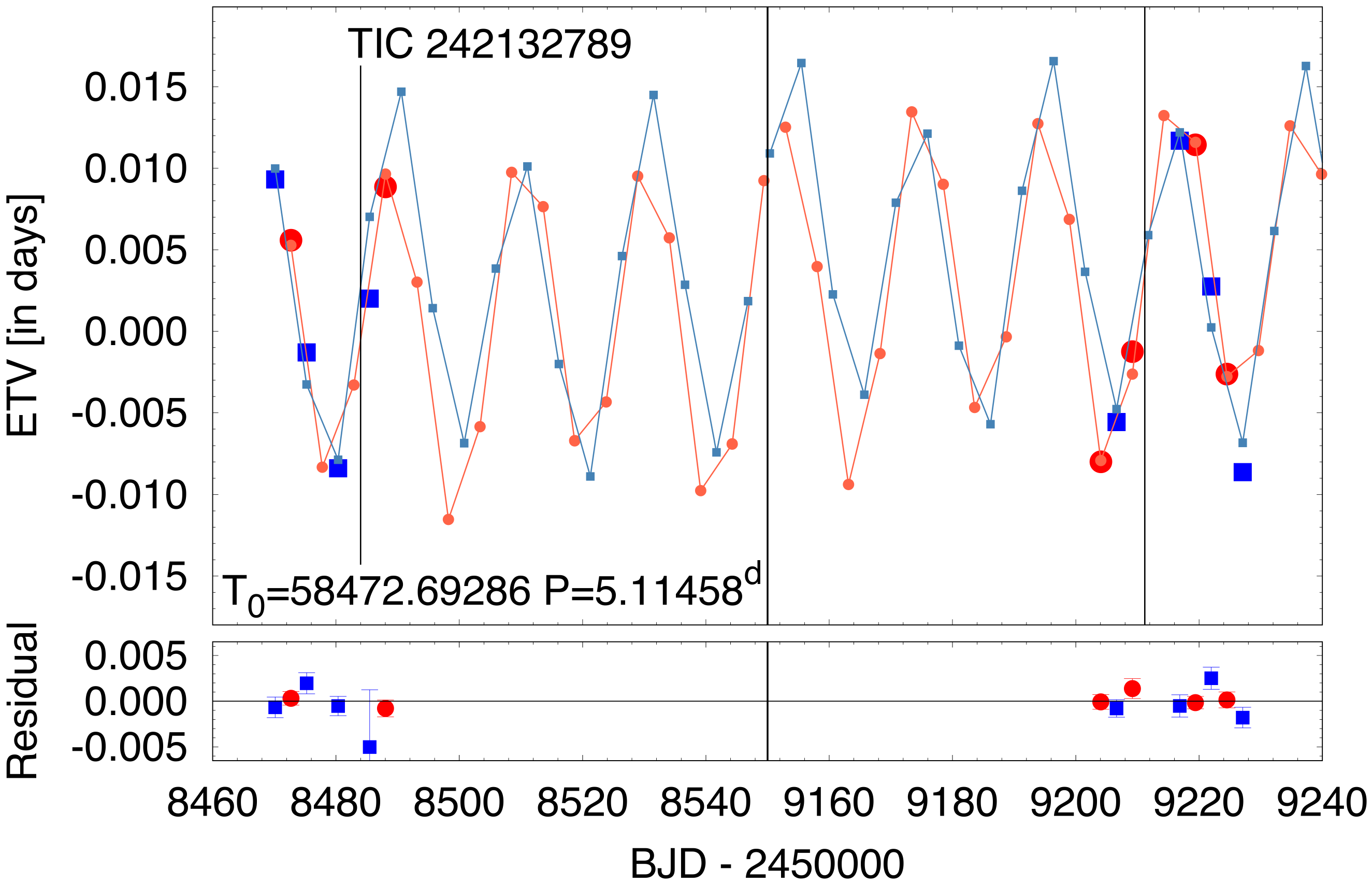}
\caption{Photodynamical fit to the {\it TESS} ETV curves for TIC 242132789.  Note the high amplitude of the ETVs and the fact that they oscillate at twice the frequency of the outer orbit (i.e., every $\sim$21 days.) The larger and darker red circles and blue squares represent the observed primary and secondary times of EB eclipses, while the smaller, lighter symbols, connected with straight lines are taken from the photodynamical model ETV curve. The two thin vertical lines denote the locations of the two third-body outer eclipses. Residuals are also shown in the lower panel, where the uncertainty on each point is also noted. }
\label{fig:242132789_ETV}
\end{center}
\end{figure}

As was mentioned in Sect.~\ref{sec:photodynamical}, the analysis of the ETV curves that are extracted from the high-precision \textit{TESS} lightcurves are inherent in our photodynamical analyses. They provide very strict constraints on the eclipsing periods of the inner EBs and, in the case of some eccentricities, the parameters $e_\mathrm{in}\cos\omega_\mathrm{in}$ are also very strictly constrained through these data. In the case of five of the six investigated systems, however, due to the very short durations of the \textit{TESS} observations, the ETV curves do not carry any useful information about the outer orbits nor, therefore, on the system configurations. The only exception is the ETV curve of TIC~242132789 which exhibits large amplitude ($\mathcal{A}_\mathrm{ETV}\sim0\fd01$), quasi-sinusoidal variations with a period which is exactly half of the outer orbital period, $P_\mathrm{out}$ (see Fig.~\ref{fig:242132789_ETV}). Here we briefly discuss the origin of this timing variation, and its implications for the analytic perturbation theories of hierarchical triple systems.

First, it is evident that this ETV cannot arise from the well-known geometric light-travel time effect (LTTE), for at least three reasons. (1) The LTTE has the same period as the outer period. (2) Since in this system the outer orbit is found to be almost circular, the third-body eclipses should have occurred at the extrema of an LTTE generated ETV curve, while, as is seen in Fig.~\ref{fig:242132789_ETV}, the \textit{TESS}-observed third-body events occurred approximately mid-way between the two extrema. Finally (3), with the use of the masses and orbital elements found from the photodynamical analysis (Table~\ref{tab:syntheticfit_TIC242132789+456194776}) one can calculate the expected amplitude of the LTTE as being $\mathcal{A}_\mathrm{LTTE}\sim10^{-3}$\,d, i.e., one order of magnitude smaller than is observed.

Second, it is also clear that the ETV cannot be the consequence of the usually considered medium period class perturbations of the tertiary.\footnote{In hierarchical triple systems the periodic perturbations have three different classes, according to their characteristic time-scales, as (i) short period ones with characteristic time-scale of $P_\mathrm{in}$, (ii) medium period ones, having time-scale of $P_\mathrm{out}$, and (iii) long period perturbations, which are effective on a time-scale of $P_\mathrm{out}^2/P_\mathrm{in}$.} It was shown by \citet{borkovitsetal03} that in a coplanar, doubly circular hierarchical triple system (like TIC 242132789) the largest amplitude, quadruple-order perturbations disappear. This finding was confirmed later with the analyses of the recently discovered, doubly circular, coplanar, triply eclipsing triple systems such as HD~181068 \citep{borkovitsetal13}, TIC~278825952 \citep{mitnyanetal20}, and TIC~193993801 \citep{borkovitsetal22}. Moreover, though it was found by \citet{borkovitsetal15} that the octuple-order perturbation terms do not vanish for such a scenario, their characteristic periods are $P_\mathrm{out}$ and/or $P_\mathrm{out}/3$, but not the half of the orbital period.

On the other hand, as was also discussed in \citet{borkovitsetal15}, for the tightest triple systems the strict hierarchical approximation no longer remains fully valid. This fact makes it necessary to include some further terms that are denoted as `$P_\mathrm{out}$ time-scale residuals of the $P_\mathrm{in}$ time-scale dynamical effects'. According to their calculations (Eqs.~20 and 21), the leading term of this expression for a doubly circular, coplanar configuration gives the following ETV contribution:
\begin{equation}
\Delta_\mathrm{short}=\frac{11}{16\pi}\frac{m_\mathrm{B}}{m_\mathrm{AB}}\frac{P_\mathrm{in}^3}{P_\mathrm{out}^2}\sin\left[2\frac{2\pi}{P_\mathrm{out}}\left(t-\mathcal{T}_\mathrm{out}^\mathrm{inf}\right)\right].
\label{Eq:ETVshort}
\end{equation}

Substituting the third-body ($m_\mathrm{B}$) and the total system ($m_\mathrm{AB}$) masses, as well as the inner and outer periods ($P_\mathrm{in,out}$) from Table~\ref{tab:syntheticfit_TIC242132789+456194776}, one  readily finds for the amplitude that $\mathcal{A}_\mathrm{short}=0\fd0064$, which is close to the observed value. Moreover, the expression above describes well not only the amplitude and period of the observed ETV, but also its phase. And, according to Eq.~(\ref{Eq:ETVshort}), both kinds of third-body eclipses should occur mid-way between the lower and upper extrema of the ETV curve, as is very nicely demonstrated in Fig.~\ref{fig:242132789_ETV}. Therefore, we may conclude that the large amplitude ETV in the case of TIC~242132789 originates from this latter effect. On the other hand, the fact that the theoretically computed amplitude is only about two thirds of the observed value, may serve as cautionary note.  And, some further more sophisticated theoretical modeling may be worthwhile for the correct analytical description of the $\sim$ month-timescale perturbations of the tightest triple star systems.

\subsection{A more in-depth analysis of TIC 456194776, including ground based RV data}
\label{app:T456194776RV}

Late in the production of this paper, we were fortunate enough to acquire a significant number of radial velocity measurements of this target.   We obtained spectroscopic observations of the target TIC 456194776 with the Tillinghast Reflector Echelle Spectrograph (TRES, \citealt{furesz08}), on the 1.5-m reflector at the Fred Lawrence Whipple Observatory (FLWO) in Arizona, USA. TRES is a high-resolution fiber-fed echelle spectrograph, with a spectral resolving power of $R = 44 000$ over the wavelength region of 3900--9100 \AA.  A total of 20 observations were obtained of TIC 456194776 between Sept 21, 2021 and Jan 26, 2022, with signal-to-noise ratios per resolution element of 23-40 in the Mgb triplet wavelength region ($\sim$5187~\AA). The spectra were extracted and reduced as per \citet{buchhave10}, with wavelength solutions derived from bracketing Th-Ar lamp exposures.  Visual inspection of the spectra revealed only the lines of the brighter tertiary (star B).  Radial velocities were derived by cross-correlation against a suitable synthetic template from a large pre-computed library based on model atmospheres by R.\ L.\ Kurucz, and a line list tuned to better match real stars \citep[see][]{Nordstrom:1994, Latham:2002}. These templates cover a limited wavelength region near the Mg b triplet. We find the tertiary to be a rapidly rotating star with an estimated $v \sin i$ of about 80~km~s$^{-1}$. 

\begin{figure}
\begin{center}
\includegraphics[width=0.99 \columnwidth]{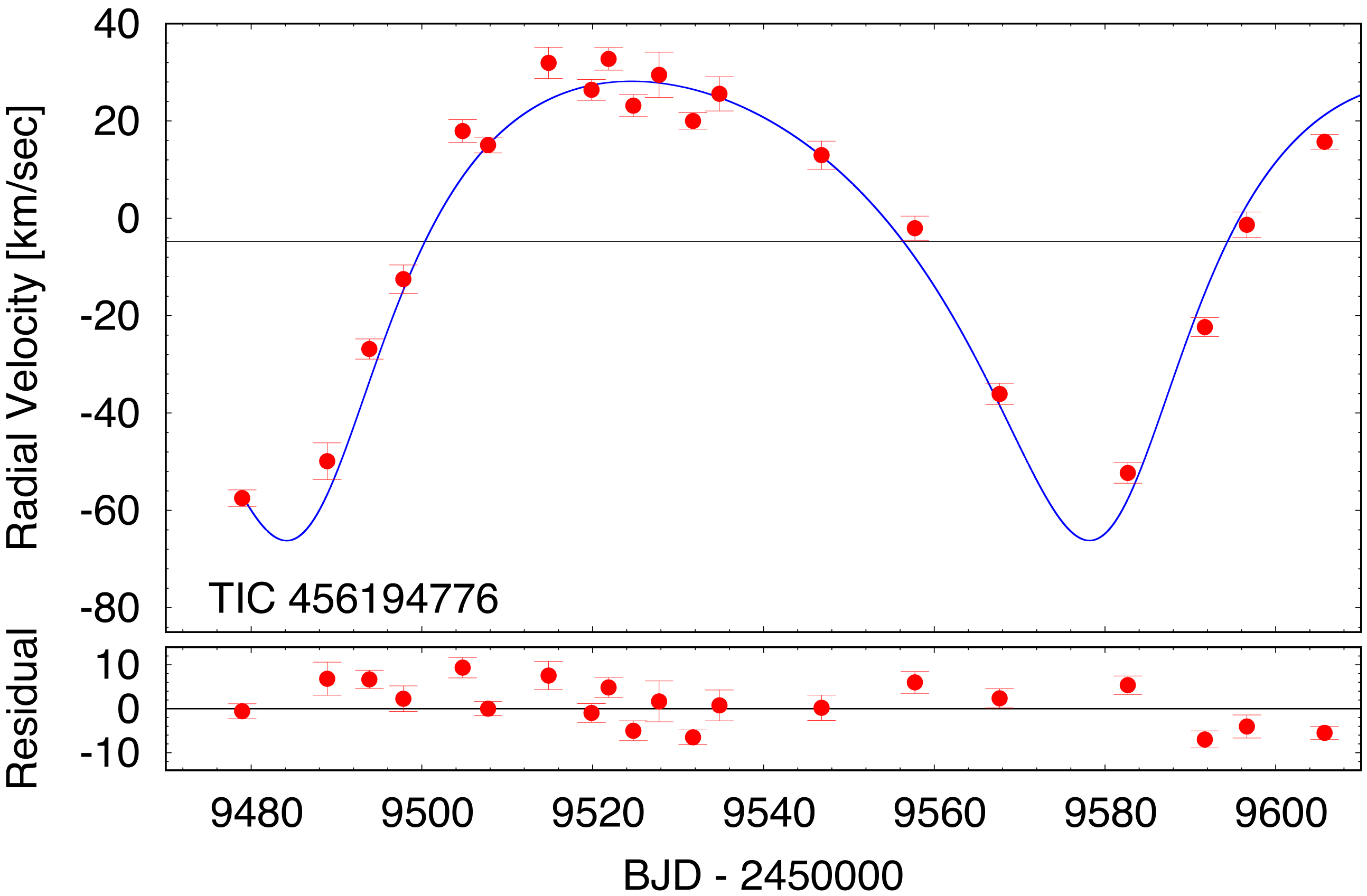}
\caption{Photodynamical fit to the 20 TRES radial velocity data points for TIC 456194776. The RVs points with uncertainties are given in Table \ref{tab:RVdata}.  The blue curve is from the spectro-photodynamical analysis described in Sect.~\ref{sec:photodynamical}.}
\label{fig:456194776_RVs}
\end{center}
\end{figure}

The 20 radial velocities and their uncertainties are given in Table \ref{tab:RVdata}, while the RV points are plotted in Fig.~\ref{fig:456194776_RVs}.  The solid blue curve is the photodynamical fit that was produced during the analysis that led up to the MDR (model-dependent-with-RVs) solution discussed in Sect.~\ref{sec:photodynamical}.

\begin{table}
\centering
\caption{Measured radial velocities of the tertiary component of TIC~456194776. The date is given as BJD -- 2\,450\,000, while the RVs and their uncertainties are in km\,s$^{-1}$.}
\label{tab:RVdata}
\begin{tabular}{lrrlrr}
\hline
\hline
Date & RV$_\mathrm{B}$ & $\sigma_\mathrm{B}$ & Date & RV$_\mathrm{B}$ & $\sigma_\mathrm{B}$ \\  
\hline
$9478.950099$ & $ -57.49 $ & $ 1.70$ & $9527.793497$ & $  29.48 $ & $ 4.65$ \\
$9488.905299$ & $ -49.90 $ & $ 3.77$ & $9531.733497$ & $  20.00 $ & $ 1.69$ \\
$9493.853498$ & $ -26.85 $ & $ 2.08$ & $9534.827997$ & $  25.57 $ & $ 3.51$ \\
$9497.825098$ & $ -12.50 $ & $ 2.92$ & $9546.797096$ & $  12.98 $ & $ 2.88$ \\
$9504.781298$ & $  17.93 $ & $ 2.34$ & $9557.734596$ & $  -2.03 $ & $ 2.48$ \\
$9507.755898$ & $  15.05 $ & $ 1.62$ & $9567.663896$ & $ -36.08 $ & $ 2.18$ \\
$9514.843597$ & $  31.94 $ & $ 3.20$ & $9582.678395$ & $ -52.31 $ & $ 2.10$ \\
$9519.872897$ & $  26.39 $ & $ 2.15$ & $9591.693095$ & $ -22.35 $ & $ 1.94$ \\
$9521.876697$ & $  32.75 $ & $ 2.31$ & $9596.620294$ & $  -1.33 $ & $ 2.62$ \\
$9524.771097$ & $  23.14 $ & $ 2.26$ & $9605.728294$ & $  15.71 $ & $ 1.51$ \\
\hline
\end{tabular}
\end{table}

\begin{table*}
 \centering
\caption{TIC 456194776: Abbreviated Parameter Comparison Between MDR and MDN Models}
 \label{tab:syntheticfit_T456194776comp}
\begin{tabular}{@{}lllllll}
\hline
 & \multicolumn{3}{c}{MDR} & \multicolumn{3}{c}{MDN} \\
\hline
\multicolumn{7}{c}{orbital elements} \\
\hline
   & \multicolumn{3}{c}{subsystem} & \multicolumn{3}{c}{subsystem}  \\
   & \multicolumn{2}{c}{Aa--Ab} & A--B & \multicolumn{2}{c}{Aa--Ab} & A--B \\
  \hline
  $a$ [R$_\odot$] & \multicolumn{2}{c}{$8.282_{-0.029}^{+0.021}$} & $143.5_{-0.32}^{+0.49}$ & \multicolumn{2}{c}{$8.287_{-0.030}^{+0.019}$} & $143.8_{-1.1}^{+0.4}$ \\
  $e$ & \multicolumn{2}{c}{$0.00437_{-0.00057}^{+0.00058}$} & $0.314_{-0.010}^{+0.011}$ & \multicolumn{2}{c}{$0.00293_{-0.00043}^{+0.00060}$} & $0.288_{-0.043}^{+0.040}$ \\
  $\omega$ [deg] & \multicolumn{2}{c}{$231.3_{-5.6}^{+4.9}$} & $197.2_{-1.5}^{+1.3}$ & \multicolumn{2}{c}{$204_{-9}^{+15}$} & $198.9_{-1.8}^{+2.0}$ \\ 
  $i$ [deg] & \multicolumn{2}{c}{$88.36_{-0.63}^{+0.28}$} & $88.544_{-0.042}^{+0.038}$ & \multicolumn{2}{c}{$89.50_{-0.85}^{+0.39}$} & $88.578_{-0.035}^{+0.035}$ \\
  $\Omega$ [deg] & \multicolumn{2}{c}{$0.0$} & $0.24_{-0.41}^{+0.50}$ & \multicolumn{2}{c}{$0.0$} & $-1.06_{-0.46}^{+0.71}$ \\
  $\gamma$ [km\,s$^-1$] & \multicolumn{3}{c}{$-4.93_{-0.20}^{+0.21}$} & \multicolumn{3}{c}{$-$} \\
   \hline  
\multicolumn{7}{c}{stellar parameters} \\
\hline
   & Aa & Ab &  B & Aa & Ab &  B \\
  \hline
 \multicolumn{7}{c}{Physical Quantities} \\
  \hline 
 $m$ [M$_\odot$] & $1.468_{-0.013}^{+0.011}$ & $1.106_{-0.014}^{+0.011}$ & $1.913_{-0.028}^{+0.026}$ & $1.464_{-0.015}^{+0.010}$ & $1.115_{-0.014}^{+0.010}$ & $1.939_{-0.056}^{+0.035}$ \\
 $R$ [R$_\odot$] & $1.666_{-0.022}^{+0.017}$ & $1.047_{-0.014}^{+0.018}$ & $4.947_{-0.072}^{+0.085}$ & $1.653_{-0.017}^{+0.017}$ & $1.055_{-0.013}^{+0.012}$ & $4.940_{-0.084}^{+0.066}$ \\
 $T_\mathrm{eff}$ [K] & $6870_{-123}^{+191}$ & $6004_{-93}^{+158}$ & $5944_{-72}^{+168}$ & $6709_{-138}^{+263}$ & $5924_{-114}^{+176}$ & $5920_{-120}^{+142}$ \\
 \hline
\multicolumn{7}{c}{Global system parameters} \\
  \hline
$\log$(age) [dex] & \multicolumn{3}{c}{$9.141_{-0.015}^{+0.007}$} & \multicolumn{3}{c}{$9.144_{-0.019}^{+0.021}$} \\
distance [pc]         & \multicolumn{3}{c}{$1609_{-25}^{+33}$} & \multicolumn{3}{c}{$1609_{-24}^{+23}$} \\  
\hline
\end{tabular}

\end{table*}

In Table  \ref{tab:syntheticfit_T456194776comp} we compare the photodynamical fits for TIC 4561944776 using both the MDN and MDR models.  Recall that in the latter we add the RV points to the analysis in addition to the photometry, ETV points, SED data, and the use of stellar evolution tracks and model atmospheres.  Of all the parameters that we compute, we limit the ones that are compared in the Table to only 5 orbital, 3 stellar, and 2 global system parameters, as illustrative and representative.  This comparison serves as a direct `calibration' as to how well we can do without the RVs.  All of the stellar and global parameters agree to within 1 mutual $\sigma$ of the two solutions.  Likewise, among the orbital elements, the $a$, $i$, and $\Omega$ values agree to within just somewhat more than 1 $\sigma$.  The two most interesting differences are in $\omega_{\rm in}$ and the eccentricities.  In particular the two models differ in $e_{\rm in}$ by 1.7 $\sigma$.  For $e_{\rm out}$, where the RV data points help most directly, the two model values differ by only 1/2 $\sigma$.  We see that the two values of $\omega_{\rm in}$ differ by 1.7 $\sigma$. Thus, overall, we find the models with and without the use of RVs to be in quite substantial agreement. Finally, it is interesting to note, but hardly surprising, that the one parameter where the error bars shrank considerably is for the outer eccentricity.

\clearpage

\section{Closer look at the distances}
\label{sec:lightcentroid}

Here we attempt to compute the contribution to the uncertainty in Gaia's parallax measurement due to the motion of the triple's center of light (`col') as the stars move around in their orbits.  Since all three stars are completely unresolved by Gaia's optics, we consider only the motion of the center of light around the triple's center of mass.  We treat the inner eclipsing binary as a single point source of light, and the tertiary as a second displaced light source in the system.  The location of the center of light is taken simply to be:
\begin{eqnarray}
\vec{r}_{\rm col}(t) & = & \frac{\mathcal{R}_B(t) L_B - \mathcal{R}_A(t) L_A}{L_A+L_B} \, \hat{r}  \\
& = & \frac{(M_A L_B - M_B L_A)}{(M_A+M_B)(L_A+L_B)} \, \mathcal{R}(t) \hat{r}
\end{eqnarray}
where $\mathcal{R}_B(t)$ and $\mathcal{R}_A(t)$ are the distances from the triple's center of mass to the tertiary star (B) and binary (A, center of mass), respectively, $L_B$ and $L_A$ are the luminosity of the tertiary and of the inner binary, respectively, and similarly for the masses $M_B$ and $M_A$.  For simplicity, here we consider only the bolometric luminosities.  The vector $\hat{r}$ is the unit vector pointing from the system center of mass to the tertiary star, and projected onto the plane of the sky.  In turn, $\vec{\mathcal{R}}(t) \equiv \mathcal{R}(t) \hat{r}$ describes the ordinary Keplerian motion of the outer orbit of the triple system.  The masses, luminosities, and orbital parameters are given in Tables \ref{tab:syntheticfit_TIC37743815+42565581}, \ref{tab:syntheticfit_TIC54060695+178010808}, and  \ref{tab:syntheticfit_TIC242132789+456194776}.

For each of our six triples, we used this prescription to compute the semi-major axis of the center of light as it orbits the center of mass of the triple system.  To generate the motion on the sky as a function of time, we used the orbital parameters for the triples given in the Tables listed above.  Because the orbits of all the triples are practically flat, and viewed nearly edge on, we simply took the motion to lie along a line on the sky.  We do not know the position angle of the orbit projected on the plane of the sky, and so we used an illustrative angle of 45$^\circ$, though after trying several different angles we realized the fact that our results are completely independent of this choice.

In Table \ref{tbl:lightcentroid} we give in columns 5, 6, and 7, the size of the semi-major axis of the center of light in micro-arcseconds, the rms deviation from the parallactic ellipse that the orbit produces, and the error that this is likely to introduce into Gaia's measurement of the parallax.  For the latter we simply use the rms deviation divided by the square root of the number of measurements Gaia makes over the 34 month duration of the Gaia eDR3 data set minus the number of astrometric fitted parameters (6).  The number of measurements is listed in the last column of Table \ref{tbl:lightcentroid} as astrometric\_matched\_transits.  We note, though, that the exact value of the distance uncertainty introduced by the motion of the center of light is also dependent on how the Gaia sampling (every few weeks) `beats' up with the outer orbital period of the triple\footnote{In this regard, for example, we note that the outer period of TIC 54060695 (60.8 days) is very close to 1/6 of a year.}.  

\begin{table*}
\caption{Details of the Distance Determinations}
 \label{tbl:lightcentroid}
\begin{tabular}{@{}lcccccccccc}
\hline
Target & Distance  & Distance & $\pi$ error$^a$ & a(col)$^b$ & rms(col)$^c$ & error(col)$^d$ & $\epsilon_i$$^e$ & $D$$^f$ & RUWE$^g$  & matched \\ 
 & This work (pc) &  Gaia (pc) & $\mu$as & $\mu$as & $\mu$as & $\mu$as & $\mu$as & ... & ... & transits  \\ 
\hline
37743815 &  $1789 \pm 78$ & $1857 \pm 39$ & 11 & 119 & 104 & 15 & 35 & 1.20 & 1.09 & 54 \\ 
42565581 &  $3150 \pm 150$ & $3281 \pm 160$ & 15 & 63 & 47 & 11 & 39 & 1.04 & 1.17 & 24 \\ 
54060695 &  $2427 \pm 34$ & $2221 \pm 50$ & 10 & 80  & 57 & 7 & 37 & 2.26 & 0.94 & 67 \\ 
178010808 & $1415 \pm 22$ & $1464\pm 30$ & 13 & 90 & 73 & 10 & 56  & 5.14 & 1.04 & 59 \\ 
242132789 & $2667 \pm 28$ & $3258 \pm 165$ & 16 & 73 & 52 & 9 & 32 & 0.82 & 1.10 & 37 \\ 
456194776 & $1690 \pm 24$ & $1590 \pm 40$ & 16 & 159 & 130 & 20 & 52 & 4.46 & 0.95 & 49 \\ 
\hline
\end{tabular}

\textit{Notes. } (a) Gaia uncertainty in the parallax (parallax\_error).  (b) Semimajor axis of the triple's center of light (`col') expressed in micro-arc seconds. (c) RMS fluctuations due to the triple's center of light motion. (d) Error contribution due to the triple's center of light motion (see text). (e) Gaia's astrometric\_excess\_noise. (f) Gaia's astrometric\_excess\_noise\_significance. (g) Gaia's renormalized unit weight error -- RUWE parameter.  (h)  The number of astrometric\_matched\_transits.  
\end{table*}

The first four columns of Table \ref{tbl:lightcentroid} are the TIC number of the triple system, the distance determined in this work as part of the photodynamical solution, the distance determined by Gaia, and the cited uncertainty in Gaia's parallax.  Columns 8, 9, and 10 in Table \ref{tbl:lightcentroid} are several Gaia measures of how well the astrometric solution fits the observations.  The parameter $\epsilon_i$ is the astrometric excess noise, which Gaia says ``measures the disagreement, expressed as an angle, between the observations of a source and the best-fitting standard astrometric model''.  The parameter $D$ is ``a dimensionless measure of the significance of the calculated astrometric\_excess\_noise ($\epsilon_i$). A value $D \gtrsim 2$ indicates that the given ($\epsilon_i$)  is probably significant.'' The parameter `RUWE' is the `renormalised unit weight error', and if large enough is sometimes taken as an indication that the source being observed consists of multiple stars.  Finally, the last column gives the astrometric\_matched\_transits, i.e., the number of astrometric visits to the target.

From a perusal of Table \ref{tbl:lightcentroid}, first we see that no value of RUWE substantially exceeds unity, indicating that the Gaia astrometric solution shows no real indication of stellar multiplicity.  Second, for the three sources which show elevated values of $D$, indicating a somewhat significant value of the astrometric\_excess\_noise parameter, the Gaia distance and our distance differ by only 3\%-9\% out of $\sim$2 kpc.  Finally, we see that the expected uncertainties introduced by the light centroid motions within the triple are typically factors of a few smaller than the cited astrometric\_excess\_noise and just comparable with the cited parallax error.  Therefore, we conclude that the motions of the center of light within the triple systems are just marginally at the level of affecting the distance measurements.  However, all the evidence (see Table \ref{tbl:lightcentroid}) suggests that the Gaia distances are not substantially affected by internal light centroid motions for our set of six sources.

The bottom line is that we generate our own independent distance measurements found as part of our photodynamical solutions.  These are generally in fine agreement with those of Gaia, however, our claimed photometric distance uncertainties are smaller than those that Gaia reports, with no reason not to believe our photodynamic results.

Interestingly, a perusal of the column in Table \ref{tbl:lightcentroid} giving the rms motions of the center of light in these systems shows that they are all in the range of 47-130 $\mu$as.  These are eminently detectable as `orbits' with Gaia in their future analyses.

\clearpage
\onecolumn

\section{Tables of determined eclipse times for all six systems}
\label{app:ToMs}

In this appendix, we tabulate the individual mid-minima times of the primary and secondary eclipses for the inner EBs of the triples considered in this study (Tables C1-C6).

\begin{table*}
\caption{Eclipse Times of TIC 37743815}
 \label{Tab:TIC_037743815_ToM}
\begin{tabular}{@{}lrllrllrl}
\hline
BJD & Cycle  & std. dev. & BJD & Cycle  & std. dev. & BJD & Cycle  & std. dev. \\ 
$-2\,400\,000$ & no. &   \multicolumn{1}{c}{$(d)$} & $-2\,400\,000$ & no. &   \multicolumn{1}{c}{$(d)$} & $-2\,400\,000$ & no. &   \multicolumn{1}{c}{$(d)$} \\ 
\hline
58469.100180 &    0.0 & 0.017417 & 58485.427452 &   18.0 & 0.000927 & 59211.083151 &  818.0 & 0.000716 \\ 
58470.005584 &    1.0 & 0.000922 & 58486.334080 &   19.0 & 0.001151 & 59211.992122 &  819.0 & 0.001195 \\ 
58470.912387 &    2.0 & 0.001245 & 58487.244723 &   20.0 & 0.000905 & 59212.899021 &  820.0 & 0.001022 \\ 
58471.820704 &    3.0 & 0.001031 & 58488.148083 &   21.0 & 0.001237 & 59213.807149 &  821.0 & 0.000824 \\ 
58472.726629 &    4.0 & 0.001354 & 58489.057499 &   22.0 & 0.000869 & 59215.619239 &  823.0 & 0.000679 \\ 
58473.638417 &    5.0 & 0.000964 & 58489.966834 &   23.0 & 0.001234 & 59216.525955 &  824.0 & 0.000809 \\ 
58474.542843 &    6.0 & 0.001260 & 59202.013203 &  808.0 & 0.000802 & 59217.431905 &  825.0 & 0.000796 \\ 
58475.448678 &    7.0 & 0.001309 & 59202.920768 &  809.0 & 0.000646 & 59218.340348 &  826.0 & 0.000910 \\ 
58476.354330 &    8.0 & 0.001386 & 59203.829011 &  810.0 & 0.000844 & 59219.247375 &  827.0 & 0.000768 \\ 
58478.169364 &   10.0 & 0.001328 & 59204.732837 &  811.0 & 0.001225 & 59220.153927 &  828.0 & 0.000794 \\ 
58479.078449 &   11.0 & 0.000895 & 59205.641488 &  812.0 & 0.000816 & 59221.062602 &  829.0 & 0.000676 \\ 
58479.983054 &   12.0 & 0.001364 & 59206.549341 &  813.0 & 0.000804 & 59221.968963 &  830.0 & 0.000851 \\ 
58480.891648 &   13.0 & 0.001021 & 59207.454035 &  814.0 & 0.001035 & 59222.874870 &  831.0 & 0.000979 \\ 
58481.797696 &   14.0 & 0.004524 & 59208.362881 &  815.0 & 0.000830 & 59225.595290 &  834.0 & 0.000982 \\ 
58482.706871 &   15.0 & 0.001242 & 59209.270764 &  816.0 & 0.000754 & 59226.503925 &  835.0 & 0.000819 \\ 
58483.612970 &   16.0 & 0.000995 & 59210.174682 &  817.0 & 0.000816 & 59227.409135 &  836.0 & 0.000712 \\ 
58484.517480 &   17.0 & 0.000866 &&&&& \\ 
\hline
\end{tabular}
\end{table*}

\begin{table*}
\caption{Eclipse Times of TIC 42565581}
 \label{Tab:TIC_042565581_ToM}
\begin{tabular}{@{}lrllrllrl}
\hline
BJD & Cycle  & std. dev. & BJD & Cycle  & std. dev. & BJD & Cycle  & std. dev. \\ 
$-2\,400\,000$ & no. &   \multicolumn{1}{c}{$(d)$} & $-2\,400\,000$ & no. &   \multicolumn{1}{c}{$(d)$} & $-2\,400\,000$ & no. &   \multicolumn{1}{c}{$(d)$} \\ 
\hline
58469.003253 &    0.0 & 0.000269 & 58486.335932 &    9.5 & 0.000267 & 59210.992711 &  407.0 & 0.000191 \\ 
58469.928359 &    0.5 & 0.000301 & 58487.234362 &   10.0 & 0.000299 & 59211.917368 &  407.5 & 0.000215 \\ 
58470.826273 &    1.0 & 0.000275 & 58488.157862 &   10.5 & 0.000259 & 59212.816157 &  408.0 & 0.000232 \\ 
58471.751681 &    1.5 & 0.000243 & 58489.057065 &   11.0 & 0.000337 & 59213.739667 &  408.5 & 0.000181 \\ 
58472.649068 &    2.0 & 0.000331 & 58489.981803 &   11.5 & 0.000265 & 59215.566622 &  409.5 & 0.000189 \\ 
58473.573118 &    2.5 & 0.000234 & 59201.881711 &  402.0 & 0.000191 & 59216.464242 &  410.0 & 0.000222 \\ 
58474.472161 &    3.0 & 0.000277 & 59202.802591 &  402.5 & 0.000198 & 59217.386432 &  410.5 & 0.000176 \\ 
58475.397381 &    3.5 & 0.000351 & 59203.701040 &  403.0 & 0.000133 & 59221.932759 &  413.0 & 0.000217 \\ 
58476.294304 &    4.0 & 0.000251 & 59204.626061 &  403.5 & 0.000239 & 59222.856814 &  413.5 & 0.000190 \\ 
58480.866051 &    6.5 & 0.000254 & 59205.525446 &  404.0 & 0.000151 & 59223.755438 &  414.0 & 0.000186 \\ 
58481.765341 &    7.0 & 0.000410 & 59206.447958 &  404.5 & 0.000160 & 59224.678757 &  414.5 & 0.000226 \\ 
58482.688807 &    7.5 & 0.000354 & 59207.347944 &  405.0 & 0.000198 & 59225.577908 &  415.0 & 0.000209 \\ 
58483.587189 &    8.0 & 0.000355 & 59208.271297 &  405.5 & 0.000221 & 59226.501008 &  415.5 & 0.000219 \\ 
58484.512607 &    8.5 & 0.000290 & 59209.169805 &  406.0 & 0.000183 & 59227.402701 &  416.0 & 0.000182 \\ 
58485.410669 &    9.0 & 0.000522 & 59210.094530 &  406.5 & 0.000197 &&& \\ 
\hline
\end{tabular}
\end{table*}

\begin{table*}
\caption{Eclipse Times of TIC 54060695}
 \label{Tab:TIC_054060695_ToM}
\begin{tabular}{@{}lrllrllrl}
\hline
BJD & Cycle  & std. dev. & BJD & Cycle  & std. dev. & BJD & Cycle  & std. dev. \\ 
$-2\,400\,000$ & no. &   \multicolumn{1}{c}{$(d)$} & $-2\,400\,000$ & no. &   \multicolumn{1}{c}{$(d)$} & $-2\,400\,000$ & no. &   \multicolumn{1}{c}{$(d)$} \\ 
\hline
58468.605940 &  -17.0 & 0.001095 & 58494.058064 &    7.0 & 0.003842 & 59205.647534 &  678.0 & 0.000588 \\ 
58469.140378 &  -16.5 & 0.005653 & 58494.590093 &    7.5 & 0.003129 & 59206.179379 &  678.5 & 0.002318 \\ 
58469.667026 &  -16.0 & 0.001367 & 58495.119416 &    8.0 & 0.000982 & 59206.706210 &  679.0 & 0.000586 \\ 
58470.197503 &  -15.5 & 0.002369 & 58495.648973 &    8.5 & 0.004338 & 59207.238507 &  679.5 & 0.001438 \\ 
58470.725605 &  -15.0 & 0.001084 & 58496.180311 &    9.0 & 0.001085 & 59207.767913 &  680.0 & 0.000569 \\ 
58471.258699 &  -14.5 & 0.002208 & 58496.715667 &    9.5 & 0.003349 & 59208.297264 &  680.5 & 0.001070 \\ 
58471.785257 &  -14.0 & 0.000656 & 58497.241219 &   10.0 & 0.001137 & 59208.828671 &  681.0 & 0.000674 \\ 
58472.315201 &  -13.5 & 0.002350 & 58497.772757 &   10.5 & 0.006593 & 59209.358054 &  681.5 & 0.001747 \\ 
58472.848220 &  -13.0 & 0.001029 & 58498.302567 &   11.0 & 0.001250 & 59209.888566 &  682.0 & 0.000546 \\ 
58473.376212 &  -12.5 & 0.002832 & 58498.830635 &   11.5 & 0.004816 & 59210.419743 &  682.5 & 0.001637 \\ 
58476.028160 &  -10.0 & 0.001878 & 58499.362499 &   12.0 & 0.001201 & 59210.950250 &  683.0 & 0.000525 \\ 
58476.559043 &   -9.5 & 0.002884 & 58499.894586 &   12.5 & 0.009955 & 59211.477981 &  683.5 & 0.002378 \\ 
58478.150295 &   -8.0 & 0.001785 & 58500.422492 &   13.0 & 0.001560 & 59212.009490 &  684.0 & 0.000576 \\ 
58478.682432 &   -7.5 & 0.003870 & 58500.955348 &   13.5 & 0.004115 & 59212.540121 &  684.5 & 0.001944 \\ 
58479.208497 &   -7.0 & 0.000615 & 58501.484125 &   14.0 & 0.000766 & 59213.070938 &  685.0 & 0.000548 \\ 
58479.740703 &   -6.5 & 0.001954 & 58502.012980 &   14.5 & 0.004056 & 59213.600270 &  685.5 & 0.001494 \\ 
58480.271792 &   -6.0 & 0.000750 & 58502.542231 &   15.0 & 0.001317 & 59215.721562 &  687.5 & 0.001951 \\ 
58480.801877 &   -5.5 & 0.001508 & 58506.260908 &   18.5 & 0.003531 & 59216.252049 &  688.0 & 0.000451 \\ 
58481.331202 &   -5.0 & 0.003872 & 58506.785416 &   19.0 & 0.000886 & 59216.784249 &  688.5 & 0.002169 \\ 
58481.863746 &   -4.5 & 0.002555 & 58507.317745 &   19.5 & 0.003490 & 59217.312569 &  689.0 & 0.000477 \\ 
58482.391152 &   -4.0 & 0.001007 & 58507.844674 &   20.0 & 0.001630 & 59217.841138 &  689.5 & 0.001451 \\ 
58482.930523 &   -3.5 & 0.003879 & 58508.378102 &   20.5 & 0.002051 & 59218.373682 &  690.0 & 0.000477 \\ 
58483.452838 &   -3.0 & 0.000904 & 58508.907052 &   21.0 & 0.000556 & 59218.907283 &  690.5 & 0.001287 \\ 
58483.984392 &   -2.5 & 0.002246 & 58509.435525 &   21.5 & 0.001646 & 59219.434531 &  691.0 & 0.000540 \\ 
58484.513781 &   -2.0 & 0.001109 & 58509.967497 &   22.0 & 0.000608 & 59219.964043 &  691.5 & 0.003971 \\ 
58485.044984 &   -1.5 & 0.011035 & 58510.498675 &   22.5 & 0.001886 & 59220.496458 &  692.0 & 0.000494 \\ 
58485.574971 &   -1.0 & 0.000736 & 58511.028526 &   23.0 & 0.015366 & 59221.027540 &  692.5 & 0.001712 \\ 
58486.107091 &   -0.5 & 0.003441 & 58511.557418 &   23.5 & 0.002134 & 59221.555636 &  693.0 & 0.000521 \\ 
58486.634619 &    0.0 & 0.000975 & 58512.087572 &   24.0 & 0.000831 & 59222.086269 &  693.5 & 0.001093 \\ 
58487.167866 &    0.5 & 0.003574 & 58512.616997 &   24.5 & 0.003283 & 59222.615951 &  694.0 & 0.000443 \\ 
58487.697342 &    1.0 & 0.001174 & 58513.148992 &   25.0 & 0.001401 & 59223.146045 &  694.5 & 0.003357 \\ 
58488.231303 &    1.5 & 0.003585 & 58513.677824 &   25.5 & 0.005850 & 59223.676496 &  695.0 & 0.000562 \\ 
58488.757146 &    2.0 & 0.000743 & 58514.209063 &   26.0 & 0.001968 & 59224.206998 &  695.5 & 0.002341 \\ 
58489.282782 &    2.5 & 0.002575 & 58514.741532 &   26.5 & 0.002921 & 59224.737511 &  696.0 & 0.000539 \\ 
58489.817359 &    3.0 & 0.000658 & 58515.268752 &   27.0 & 0.001399 & 59225.270801 &  696.5 & 0.001181 \\ 
58491.937855 &    5.0 & 0.001191 & 58515.801427 &   27.5 & 0.005208 & 59225.798902 &  697.0 & 0.000578 \\ 
58492.471299 &    5.5 & 0.001919 & 59204.060043 &  676.5 & 0.002236 & 59226.326089 &  697.5 & 0.001719 \\ 
58492.998429 &    6.0 & 0.001038 & 59204.584860 &  677.0 & 0.000512 & 59226.857789 &  698.0 & 0.000477 \\ 
58493.528706 &    6.5 & 0.006569 & 59205.118376 &  677.5 & 0.001569 & 59227.389174 &  698.5 & 0.001343 \\ 
\hline
\end{tabular}
\end{table*}

\begin{table*}
\caption{Eclipse Times of TIC 178010808}
 \label{Tab:TIC_178010808_ToM}
\begin{tabular}{@{}lrllrllrl}
\hline
BJD & Cycle  & std. dev. & BJD & Cycle  & std. dev. & BJD & Cycle  & std. dev. \\ 
$-2\,400\,000$ & no. &   \multicolumn{1}{c}{$(d)$} & $-2\,400\,000$ & no. &   \multicolumn{1}{c}{$(d)$} & $-2\,400\,000$ & no. &   \multicolumn{1}{c}{$(d)$} \\ 
\hline
58492.073256 &    0.0 & 0.000530 & 58508.999380 &   20.5 & 0.000173 & 59237.301079 &  902.5 & 0.000122 \\ 
58492.486382 &    0.5 & 0.000444 & 58509.412402 &   21.0 & 0.000182 & 59237.713594 &  903.0 & 0.000113 \\ 
58492.898591 &    1.0 & 0.000428 & 58509.825647 &   21.5 & 0.000180 & 59238.126559 &  903.5 & 0.000106 \\ 
58493.312092 &    1.5 & 0.000442 & 58510.238029 &   22.0 & 0.000250 & 59238.539138 &  904.0 & 0.000120 \\ 
58493.724843 &    2.0 & 0.000418 & 58510.650991 &   22.5 & 0.000185 & 59238.952447 &  904.5 & 0.000108 \\ 
58494.138138 &    2.5 & 0.000468 & 58511.063326 &   23.0 & 0.000173 & 59239.365005 &  905.0 & 0.000105 \\ 
58494.550590 &    3.0 & 0.000438 & 58511.476705 &   23.5 & 0.000185 & 59239.778311 &  905.5 & 0.000138 \\ 
58494.963278 &    3.5 & 0.000427 & 58511.889525 &   24.0 & 0.000203 & 59240.190499 &  906.0 & 0.000115 \\ 
58495.376087 &    4.0 & 0.000418 & 58512.303034 &   24.5 & 0.000284 & 59240.604009 &  906.5 & 0.000136 \\ 
58495.788552 &    4.5 & 0.000385 & 58513.128904 &   25.5 & 0.000591 & 59242.253736 &  908.5 & 0.000663 \\ 
58496.201972 &    5.0 & 0.000364 & 58513.540884 &   26.0 & 0.000258 & 59242.667866 &  909.0 & 0.000208 \\ 
58496.614783 &    5.5 & 0.000375 & 58513.953829 &   26.5 & 0.000315 & 59243.081154 &  909.5 & 0.000205 \\ 
58497.027188 &    6.0 & 0.000319 & 58514.367120 &   27.0 & 0.000291 & 59243.493740 &  910.0 & 0.000206 \\
58497.440269 &    6.5 & 0.000421 & 58514.780110 &   27.5 & 0.000308 & 59244.319388 &  911.0 & 0.000138 \\ 
58497.852818 &    7.0 & 0.000286 & 58515.192856 &   28.0 & 0.000288 & 59244.732030 &  911.5 & 0.000150 \\ 
58498.266191 &    7.5 & 0.000291 & 58515.605843 &   28.5 & 0.000348 & 59245.145086 &  912.0 & 0.000124 \\ 
58498.679015 &    8.0 & 0.000260 & 58516.018403 &   29.0 & 0.000479 & 59245.557762 &  912.5 & 0.000124 \\ 
58499.091466 &    8.5 & 0.000334 & 59229.043614 &  892.5 & 0.000326 & 59245.970826 &  913.0 & 0.000122 \\ 
58499.503950 &    9.0 & 0.000337 & 59229.456515 &  893.0 & 0.000185 & 59246.383649 &  913.5 & 0.000158 \\ 
58499.917184 &    9.5 & 0.000470 & 59229.869065 &  893.5 & 0.000198 & 59246.796734 &  914.0 & 0.000132 \\ 
58500.329864 &   10.0 & 0.000232 & 59230.282389 &  894.0 & 0.000156 & 59247.209664 &  914.5 & 0.000173 \\ 
58500.742488 &   10.5 & 0.000299 & 59230.695285 &  894.5 & 0.000179 & 59247.622092 &  915.0 & 0.000159 \\ 
58501.155265 &   11.0 & 0.000342 & 59231.108066 &  895.0 & 0.000177 & 59248.034725 &  915.5 & 0.000214 \\ 
58501.568457 &   11.5 & 0.000229 & 59231.520912 &  895.5 & 0.000178 & 59248.447843 &  916.0 & 0.000164 \\ 
58501.981398 &   12.0 & 0.000379 & 59231.933777 &  896.0 & 0.000164 & 59248.860334 &  916.5 & 0.000177 \\ 
58502.394085 &   12.5 & 0.000288 & 59232.346900 &  896.5 & 0.000162 & 59249.273465 &  917.0 & 0.000205 \\ 
58502.806852 &   13.0 & 0.000206 & 59232.759717 &  897.0 & 0.000152 & 59249.686580 &  917.5 & 0.000229 \\ 
58504.870813 &   15.5 & 0.000309 & 59233.172386 &  897.5 & 0.000157 & 59250.099235 &  918.0 & 0.000185 \\ 
58505.283323 &   16.0 & 0.000199 & 59233.584727 &  898.0 & 0.000136 & 59250.512084 &  918.5 & 0.000209 \\ 
58505.696737 &   16.5 & 0.000183 & 59233.998317 &  898.5 & 0.000140 & 59250.924619 &  919.0 & 0.000227 \\ 
58506.109155 &   17.0 & 0.000115 & 59234.410873 &  899.0 & 0.000138 & 59251.337762 &  919.5 & 0.000200 \\ 
58506.522169 &   17.5 & 0.000172 & 59234.823729 &  899.5 & 0.000178 & 59251.750491 &  920.0 & 0.000208 \\ 
58506.934614 &   18.0 & 0.000119 & 59235.236549 &  900.0 & 0.000127 & 59252.163121 &  920.5 & 0.000234 \\ 
58507.347889 &   18.5 & 0.000220 & 59235.649505 &  900.5 & 0.000143 & 59252.576219 &  921.0 & 0.000220 \\ 
58507.760879 &   19.0 & 0.000144 & 59236.062234 &  901.0 & 0.000122 & 59252.988873 &  921.5 & 0.000261 \\ 
58508.173585 &   19.5 & 0.000181 & 59236.475241 &  901.5 & 0.000138 & 59253.401911 &  922.0 & 0.000244 \\ 
58508.586585 &   20.0 & 0.000282 & 59236.888391 &  902.0 & 0.000115 & 59253.814624 &  922.5 & 0.000293 \\ 
\hline
\end{tabular}
\end{table*}

\begin{table*}
\caption{Eclipse Times of TIC 242132789}
 \label{Tab:TIC_242132789_ToM}
\begin{tabular}{@{}lrllrllrl}
\hline
BJD & Cycle  & std. dev. & BJD & Cycle  & std. dev. & BJD & Cycle  & std. dev. \\ 
$-2\,400\,000$ & no. &   \multicolumn{1}{c}{$(d)$} & $-2\,400\,000$ & no. &   \multicolumn{1}{c}{$(d)$} & $-2\,400\,000$ & no. &   \multicolumn{1}{c}{$(d)$} \\ 
\hline
58470.130373 & -143.5 & 0.001382 & 58488.030554 & -140.0 & 0.001011 & 59219.419094 &    3.0 & 0.000797 \\ 
58472.684648 & -143.0 & 0.000789 & 59204.057046 &    0.0 & 0.000937 & 59221.968075 &    3.5 & 0.001003 \\ 
58475.235600 & -142.5 & 0.001309 & 59206.617194 &    0.5 & 0.000948 & 59224.520611 &    4.0 & 0.000908 \\ 
58480.343817 & -141.5 & 0.000752 & 59209.178049 &    1.0 & 0.001010 & 59227.074177 &    4.5 & 0.001134 \\ 
58482.906986 & -141.0 & 0.000809 & 59216.861384 &    2.5 & 0.001011 &&& \\ 
\hline
\end{tabular}
\end{table*}

\begin{table*}
\caption{Eclipse Times of TIC 456194776}
 \label{Tab:TIC_456194776_ToM}
\begin{tabular}{@{}lrllrllrl}
\hline
BJD & Cycle  & std. dev. & BJD & Cycle  & std. dev. & BJD & Cycle  & std. dev. \\ 
$-2\,400\,000$ & no. &   \multicolumn{1}{c}{$(d)$} & $-2\,400\,000$ & no. &   \multicolumn{1}{c}{$(d)$} & $-2\,400\,000$ & no. &   \multicolumn{1}{c}{$(d)$} \\ 
\hline
58790.694887 &   -0.5 & 0.007130 & 58803.585185 &    7.0 & 0.000549 & 58813.901513 &   13.0 & 0.000773 \\
58791.550543 &    0.0 & 0.002673 & 58804.448028 &    7.5 & 0.000535 & 58814.766254 &   13.5 & 0.009046 \\ 
58792.414562 &    0.5 & 0.002784 & 58805.303613 &    8.0 & 0.000447 & 59098.452190 &  178.5 & 0.000268 \\
58793.270358 &    1.0 & 0.001071 & 58806.166199 &    8.5 & 0.000500 & 59104.469630 &  182.0 & 0.000269 \\ 
58794.133511 &    1.5 & 0.001933 & 58807.024167 &    9.0 & 0.000422 & 59116.511643 &  189.0 & 0.000049 \\ 
58794.987072 &    2.0 & 0.002167 & 58807.888026 &    9.5 & 0.001142 & 59159.493484 &  214.0 & 0.000048 \\
58795.852371 &    2.5 & 0.001374 & 58808.742542 &   10.0 & 0.000294 & 59168.946719 &  219.5 & 0.000157 \\ 
58796.708543 &    3.0 & 0.001245 & 58810.463083 &   11.0 & 0.000740 & 59276.410632 &  282.0 & 0.000140 \\ 
58797.570872 &    3.5 & 0.001705 & 58811.324887 &   11.5 & 0.001979 & 59515.405227 &  421.0 & 0.000097 \\  
58798.426856 &    4.0 & 0.000800 & 58812.181601 &   12.0 & 0.000487 & 59527.433741 &  428.0 & 0.000100 \\  
58799.290589 &    4.5 & 0.001175 & 58813.045175 &   12.5 & 0.001882 & 59539.466978 &  435.0 & 0.000116 \\  
58800.146908 &    5.0 & 0.000683 &&&&&&\\
\hline
\end{tabular}

\textit{Notes.} Eclipses between cycle numbers $-0.5$ and $13.5$ was observed with \textit{TESS}, while the last 9 events were observed in the frame of our ground-based follow up campaign.

\end{table*}

\end{document}